\documentclass[11pt, a4paper]{article}
\usepackage{styleBuding}
\usepackage{bm}
\usepackage{color}
\usepackage[usenames,dvipsnames,svgnames,table]{xcolor}
\usepackage{amsmath}
\usepackage{amssymb}
\usepackage{graphicx}
\usepackage{adjustbox}
\usepackage{slashed}
\usepackage{soul}
\usepackage{multirow}
\usepackage{subfigure}
\usepackage{indentfirst}
\usepackage{siunitx} 
\usepackage{longtable}
\usepackage{caption}
\usepackage[utf8]{inputenc}

\pdfoutput=1

\begin{document}

\title{\boldmath 95 GeV excesses in the $\mathbb{Z}_3$-symmetric next-to minimal supersymmetric standard model}

\author[1,2]{Jingwei Lian}

\affiliation[1]{Postdoctoral Innovation and Development Station, Henan Institute of Science and Technology, Xinxiang 453003, China}
\affiliation[2]{Department of Physics, Henan Normal University, Xinxiang 453007, China}


\emailAdd{lianjw@hist.edu.cn}

\abstract{
Recent analyses by CMS and ATLAS suggest a deviation in the di-photon channel at approximately 95 GeV, alongside a previously observed excess in $b\bar{b}$ signals at a similar mass by LEP, potentially hinting at a new scalar particle.
This study explores this possibility within the framework of the $\mathbb{Z}_3$-symmetric Next-to-Minimal Supersymmetric Standard Model. A comprehensive parameter scan was conducted, integrating constraints from dark matter relic density, direct detection experiments, and the properties of the observed 125 GeV Higgs boson. The results demonstrate that the model can accommodate the observed excesses with a singlet-dominated CP-even scalar boson near 95 GeV. The model accurately predicts signal strengths of the di-photon and $b\bar{b}$ channels at a level of $1\sigma$. Furthermore, it accounts for the measured dark matter relic abundance through Bino-dominated neutralinos co-annihilation with Wino-like electroweakinos, all while remaining consistent with existing LHC constraints. These findings pave the way for future validation at the high-luminosity LHC and linear colliders, which may offer crucial tests of the model's predictions. }

\maketitle

\section{Introduction\label{sec:intro}}
The Higgs boson plays a pivotal role in the Standard Model (SM) of particle physics,  serving as the primary agent responsible for electroweak symmetry breaking (EWSB). It is intimately connected to a multitude of fundamental inquiries concerning the nature of the Universe. Beyond the SM (BSM), various well-motivated theories necessitate expanded scalar sectors to grapple with unresolved issues inherent to the SM, such as neutrino mass, dark matter (DM) origin and baryon number asymmetries. These extended scalar sectors, while offering potential resolutions to existing SM enigmas, may concurrently introduce notable disparities compared to SM predictions. Following the landmark discovery of the Higgs boson with a mass of approximately 125 GeV at the Large Hadron Collider (LHC) by the ATLAS and CMS collaborations~\cite{ATLAS:2012yve, CMS:2012qbp}, the relentless search for additional scalar bosons has been a primary objective of the collider's scientific agenda.  Although conclusive evidence of BSM remains elusive, the persisting uncertainty within the measurements of Higgs boson couplings, fluctuating by less than approximately 20\%, affords an intriguing margin for the exploration of BSM explanations~\cite{ATLAS:2019nkf}.  

In the pursuit of new scalar bosons at the LHC, the di-photon signal stands prominently as the ``golden channel", esteemed for its capacity to provide precise measurements and clean experimental signatures. In a recent update, the CMS collaboration has released their latest analysis focused on the search for a light Higgs boson. This investigation reaffirms a previously reported excess observed at $m_{\gamma\gamma} = 95.4$ GeV with a local significance of 2.9$\sigma$ by using advanced analysis techniques to 13 TeV data amassed over the first, second, and third years of Run-2~\cite{CMS:2023yay}.
Besides, the ATLAS collaboration has also published an analysis of the full Run-2 LHC dataset amounting to 140~fb$^{-1}$ targeting the di-photon signals within the mass range of 66~GeV to 110~GeV~\cite{Arcangeletti}. This analysis uncovers an excess at an invariant mass around 95 GeV with a local significance of $1.7\sigma$. Notably, the renormalized di-photon production rates, as observed independently by both collaborations, align within uncertainties for the same mass value, suggesting the plausible emergence of the excesses from the production of a singular new particle. If it is further validated, this could signify the inaugural hint of new physics in the Higgs-boson sector. The combined signal strength yields a 3.1$\sigma$ local excess~\cite{Biekotter:2023oen}:
\begin{equation}
\mu^{\rm exp}_{\gamma\gamma} \equiv  \mu^{\rm ATLAS + CMS}_{\gamma \gamma} = \frac{\sigma(p p \to \phi \to \gamma\gamma)}{\sigma_{\rm SM} (p p \to H_{\rm SM} \to \gamma\gamma)} = 0.24^{+ 0.09}_{-0.08},  \label{diphoton-rate}
\end{equation}
where, $\phi$ represents a hypothetically non-Standard scalar with a mass of $m_\phi = 95.4~{\rm GeV}$ responsible for the di-photon excess, and $\sigma_{\rm SM}$ denotes the cross section for a hypothetical SM-like Higgs boson $H_{\rm SM}$ at the same mass. Moreover, the existence of a light $\phi$ finds additional support from the measurement of bottom quark pair events at the Large Electron-Positron (LEP) collider. The combined data hinted at the production of a scalar particle $\phi$ with an invariant mass of approximately 98 GeV through the process $e^+e^- \rightarrow Z\phi \rightarrow Z (b\bar{b})$, with a signal strength of $\mu_{b \bar b}^\text{exp} = 0.117 \pm 0.057$ and a local significance of $2.3\sigma$~\cite{LEPWorkingGroupforHiggsbosonsearches:2003ing,Azatov:2012bz,Cao:2016uwt}. 
Given the limited mass resolution for dijets at LEP, this $b\bar{b}$ excess might correspond to the same particle responsible for the di-photon excess.
Furthermore, the scalar around 95 GeV is also supported by an excess in the di-$\tau$ channel observed by CMS~\cite{CMS:2022goy} and in the $WW$ channel, as discussed in Ref.~\cite{Coloretti:2023wng}. Collectively, these observations strengthen the hypothesis that the 95 GeV scalar could be a crucial piece in the quest for new physics.

The conjecture surrounding the observed excesses has ignited a flurry of investigations into the plausibility of interpreting both phenomena within BSM theories~\cite{Ashanujjaman:2023etj,Aguilar-Saavedra:2020wrj, Kundu:2019nqo, Fox:2017uwr, Belyaev:2023xnv, Azevedo:2023zkg, Benbrik:2022dja, Benbrik:2022azi,Haisch:2017gql,Biekotter:2019mib,Biekotter:2019kde,Biekotter:2020cjs,Biekotter:2021ovi,Heinemeyer:2021msz,Biekotter:2022jyr,Li:2023hsr,Biekotter:2023oen,Biekotter:2023jld, Aguilar-Saavedra:2023tql, Banik:2023ecr,Dutta:2023cig,Sachdeva:2019hvk, Vega:2018ddp,Borah:2023hqw,Arcadi:2023smv,Ahriche:2023wkj,
Chen:2023bqr,Dev:2023kzu,Wang:2024bkg}. This intellectual fervor finds a particularly rich ground in the low-energy supersymmetric (SUSY) extensions of the SM as they provide a versatile platform for accommodating various particle sectors, allowing for nuanced examinations of their interplay and potential collective contributions to the observed phenomena~\cite{Fan:2013gjf, Cao:2016uwt, Biekotter:2017xmf, Beskidt:2017dil, Heinemeyer:2018wzl, Heinemeyer:2018jcd, Wang:2018vxp,Domingo:2018uim, Cao:2019ofo,Biekotter:2019gtq, Choi:2019yrv, Abdelalim:2020xfk,Hollik:2020plc, Biekotter:2021qbc, Li:2022etb, Ellwanger:2023zjc,Cao:2023gkc,Ahriche:2023hho,Ellwanger:2024txc,Liu:2024cbr,Ellwanger:2024vvs}.
Among these,  the Next-to-Minimal Supersymmetric Standard Model (NMSSM)~\cite{Ellwanger:2009dp} has attracted considerable attention in recent years. Distinguishing itself from the well-known Minimal Supersymmetric Standard Model (MSSM), the NMSSM augments the latter by introducing a gauge-singlet Higgs field. The allure of this model lies not only in the natural solution to the so-called $\mu$-problem inherent in the MSSM but also in its capacity to furnish more viable DM candidates~\cite{Ellwanger:2014hia,Cao:2021ljw,Cao:2019qng}. In the context of elucidating the observed excesses, especially within the scenario involving a light CP-even Higgs, the model's distinctive features contribute to a sizeable lift in the mass of the SM-like Higgs boson. This elevation is attributed to the combined effects of an additional tree-level contribution and the singlet-doublet mixing~\cite{Ellwanger:2011aa,Badziak:2013bda,Cao:2012fz}, effectively mitigating the need for substantial radiative corrections from top/stop loops. Such alleviation is crucial for aligning with the experimental measurement of the SM-like Higgs boson mass at approximately 125 GeV, thus enhancing the model's explanatory power.  

The NMSSM with a $\mathbb{Z}_3$ discrete symmetry ($\mathbb{Z}_3$-NMSSM) remains a popular framework due to its economical structure. Numerous investigations into the Higgs sector phenomenology of this model have largely focused on collider measurements without considering other experimental constraints~\cite{Heinemeyer:2018wzl, Heinemeyer:2018jcd, Domingo:2018uim, Biekotter:2021qbc, Li:2022etb, Ellwanger:2023zjc}. These studies demonstrate that the $\mathbb{Z}_3$-NMSSM can accommodate a resonant production of a light scalar boson near 95 GeV in the di-photon channel based on preceding data, as well as a possible resonance in the $b \bar{b}$ channel. Moreover, it is capable of simultaneously fitting a heavier resonance at either $\sim$ 400 GeV~\cite{Biekotter:2021qbc} or $\sim$ 650 GeV~\cite{Ellwanger:2023zjc}. However, when the excesses are analyzed in combination with DM constraints, previous works, such as~\cite{Fan:2013gjf, Cao:2016uwt, Wang:2018vxp, Ellwanger:2024txc, Ellwanger:2024vvs}, encounter difficulties in predicting both signal strengths while satisfying DM experimental bounds. For instance, the study in~\cite{Cao:2016uwt} reports that the $\mathbb{Z}_3$-NMSSM struggles to produce elevated di-photon rates at the $1\sigma$ level after accounting for correct relic density and direct detection constraints. Furthermore, recent results in~\cite{Ellwanger:2024txc} indicate challenges in reproducing the central values or $1\sigma$ ranges for $\mu_{\gamma\gamma}^{\rm exp}$ and $\mu_{b \bar{b}}^{\rm exp}$ simultaneously (see Table 4 for benchmark points and compare Fig. 1 and 2 of~\cite{Ellwanger:2024txc}).\footnote{This difficulty may arise from assumptions or sampling strategies emphasized in the present work. In addition, their findings suggest a preference for electroweakinos with masses not far above 100 GeV, which could face stringent constraints from ongoing SUSY searches at the LHC.} Besides, the study in~\cite{Ellwanger:2024vvs} explores a compressed Higgsino-like triplet scenario, which only predicts di-photon rates of less than 0.1 (see Table 4 and Fig. 2 of~\cite{Ellwanger:2024vvs}).

In essence, the $\mathbb{Z}_3$-NMSSM manifested a distinct trait by employing six input parameters for both the Higgs sector and the neutralino sector. Among these parameters, $\lambda$, $\kappa$, $\tan \beta$, and $\mu$ were shared between both sectors, engendering an inherent entanglement between the theory's Higgs and DM physics. This entanglement introduced a notable complexity, exerting an influence on the maximal reach of the di-photon signal rate.
Though efforts to disentangle the intricacies of the theoretical framework, including extensions of the $\mathbb{Z}_3$-NMSSM with seesaw mechanism~\cite{Cao:2019ofo}, and the general version of the NMSSM with $\mathbb{Z}_3$-symmetry breaking~\cite{Choi:2019yrv, Cao:2023gkc, Cao:2024axg} , have proven promising in addressing the complexity, there remains merit in conducting a comprehensive reinvestigation into the capacity of the popular $\mathbb{Z}_3$-NMSSM to fully accommodate the observed excesses while adhering to various experimental constraints. This reexamination becomes particularly pertinent in light of the decrease in the di-photon rate as indicated by recent reports and the notable improvement in sensitivities of DM direct detection experiments by more than one order of magnitude compared to prior studies. 
Through utilization of analytic expressions and a novel sampling strategy, the present study endeavors to demonstrate that the central values of both signal rates can be simultaneously accommodated in this cost-effective model.  Under the circumstances, the DM relic abundance is entirely accounted for by the Bino-like neutralinos, while the electroweakinos exhibit sufficient mass to be discernible by the LHC experiments.\footnote{Note that the analyses of DM annihilation mechanisms in~\cite{Ellwanger:2024txc} and~\cite{Ellwanger:2024vvs} focus primarily on the Singlino-like neutralino. Additionally, in~\cite{Ellwanger:2024txc}, the mass of the Bino-like lightest supersymmetric particle (LSP) is predicted to be relatively light, approximately 100 GeV. These aspects differ significantly from those explored in the present work. }

The paper is organized as follows. Sec.\ref{Section-theory} briefly recapitulates the basics of the $\mathbb{Z}_3$-NMSSM framework and the formulas of the two excesses. In Sec.\ref{Section-excess}, a sophisticated scan over the model parameter space is performed, and the numerical results are presented in both figures and tables. Finally, the conclusion and comments are made in Sec.\ref{conclusion}.

\section{Theoretical preliminaries}  \label{Section-theory}

\subsection{Basics of the $\mathbb{Z}_3$-NMSSM} \label{Section-Model}

The NMSSM is konwn as a straightforward extension of the MSSM with the inclusion of a gauge-singlet chiral superfield. Upon imposing a discrete $\mathbb{Z}_3$ symmetry, the scale-invariant superpotential and its associated soft SUSY-breaking terms of the Higgs scalars can be expressed as follows \cite{Ellwanger:2009dp}:
\begin{eqnarray} \label{NMSSMsp}
  W &=& W_{\rm Yukawa} + \lambda \hat{S} \hat{H_u} \cdot \hat{H_d}+\frac{1}{3} \kappa \hat{S}^3, \\
  -\mathcal{L}_{soft} & = &( A_{\lambda}\lambda S H_u \cdot H_d + \frac{1}{3} A_{\kappa} \kappa S^3 +h.c. )  \nonumber \\
   &&+ m^2_{H_u}|H_u|^2 + m^2_{H_d}|H_d|^2 + m^2_{s}|S|^2. \label{NMSSMsoftL}
  \end{eqnarray}
Here, $SU(2)_L$ indices are disregarded here for the sake of brevity. $W_{\rm Yukawa}$ includes all Yukawa coupling terms identical to those in the MSSM. $\hat{H}_u=(\hat{H}_u^+,\hat{H}_u^0)^T,\hat{H}_d=(\hat{H}_d^0,\hat{H}_d^-)^T$ represent the usual $SU(2)_L$ doublet Higgs superfields. $\hat{S}$ is the superfield that transforms as a singlet under the SM gauge group. 

The Higgs sector of the NMSSM comprises the scalar components of the Higgs superfields, denoted by $H_u$, $H_d$ and $S$ as expressed in Eq.(\ref{NMSSMsoftL}). The interactions of these Higgs fields are parametrized by the dimensionless coefficients $\lambda$ and $\kappa$ in Eq.(\ref{NMSSMsp}), along with the soft-breaking trilinear coefficients $A_\lambda$ and $A_\kappa$ in Eq.(\ref{NMSSMsoftL}). The SUSY-breaking mass parameters $m^2_{H_u},m^2_{H_d},m^2_{s}$ in Eq.(\ref{NMSSMsoftL}) can be determined by solving the conditional equations for minimizing the scalar potential and expressed in terms of the vacuum expectation values (vevs) of the Higgs fields, namely, $\left\langle H_u^0 \right\rangle = v_u/\sqrt{2}$, $\left\langle H_d^0 \right\rangle = v_d/\sqrt{2}$, and $\left\langle S \right\rangle = v_s/\sqrt{2}$, with $v = \sqrt{v_u^2+v_d^2}\simeq 246~\mathrm{GeV}$. Consequently, the Higgs sector is characterized by the following six input parameters: 
\begin{eqnarray}
\lambda,\, \kappa,\, A_\lambda,\, A_\kappa,\, \tan{\beta},\, \mu,
\label{Higgs_pars}
 \end{eqnarray}
where $\tan{\beta}$ is defined as the ratio of the vevs, $\tan{\beta} \equiv v_u/v_d$, and $\mu$ is generated dynamically by $\mu \equiv \lambda v_s/\sqrt{2}$.  Throughout this work, all of these parameters are treated as real numbers, as the $CP$ conservation is assumed.

It is advantageous to define the combinations of Higgs fields: $H_{\rm SM} \equiv   \sqrt{2} {\rm Re} (\sin\beta  H_u^0 + \cos\beta H_d^0)$, $H_{\rm NSM} \equiv \sqrt{2} {\rm Re} ( \cos\beta H_u^0 - \sin\beta H_d^0)$, and $A_{\rm NSM} \equiv \sqrt{2} {\rm Im} ( \cos\beta H_u^0 - \sin\beta H_d^0)$. Here, $H_{\rm SM}$ has the same couplings to SM particles as a SM Higgs boson, while $H_{\rm NSM}$ and $A_{\rm NSM}$ represent non-SM-like $CP$-even and $CP$-odd states respectively. The singlet field is left unrotated, denoted by $\sqrt{2} S \equiv H_{\rm S} + i A_S $. With this definition, $H_{\rm SM}$ would acquire the total SM-like Higgs field vev, i.e. $\left\langle H_{\rm SM} \right\rangle = v$ and $\left\langle H_{\rm NSM} \right\rangle = 0$. In the basis $\left(H_{\rm NSM}, H_{\rm SM}, H_{\rm S} \right)$, the elements of the $CP$-even Higgs boson mass matrix $\mathcal{M}_S^2$  can be written as~\cite{Ellwanger:2009dp,Miller:2003ay}
\begin{equation}
\begin{aligned}
  M^2_{11} &=  M^2_A + (m^2_Z  - \frac{1}{2}\lambda^2 v^2) \sin^2 2\beta,  \\
  M^2_{12} &=  -\frac{1}{2}(m^2_Z - \frac{1}{2}\lambda^2 v^2)\sin4\beta,  \\
  M^2_{13} &=  \sqrt{2} \lambda \mu v (1-\delta) \cot{2\beta},  \\
  M^2_{22} &=  m_Z^2\cos^2 2\beta + \frac{1}{2} \lambda^2v^2\sin^2 2\beta,  \\
  M^2_{23} &=  \sqrt{2} \lambda \mu v \delta ,  \\
  M^2_{33} &=  \frac{\lambda^2 v^2 A_{\lambda}{\sin}2\beta}{4\mu}+\frac{\mu}{\lambda}(\kappa A_{\kappa}+4\kappa^2 \frac{\mu}{\lambda}),
  \label{Mass-CP-even-Higgs}
\end{aligned}
\end{equation}
where $M^2_A \equiv \frac{2\mu}{\sin2\beta}(A_{\lambda}+\frac{\kappa}{\lambda} \mu )$ is the typical definition of the squared mass of the doublet-dominated pseudoscalar Higgs boson in the $S$ decoupling limit. It is commonly employed to substitute the soft-breaking coefficients $A_{\lambda}$ as an input parameter. Inspired by the theoretically well-motivated scenario in the alignment without decoupling limit~\cite{Carena:2015moc, Biekotter:2021qbc}, this work introduces a $\delta$ factor, defined as: 
\begin{eqnarray}
\delta \equiv 1-(\frac{M_A}{2\mu/\sin2\beta})^2 -\frac{\kappa}{2\lambda}\sin2\beta, \label{definition-delta}
 \end{eqnarray}
to simplify the expression and allow for direct manipulation of the mixings of $H_{\rm S}$ with $H_{\rm SM}$ and $H_{\rm NSM}$. As illustrated in Sec.\ref{Section-excess}, adopting the $\delta$ factor instead of $M^2_A$ as an input parameter proves beneficial in obtaining solutions consistent with the observed excesses.
Moreover, the equation for $M^2_{22}$ in Eq.(\ref{Mass-CP-even-Higgs}) implies that the mass of the $H_{\rm SM}$-dominated Higgs boson can be increased by an additional contribution $\frac{1}{2} \lambda^2 v^2 \sin^2 2\beta$ compared to that in the MSSM. Additionally, the mixing between $H_{\rm SM}$ and $H_{\rm S}$ can further augment the $H_{\rm SM}$-dominated boson's mass when ${M}^2_{22} > {M}^2_{33}$. These increments facilitate the attainment of a SM-like Higgs boson with a mass around 125 GeV, mitigating the need for substantial radiative corrections originating from stop loops.  

The model under consideration predicts the existence of three $CP$-even Higgs mass eigenstates, denoted as $h_i=\{h,H,h_{\rm s}\}$, which arise from the mixings of the interaction eigenstates $\left(H_{\rm NSM}, H_{\rm SM}, H_{\rm S} \right)$,
  \begin{eqnarray}
    h_i & = & V_{h_i}^{\rm NSM} H_{\rm NSM}+V_{h_i}^{\rm SM} H_{\rm SM}+V_{h_i}^{\rm S} H_{\rm S}. 
   \label{Vij}
  \end{eqnarray} 
Here, the mixing angles $V^j_{h_i} $ constitute a unitary matrix that diagonalizes the squared mass matrix delineated in Eq.(\ref{Mass-CP-even-Higgs}). In addition, the imaginary components $A_S$ and $A_{\rm NSM}$ mix into two $CP$-odd Higgs mass eigenstates $a_i=\{A_s, A_H\}$, while the charged components give rise to a pair of charged Higgs bosons $H^\pm$. Current experimental limitations provide crucial insights into the characteristics of the Higgs sector:
\begin{itemize}
 \item The $CP$-even mass eigenstate $h$ aligns with the 125 GeV SM-like Higgs boson observed at the LHC. This state is predominantly composed of the $H_{\rm SM}$ component, with contributions from $H_{\rm NSM}$ and Re$[S]$ components restricted to be less than about 10\% \cite{ATLAS:2022vkf,CMS:2022dwd}, i.e., $\sqrt{\left (V_h^{\rm NSM} \right )^2 + \left ( V_h^{\rm S} \right )^2} \lesssim 0.1$ and $|V_h^{\rm SM}| \sim 1$.  
 \item The heavy $CP$-even boson $H$ exhibits a nearly degenerate mass with the $CP$-odd boson $A_H$ and the charged Higgs bosons $H^\pm$. Searches for additional Higgs bosons at the LHC, coupled with indirect constraints from $B$ physics, strongly suggest substantial masses for these bosons, e.g., $m_{H} \gtrsim 0.5~{\rm TeV}$ \cite{ATLAS:2020zms,CMS:2022goy}. 
 \item Collider data currently available allow the singlet-dominated scalar states to possess moderate masses while retaining significant doublet components~\cite{Cao:2013gba}. This particular characteristic forms the cornerstone of the present study.
  \end{itemize}
  
The neutralino sector within the NMSSM framework is comprised of the Bino field $\tilde{B}$, the Wino field $\tilde{W}$, the Higgsino fields $\tilde{H}_d^0$ and $\tilde{H}_u^0$, and the Singlino field $\tilde{S}$. In the basis $\psi \equiv (\tilde{B},\tilde{W},\tilde{H}_d^0,\tilde{H}_u^0,\tilde{S})$, the symmetric neutralino mass matrix is formulated as follows~\cite{Ellwanger:2009dp}:
  \begin{equation}
    {\cal M} = \left(
    \begin{array}{ccccc}
    M_1 & 0 & -m_Z \sin \theta_W \cos \beta & m_Z \sin \theta_W \sin \beta & 0 \\
      & M_2 & m_Z \cos \theta_W \cos \beta & - m_Z \cos \theta_W \sin \beta &0 \\
    & & 0 & -\mu & - \frac{1}{\sqrt{2}} \lambda v \sin \beta \\
    & & & 0 & -\frac{1}{\sqrt{2}} \lambda v \cos \beta \\
    & & & & \frac{2\kappa}{\lambda} \mu
    \end{array}
    \right), \label{eq:MN}
    \end{equation}
where $\theta_W$ represents the weak mixing angle, and $M_1$ and $M_2$ denote the soft-breaking masses of the Bino and Wino fields, respectively. Diagonalizing $\cal{M}$ by a rotation matrix $N$ results in five mass eigenstates:
\begin{eqnarray}
\tilde{\chi}_i^0 = N_{i1} \psi^0_1 +   N_{i2} \psi^0_2 +   N_{i3} \psi^0_3 +   N_{i4} \psi^0_4 +   N_{i5} \psi^0_5.
\end{eqnarray}
Here, $\tilde{\chi}_i^0\,(i=1,2,3,4,5)$ are labeled in a mass-ascending order, and the matrix element $N_{ij}$ parametrizes the component of the field $\psi^0_j$ in $\tilde{\chi}_i^0$. Under the assumption of R-parity conservation, the lightest neutralino $\tilde{\chi}_1^0$ emerges as a viable DM candidate if it also serves as the Lightest Supersymmetric Particle (LSP). To reconcile the total measured DM relic density with the absence of direct detection signals, the composition of $\tilde{\chi}_1^0$  typically necessitates a dominance of either the $\tilde{B}$ component or the $\tilde{S}$ component ~\cite{Baum:2017enm, Cao:2019qng}. Indeed, the $\tilde{B}$-dominated $\tilde{\chi}_1^0$ in the NMSSM bears a clear resemblance to that in the MSSM.~\footnote{One  distinction lies in the fact that within the NMSSM framework, the $\tilde{B}$-dominated $\tilde{\chi}_1^0$s may co-annihilate with $\tilde{S}$-dominated neutralinos to obtain the observed abundance. Nevertheless, this scenario is confined to a rather narrow parameter space characterized by $|M_1| \simeq |2\kappa\mu/\lambda|$, necessitating sufficiently large values of $\lambda$, $\kappa$, and $\mu$.} The $\tilde{\chi}_1^0$ DM in this case can annihilate through various channels, including the SM-like Higgs funnel, $Z$ funnel or co-annihilation with electroweakinos, squarks, sleptons and gluinos. 
Unfortunately, due to strong constraints from current LHC searches for SUSY particles, only co-annihilation with Wino-like electroweakinos becomes the dominant mechanism to achieve the correct relic abundance. 
Regarding the $\tilde{S}$-dominated $\tilde{\chi}_1^0$, the typical annihilation channels, $\tilde{\chi}_1^0 \tilde{\chi}_1^0 \to t \bar{t}$, $h_s A_s$, and $h A_s$ (where $t$ represents the top quark) can not fully account for the measured abundance. This limitation arises because a relatively large value of $\lambda$ is disfavored by the null results from direct detection experiments~\cite{Cao:2019qng}. Consequently, the $\tilde{S}$-dominated $\tilde{\chi}_1^0$ is more likely to attain the correct abundance through co-annihilation with electroweakinos. However, within the context of the $\mathbb{Z}_3$-NMSSM, this scenario suffers from a suppressed Bayesian evidence due to the fine-tuning required in a narrow region of parameter space, e.g. $|\kappa| \leq \lambda/2 $ with $\lambda \lesssim 0.06$~\cite{Zhou:2021pit}. Furthermore, the specific conditions for $V_{h_s}^{\rm SM}$ and $V_{h_s}^{\rm NSM}$ to match the central values of the 95 GeV excesses would constrain the model's capability to predict compatible  $\tilde{\chi}_1^0$-nucleon cross sections. In summary, as demonstrated later in this work, the interpretation of the combined excess favors the neutralino DM being dominated by the $\tilde{B}$-component. 

\subsection{\texorpdfstring{$\gamma\gamma$}{} and \texorpdfstring{$b\bar{b}$}{} signals}
\label{subsection-signals}
The observed $\gamma\gamma$ excess can be attributed to either a CP-even or CP-odd Higgs boson, or a combination of both. However, explaining the $b\bar{b}$ excess specifically requires the involvement of a CP-even Higgs boson, as CP-odd Higgs bosons lack couplings to the Z boson, precluding them from contributing significantly to this signal. Given this, the present study focuses on the resonant production of the lightest singlet-dominated CP-even Higgs, $h_s$, to fully account for both the $\gamma\gamma$ and $b\bar{b}$ excesses. In the narrow width approximation, the signal strengths normalized to their SM predictions are calculated as follows~\cite{Cao:2023gkc}:
\begin{eqnarray}
	\mu_{\gamma\gamma}|_{m_{h_s} = 95.4~{\rm GeV}} &=&
  \frac{\sigma_{\rm SUSY}(p p \to h_s)}
       {\sigma_{\rm SM}(p p \to h_s )} \times
       \frac{{\rm Br}_{\rm SUSY}(h_s \to \gamma \gamma)}
       {{\rm Br}_{\rm SM}(h_s \to \gamma \gamma)} \nonumber  \\
& \simeq & |C_{h_s g g}|^2 \times |C_{h_s \gamma \gamma}|^2  \times {\rm R}_{\rm width},  \label{muCMS} \\
  	\mu_{b\bar{b}}|_{m_{h_s} = 95.4~{\rm GeV}} &=&
  \frac{\sigma_{\rm SUSY}(e^+e^-\to Z h_s)}
       {\sigma_{\rm SM}(e^+e^-\to Z h_s)} \times
       \frac{{\rm Br}_{\rm SUSY}(h_s\to b\bar{b})}
       {{\rm Br}_{\rm SM}(h_s \to b\bar{b})} \nonumber \\
  & \simeq & \left|C_{h_s V V}\right|^2 \times |C_{h_s b \bar{b}}|^2 \times {\rm R}_{\rm width},    \label{muLEP}
  \end{eqnarray}
where the mass of $h_s$ is fixed at $95.4~{\rm GeV}$. The production rate $\sigma( p p \to h_s)$ and the decay branching ratio ${\rm Br} (h_s \to \gamma \gamma)$ labeled with the subscript `SUSY' refer to the predictions from the model, whereas those with the subscript `SM' assume SM couplings for $h_s$. As the gluon fusion process is the primary contributor to Higgs production, the ratio of production rates, $\sigma_{\rm SUSY}(pp \to h_s) / \sigma_{\rm SM}(pp \to h_s)$, can be approximately expressed as $|C_{h_s gg}|^2$ at leading order. Here, $C_{h_s gg} \equiv {\cal{A}}_{\rm SUSY}^{h_s gg} / {\cal{A}}_{\rm SM}^{h_s gg}$ represents the ratio of the coupling strength between $h_s$ and gluons predicted by the SUSY model $ {\cal{A}}_{\rm SUSY}^{h_s gg}$ to the corresponding coupling in the SM ${\cal{A}}_{\rm SM}^{h_s gg}$. \footnote{ The SUSY coupling strength ${\cal{A}}_{\rm SUSY}^{h_s g g}$ is contributed by the loops mediated by quarks and squarks~\cite{King:2012tr}. It is computed in the \textsf{SPheno-4.0.5} code~\cite{Porod2003SPheno,Porod2011SPheno3} using the following formula:
\begin{eqnarray}
{\cal{A}}_{\rm SUSY}^{h_s g g} &=& \sum_q C_{h_s q \bar{q}}^{tree} \times \frac{m_q(Q)}{m_q(pole)} \times {\cal{A}}_{\rm SM}^{h_s g g, q} + \sum_{\tilde{q}} {\cal{A}}_{\rm SUSY}^{h_s g g, \tilde{q}}, \nonumber
\end{eqnarray}
where ${\cal{A}}_{\rm SUSY}^{h_s g g, q}$ and ${\cal{A}}_{\rm SUSY}^{h_s g g, \tilde{q}}$ represent the contributions to the $h_s g g$ coupling from quark loops and squark loops, respectively~\cite{Djouadi:2005gj}. ${\cal{A}}_{\rm SM}^{h_s g g, q}$ denotes the quark loop contribution in the SM with the formulation available in Ref.~\cite{Djouadi:2005gi}. $C_{h_s q \bar{q}}^{tree}$ is the tree-level coupling between $h_s$ and quarks. $m_q (Q)$ and $m_q (pole)$ represent the running mass of the quark at scale $Q \approx m_{h_s}$ and the quark pole mass, respectively. Higher-order QCD corrections are applied to these calculations based on the methods detailed in Refs.~\cite{Spira:1995rr, Staub:2016dxq}.} Similarly, the normalized coupling strength of $h_s$ to photons, denoted by $C_{h_s \gamma \gamma}$, is defined in a comparable manner, including additional contributions from loops involving the W boson, charged Higgs bosons, and charginos~\cite{King:2012tr}. However, the SUSY contributions to $C_{h_s gg}$ and $C_{h_s \gamma \gamma}$ are not significant in this study. Specifically, given the substantial masses of the squarks, their contributions to these couplings are typically negligible, amounting to just a few thousandths of the SM contributions. The contribution from the charged Higgs boson to $C_{h_s \gamma \gamma}$ is even smaller, typically on the order of 0.001\%, again owing to the large mass.  The contribution from charginos to $C_{h_s \gamma \gamma}$ may reach up to $10\%$ under optimal conditions, particularly when the Higgsino mass is relatively light and $\lambda$ is sufficiently large~\cite{Choi:2012he}. Overall, while the ratios $C_{h_s gg}$ and $C_{h_s \gamma \gamma}$ are key to understanding the production and decay processes of the singlet-dominated CP-even Higgs, the contributions from squarks and charged Higgs bosons generally have a limited impact in this context. The chargino loops might emerge as a slight effect when comparing distinctive scenarios.

The coefficient ${\rm R}_{\rm width}$ appearing in Eq.( \ref{muCMS}) and Eq.( \ref{muLEP}) is derived as the ratio of the SM prediction for the total decay width of $h_s$ to its width in the SUSY model. The reciprocal of this ratio has the following expression ~\cite{Cao:2023gkc}:
\begin{eqnarray}
 1/{\rm R}_{\rm width} & = & |C_{h_s b \bar{b}}|^2 \times {\rm Br}_{\rm SM}(h_s \to b\bar{b}) + |C_{h_s \tau \bar{\tau}}|^2 \times {\rm Br}_{\rm SM}(h_s \to \tau \bar{\tau})  \nonumber \\
&& + |C_{h_s c \bar{c}}|^2 \times {\rm Br}_{\rm SM}(h_s \to c \bar{c}) +  |C_{h_s g g}|^2 \times {\rm Br}_{\rm SM}(h_s \to g g) + \cdots,  \\
& \simeq&  0.801 \times |C_{h_s b \bar{b}}|^2 + 0.083 \times |C_{h_s \tau \bar{\tau}}|^2 + 0.041 \times |C_{h_s c \bar{c}}|^2 + 0.067 \times |C_{h_s g g}|^2, \nonumber
 \label{Rwidth}
\end{eqnarray}
where $C_{h_s f \bar{f}}$ ($f=b$, $\tau$, $c$) are the normalized couplings of $h_s$ to the fermion pairs $f \bar{f}$. ${\rm Br}_{\rm SM}(h_s \to f\bar{f})$ are the corresponding SM branching ratios for $h_s$, obtained from the LHC Higgs Cross Section Working Group, which incorporates all known higher-order QCD corrections~\cite{LHCHiggsCrossSectionWorkingGroup:2013rie}.

In the absence of significant SUSY contributions, the normalized couplings of $h_s$ to fermions and W or Z boson pairs can be expressed in terms of the rotation matrix elements $V^i_j$ as defined in Eq.~(\ref{Vij})~\cite{Ellwanger:2009dp},
\begin{eqnarray}
C_{h_s t \bar{t}} &=&  V_{h_s}^{\rm SM} + V_{h_s}^{\rm NSM} \cot \beta  \simeq V_{h_s}^{\rm SM}, \quad C_{h_s b \bar{b}} =  V_{h_s}^{\rm SM} - V_{h_s}^{\rm NSM} \tan \beta,  \quad C_{h_s V V} = V_{h_s}^{\rm SM},  \nonumber \\
C_{h_s c \bar{c}} &=& C_{h_s t \bar{t}}, \quad \quad C_{h_s \tau \bar{\tau}} = C_{h_s b \bar{b}}, \quad \quad C_{h_s g g} \simeq C_{h_s t \bar{t}}, \quad \quad C_{h_s \gamma \gamma} \simeq V_{h_s}^{\rm SM}, 
\label{hs-couplings}
\end{eqnarray}
Based on these relations, one can derive values for $V^i_j$ required to obtain the central values of $\mu_{\gamma \gamma}$ and $\mu_{b \bar{b}}$: $V_{h_s}^{\rm SM} \simeq 0.36$ and $(V_{h_s}^{\rm SM} - V_{h_s}^{\rm NSM} \tan \beta) \simeq 0.70 \times V_{h_s}^{\rm SM} \simeq 0.25 $. These lead to the branching ratios ${\rm Br}_{\rm SUSY} (h_s \to \gamma \gamma) \simeq 1.86 \times {\rm Br}_{\rm SM} (h_s \to \gamma \gamma) \simeq 2.58 \times 10^{-3}$ and ${\rm Br}_{\rm SUSY} (h_s \to b \bar{b}) \simeq 0.90 \times {\rm Br}_{\rm SM} (h_s \to  b \bar{b}) \simeq 72.6\%$.
However, given the use of the exact formulas for $C_{h_s g g}$ and $C_{h_s \gamma \gamma}$ through out this study, these values could deviate slightly from $C_{h_s t \bar{t}}$ by about $4\%$ and $14\%$, respectively. Consequently, the updated values for the Higgs mixing angles would be $V_{h_s}^{\rm SM} \simeq 0.35$, $(V_{h_s}^{\rm SM} - V_{h_s}^{\rm NSM} \tan \beta) \simeq 0.83 \times V_{h_s}^{\rm SM} \simeq 0.291$. As a result, the corresponding branching ratios would change to ${\rm Br}_{\rm SUSY} (h_s \to \gamma \gamma) \simeq 1.81 \times {\rm Br}_{\rm SM} (h_s \to \gamma \gamma) \simeq 2.52 \times 10^{-3}$, and ${\rm Br}_{\rm SUSY} (h_s \to b \bar{b}) \simeq 0.96 \times {\rm Br}_{\rm SM} (h_s \to  b \bar{b}) \simeq 76.9\%$. These results underscore the importance of achieving an appropriate balance between the coupling $C_{h_s t \bar{t}}$ and a relatively suppressed $C_{h_s b \bar{b}}$ for adequately explaining the observed excesses~\cite{Barbieri:2013nka, Cao:2016uwt}. Notably, small deviations in $C_{h_s g g}$ and $C_{h_s \gamma \gamma}$ from $C_{h_s t \bar{t}}$ can significantly reduce the required value of $V_{h_s}^{\rm NSM} \tan \beta$ to match the central values of the observed signals while having little effect on $V_{h_s}^{\rm SM}$. Additionally, the approximations indicate that reducing $V_{h_s}^{\rm NSM} \tan \beta$ can enhance $C_{h_s b \bar{b}}$ resulting in a higher $\mu_{b \bar{b}}$ and a lower $\mu_{\gamma \gamma}$ assuming other couplings remain constant~\cite{Cao:2016uwt}.

The explanation for the observed di-photon and $b \bar{b}$ excesses relies heavily on the parameters in the Higgs sector. By using eigenvalue equations of the mass eigenstates $h$ and $h_s$ and recognizing that $m_{h_s}, m_h \ll m_A$ indicates minor mixings of $H_{\rm NSM}$ with $H_{\rm SM}$ and $H_{\rm S}$, the following tree-level approximations can be derived:
\begin{eqnarray}
&&m_B^2 \equiv M^2_{33}  \simeq  m_{h_s}^2 |V_{h_s}^S|^2 + m_h^2 |V_{h_s}^{\rm SM}|^2, \label{mB2} \\
&&V_{h_s}^{\rm SM} \simeq \frac{\sqrt{2}\delta\lambda\mu v}{m_{h}^2 - m_B^2}V_{h_s}^S,  \label{VSM} \\
&&V_{h_s}^{\rm NSM} \simeq  \frac{\sqrt{2}(\delta-1)\lambda\mu v\cot{2\beta}}{m_A^2 - m_{h_s}^2} V_{h_s}^S, \label{VNSM}
\end{eqnarray}
Here, $m_B^2$ denotes the squared mass element of the singlet field.
These expressions suggest that the observed excesses constrain $m_B$ within a narrow range (approximately 100 GeV, as will be discussed later). The relations also imply that the following condition must be met to achieve the central values of the observed excesses:
\begin{eqnarray}
\frac{1}{\delta} = 1+ \frac{m_A^2 - m_{h_s}^2}{m_{h}^2 - m_B^2} \times \frac{2}{5(\tan^2\beta -1)}.  \label{Approximation-delta}
\end{eqnarray}
This condition suggests $\tan\beta \lesssim 10 $ for $m_A \simeq 2~{\rm TeV}$ and $m_B \simeq 100~{\rm GeV}$ when $\delta$ is required to be  smaller than 0.3. Additionally, the following approximation can be inferred:
\begin{eqnarray}
\delta\lambda \simeq 0.03 \times \left ( \frac{\mu}{200~{\rm GeV}} \right )^{-1},  \label{delta_lam}
\end{eqnarray}
to predict $V_{h_s}^{\rm NSM} \tan \beta \simeq 0.07$. Given that LHC searches for electroweakinos have established $\mu_{tot} \gtrsim 200~{\rm GeV}$~\cite{ATLAS:2021moa}, the explanation points toward $\delta\lambda \lesssim 0.03$.

\section{Explanation of the excesses}  \label{Section-excess}
This section introduces the sampling strategy and presents detailed numerical results that interpret the observed di-photon and $b\bar{b}$ excesses while respecting constraints from other experiments, especially those related to DM detections. The numerical analysis comprises several key steps. First, the process begins with utilizing the package \textsf{SARAH-4.14.3}~\cite{SARAH_Staub2008,SARAH3_Staub2012,SARAH4_Staub2013,SARAH_Staub2015} to build the model routines of the $\mathbb{Z}_3$-NMSSM. Second, the codes \textsf{SPheno-4.0.5}~\cite{Porod2003SPheno,Porod2011SPheno3} and \textsf{FlavorKit}~\cite{Porod:2014xia} are employed to generate particle spectrum and compute low energy flavor observables, respectively. Subsequently, the DM physics observables are calculated with the package \textsf{MicrOMEGAs-5.0.4}~\cite{Belanger2002,Belanger2004,Belanger2005,Belanger2006,BelangerRD2006qa,Belanger2008,Belanger2010pz,Belanger2013,Barducci2016pcb,Belanger2018}. Finally, resulting samples are analyzed using both Bayesian inference, focusing on the posterior probability density function (PDF), and Frequentist statistics, emphasizing the profile likelihood (PL)~\cite{Fowlie:2016hew}. These statistical approaches allow for a comprehensive understanding of the parameter space, facilitating a nuanced interpretation of the experimental results.

\subsection{Strategy in scanning the parameter space} \label{scanStrategy}
To investigate the properties of Higgs bosons within the present framework, this study employs a specialized parameter scan strategy. The scan focuses on the parameters defined in Eq.(\ref{Higgs_pars}), but substitutes $A_{\lambda}$ and $A_{\kappa}$ with $\delta$ and $m^2_B$, respectively. This reconfiguration streamlines the sampling process without compromising the key aspects of the investigation. The soft trilinear coefficients for the third-generation squarks, $A_t$ and $A_b$, are also considered free parameters due to their significant influence on the SM-like Higgs boson mass via radiative corrections. Additionally, the scan examines the soft-breaking masses for Bino and Wino $M_{1/2}$ to accommodate the possibility of a Bino-dominated $\tilde{\chi}_1^0$ as the DM candidate. In total, the scan explores a 9-dimensional parameter space as delineated in Table~\ref{tab:scan}. The MultiNest algorithm is employed for this multi-dimensional scan with a high number of live points ${\it{nlive}} = 8000$ to ensure a comprehensive survey of the parameter space. \footnote{In the context of nested sampling algorithms like MultiNest, \textit{nlive} refers to the number of active or live points that is utilized in each iteration to define the iso-likelihood contour~\cite{MultiNest2009,Importance2019}. A larger \textit{nlive} value translates to a more detailed scan of the parameter space.}
\begin{table}[tbp]
\caption{Ranges of specific input parameters. All parameters are assigned a flat prior probability distribution due to their well-defined physical interpretation. To streamline the analysis, the soft trilinear coefficients for the third-generation squarks are unified such that $A_t = A_b$. Dimensional parameters not directly relevant to this study are fixed to specific values for simplicity and alignment with experimental constraints. The gluino mass is set to $M_3 = 3~{\rm TeV}$, and a common value of $2~{\rm TeV}$ is assigned to other unspecified parameters to ensure consistency with limits from LHC searches for new physics. All parameters are defined at the renormalization scale $Q_{input} = 1~{\rm TeV}$. The parameter space is acquired based on the analysis in the previous section and validated through multiple trial scans over much broader ranges to ensure comprehensive coverage and effective exploration of the potential regions of interest.
\label{tab:scan}}
\centering
\vspace{0.3cm}
\resizebox{0.7\textwidth}{!}{
\begin{tabular}{c|c|c|c|c|c}
\hline\hline
Parameter & Prior & Range & Parameter & Prior & Range   \\
\hline
$\mu/{\rm TeV}$ & Flat & $0.3 \sim 1.0 $ & $\lambda$ & Flat & $0.001 \sim 0.5$ \\
$m_B/{\rm GeV}$ & Flat & $80 \sim 120$ & $\kappa$ & Flat & $-0.5 \sim 0.5$ \\
$A_t/{\rm TeV}$ & Flat & $2.0 \sim 5.0$ &  $\delta$ & Flat & $0.0 \sim 0.3$ \\
$M_1/{\rm TeV}$ & Flat & $-1.0 \sim -0.3$ & $\tan \beta$ & Flat & $1.0 \sim 10$ \\
$M_2/{\rm TeV}$ & Flat & $0.3 \sim 1.0$ & & \\
\hline\hline
\end{tabular}}
\end{table}

\begin{table}[]
\caption{Experimental analyses of the electroweakino production processes considered in this study categorizing by the topologies of the SUSY signals.}
\label{tab:LHC}
	\vspace{0.2cm}
	\resizebox{0.97\textwidth}{!}{
		\begin{tabular}{llll}
			\hline\hline
			\texttt{Scenario} & \texttt{Final State} &\multicolumn{1}{c}{\texttt{Name}}\\\hline
			\multirow{6}{*}{$\tilde{\chi}_{2}^0\tilde{\chi}_1^{\pm}\rightarrow WZ\tilde{\chi}_1^0\tilde{\chi}_1^0$}&\multirow{6}{*}{$n\ell (n\geq2) + nj(n\geq0) + \text{E}_\text{T}^{\text{miss}}$}&\texttt{CMS-SUS-20-001($137fb^{-1}$)}~\cite{CMS:2020bfa}\\&&\texttt{ATLAS-2106-01676($139fb^{-1}$)}~\cite{ATLAS:2021moa}\\&&\texttt{CMS-SUS-17-004($35.9fb^{-1}$)}~\cite{CMS:2018szt}\\&&\texttt{CMS-SUS-16-039($35.9fb^{-1}$)}~\cite{CMS:2017moi}\\&&\texttt{ATLAS-1803-02762($36.1fb^{-1}$)}~\cite{ATLAS:2018ojr}\\&&\texttt{ATLAS-1806-02293($36.1fb^{-1}$)}~\cite{ATLAS:2018eui}\\\\
			\multirow{6}{*}{$\tilde{\chi}_{2}^0\tilde{\chi}_1^{\pm}\rightarrow Wh\tilde{\chi}_1^0\tilde{\chi}_1^0$}&\multirow{6}{*}{$n\ell(n\geq1) + nb(n\geq0) + nj(n\geq0) + \text{E}_\text{T}^{\text{miss}}$}&\texttt{ATLAS-1909-09226($139fb^{-1}$)}~\cite{ATLAS:2020pgy}\\&&\texttt{CMS-SUS-17-004($35.9fb^{-1}$)}~\cite{CMS:2018szt}\\&&\texttt{CMS-SUS-16-039($35.9fb^{-1}$)}~\cite{CMS:2017moi}\\
			&&\texttt{ATLAS-1812-09432($36.1fb^{-1}$)}\cite{ATLAS:2018qmw}\\&&\texttt{CMS-SUS-16-034($35.9fb^{-1}$)}\cite{CMS:2017kxn}\\&&\texttt{CMS-SUS-16-045($35.9fb^{-1}$)}~\cite{CMS:2017bki}\\\\
			\multirow{2}{*}{$\tilde{\chi}_1^{\mp}\tilde{\chi}_1^{\pm}\rightarrow WW\tilde{\chi}_1^0 \tilde{\chi}_1^0$}&\multirow{2}{*}{$2\ell + \text{E}_\text{T}^{\text{miss}}$}&\texttt{ATLAS-1908-08215($139fb^{-1}$)}~\cite{ATLAS:2019lff}\\&&\texttt{CMS-SUS-17-010($35.9fb^{-1}$)}~\cite{CMS:2018xqw}\\\\
			\multirow{1}{*}{$\tilde{\chi}_2^{0}\tilde{\chi}_1^{\pm}\rightarrow ZW\tilde{\chi}_1^0\tilde{\chi}_1^0$}&\multirow{2}{*}{$2j(\text{large}) + \text{E}_\text{T}^{\text{miss}}$}&\multirow{2}{*}{\texttt{ATLAS-2108-07586($139fb^{-1}$)}~\cite{ATLAS:2021yqv}}\\
			{$\tilde{\chi}_1^{\pm}\tilde{\chi}_1^{\mp}\rightarrow WW\tilde{\chi}_1^0\tilde{\chi}_1^0$}&&\\\\
			\multirow{1}{*}{$\tilde{\chi}_2^{0}\tilde{\chi}_1^{\pm}\rightarrow (h/Z)W\tilde{\chi}_1^0\tilde{\chi}_1^0$}&\multirow{2}{*}{$j(\text{large}) + b(\text{large}) + \text{E}_\text{T}^{\text{miss}}$}&\multirow{2}{*}{\texttt{ATLAS-2108-07586($139fb^{-1}$)}~\cite{ATLAS:2021yqv}}\\
			{$\tilde{\chi}_2^{0}\tilde{\chi}_3^{0}\rightarrow (h/Z)Z\tilde{\chi}_1^0\tilde{\chi}_1^0$}&&\\\\
			$\tilde{\chi}_2^{0}\tilde{\chi}_1^{\mp}\rightarrow h/ZW\tilde{\chi}_1^0\tilde{\chi}_1^0,\tilde{\chi}_1^0\rightarrow \gamma/Z\tilde{G}$&\multirow{2}{*}{$2\gamma + n\ell(n\geq0) + nb(n\geq0) + nj(n\geq0) + \text{E}_\text{T}^{\text{miss}}$}&\multirow{2}{*}{\texttt{ATLAS-1802-03158($36.1fb^{-1}$)}~\cite{ATLAS:2018nud}}\\$\tilde{\chi}_1^{\pm}\tilde{\chi}_1^{\mp}\rightarrow WW\tilde{\chi}_1^0\tilde{\chi}_1^0,\tilde{\chi}_1^0\rightarrow \gamma/Z\tilde{G}$&&\\\\
			$\tilde{\chi}_2^{0}\tilde{\chi}_1^{\pm}\rightarrow ZW\tilde{\chi}_1^0\tilde{\chi}_1^0,\tilde{\chi}_1^0\rightarrow h/Z\tilde{G}$&\multirow{4}{*}{$n\ell(n\geq4) + \text{E}_\text{T}^{\text{miss}}$}&\multirow{4}{*}{\texttt{ATLAS-2103-11684($139fb^{-1}$)}~\cite{ATLAS:2021yyr}}\\$\tilde{\chi}_1^{\pm}\tilde{\chi}_1^{\mp}\rightarrow WW\tilde{\chi}_1^0\tilde{\chi}_1^0,\tilde{\chi}_1^0\rightarrow h/Z\tilde{G}$&&\\$\tilde{\chi}_2^{0}\tilde{\chi}_1^{0}\rightarrow Z\tilde{\chi}_1^0\tilde{\chi}_1^0,\tilde{\chi}_1^0\rightarrow h/Z\tilde{G}$&&\\$\tilde{\chi}_1^{\mp}\tilde{\chi}_1^{0}\rightarrow W\tilde{\chi}_1^0\tilde{\chi}_1^0,\tilde{\chi}_1^0\rightarrow h/Z\tilde{G}$&&\\\\
			\multirow{3}{*}{$\tilde{\chi}_{i}^{0,\pm}\tilde{\chi}_{j}^{0,\mp}\rightarrow \tilde{\chi}_1^0\tilde{\chi}_1^0+\chi_{soft}\rightarrow ZZ/H\tilde{G}\tilde{G}$}&\multirow{3}{*}{$n\ell(n\geq2) + nb(n\geq0) + nj(n\geq0) + \text{E}_\text{T}^{\text{miss}}$}&\texttt{CMS-SUS-16-039($35.9fb^{-1}$)}~\cite{CMS:2017moi}\\&&\texttt{CMS-SUS-17-004($35.9fb^{-1}$)}~\cite{CMS:2018szt}\\&&\texttt{CMS-SUS-20-001($137fb^{-1}$)}~\cite{CMS:2020bfa}\\\\
			\multirow{2}{*}{$\tilde{\chi}_{i}^{0,\pm}\tilde{\chi}_{j}^{0,\mp}\rightarrow \tilde{\chi}_1^0\tilde{\chi}_1^0+\chi_{soft}\rightarrow HH\tilde{G}\tilde{G}$}&\multirow{2}{*}{$n\ell(n\geq2) + nb(n\geq0) + nj(n\geq0) + \text{E}_\text{T}^{\text{miss}}$}&\texttt{CMS-SUS-16-039($35.9fb^{-1}$)}~\cite{CMS:2017moi}\\&&\texttt{CMS-SUS-17-004($35.9fb^{-1}$)}~\cite{CMS:2018szt}\\\\
			$\tilde{\chi}_{2}^{0}\tilde{\chi}_{1}^{\pm}\rightarrow W^{*}Z^{*}\tilde{\chi}_1^0\tilde{\chi}_1^0$&$3\ell + \text{E}_\text{T}^{\text{miss}}$&\texttt{ATLAS-2106-01676($139fb^{-1}$)}~\cite{ATLAS:2021moa}\\\\
			\multirow{3}{*}{$\tilde{\chi}_{2}^{0}\tilde{\chi}_{1}^{\pm}\rightarrow Z^{*}W^{*}\tilde{\chi}_1^0\tilde{\chi}_1^0$}&\multirow{2}{*}{$2\ell + nj(n\geq0) + \text{E}_\text{T}^{\text{miss}}$}&\texttt{ATLAS-1911-12606($139fb^{-1}$)}~\cite{ATLAS:2019lng}\\&&\texttt{ATLAS-1712-08119($36.1fb^{-1}$)}~\cite{ATLAS:2017vat}\\&&\texttt{CMS-SUS-16-048($35.9fb^{-1}$)}~\cite{CMS:2018kag}\\\\
			\multirow{3}{*}{$\tilde{\chi}_{2}^{0}\tilde{\chi}_{1}^{\pm}+\tilde{\chi}_{1}^{\pm}\tilde{\chi}_{1}^{\mp}+\tilde{\chi}_{1}^{\pm}\tilde{\chi}_{1}^{0}$}&\multirow{3}{*}{$2\ell + nj(n\geq0) + \text{E}_\text{T}^{\text{miss}}$}&\texttt{ATLAS-1911-12606($139fb^{-1}$)}~\cite{ATLAS:2019lng}\\&&\texttt{ATLAS-1712-08119($36.1fb^{-1}$)}~\cite{ATLAS:2017vat}\\&&\texttt{CMS-SUS-16-048($35.9fb^{-1}$)}~\cite{CMS:2018kag}\\\hline 
\end{tabular}}
\end{table}

The likelihood function that guides the scan process is constructed as $\mathcal{L} \equiv  \mathcal{L}_{\gamma \gamma + b \bar{b}} \times \mathcal{L}_{\rm Res}$. The term $\mathcal{L}_{\gamma \gamma + b \bar{b}} = \exp ( -\frac{1}{2}\chi^2_{\gamma \gamma + b\bar{b}} )$ quantify the compatibility with the experimental observations of $\gamma\gamma$ and $b\bar{b}$ excesses at $95{\rm GeV}$ by defining the following $\chi^2$ function:
\begin{equation}
\begin{split}
\chi^2_{\gamma \gamma + b\bar{b}} = \left( \frac{\mu_{\gamma\gamma} - 0.24}{0.08}\right)^2 - \left( \frac{\mu_{b\bar{b}} - 0.117}{0.057}\right)^2 .   
\end{split}
\label{chi2-excesses}
\end{equation}
The $\mathcal{L}_{\rm Res}$ term represents additional restrictions from relevant experiments on the theory, defined such that $\mathcal{L}_{\rm Res} = 1$ if conditions are met, and $\mathcal{L}_{\rm Res} = \exp\left [-100 \right ]$ otherwise. These restrictions include:
\begin{itemize}
\item \textbf{Mass of the light Higgs:} The mass of $h_s$ is expected to be around $95.4~{\rm GeV}$ to account for the excesses. Considering theoretical and experimental uncertainties of  $1~{\rm GeV}$, the acceptable range is $94.4~{\rm GeV} \leq m_{h_s} \leq 96.4~{\rm GeV}$.
\item \textbf{Higgs data fit:} The properties of $h$ that corresponds to the discovered Higgs boson at the LHC, must align with measurements by the ATLAS and CMS collaborations at the $95\%$ confidence level. This condition is validated using the code \textsf{HiggsSignals-2.6.2}~\cite{HS2013xfa,HSConstraining2013hwa,HS2014ewa,HS2020uwn}, with a criterion for acceptance being a p-value greater than 0.05, indicating compatibility between observed data and theoretical predictions. 

\item \textbf{Extra Higgs searches:} The signal rates for additional Higgs bosons must comply with cross-section limits based on experimental data at LEP, Tevatron, and LHC. This requirement is implemented using the \textsf{HiggsBounds-5.10.2} code~\cite{HB2008jh,HB2011sb,HBHS2012lvg,HB2013wla,HB2020pkv} and further refined through the \textsf{HiggsTools} code~\cite{Bahl:2022igd}.
\item \textbf{DM relic density:} The central value for the DM relic density, $\Omega {h^2}=0.120$, is derived from the Planck-2018 data~\cite{Planck:2018vyg}. To account for a theoretical uncertainty of $20\%$, the acceptable range $0.096 \leq \Omega {h^2} \leq 0.144$ is applied.
\item \textbf{DM detections:} The spin-independent (SI) and spin-dependent (SD) DM-nucleon scattering cross-sections must be below the upper bounds imposed by the LZ experiments~\cite{LZ:2022lsv}.\footnote{The LZ collaboration has recently reported the latest 2024 results~\cite{LZ2024slides}. The preliminary combined results will be displayed against the 2022 limits used in the likelihood function, along with the projected sensitivity for future studies~\cite{LZ:2018qzl} in Fig.\ref{Fig3}.}  The constraints of DM indirect searches from Fermi-LAT on gamma-ray emissions are not considered, as they do not impose restrictions for \( |m_{\tilde{\chi}_1^0}| \gtrsim 100~{\rm GeV} \)~\cite{Fermi-LAT:2015att}.
\item \textbf{$B$-physics observables:} The branching ratios of $B_s \to \mu^+ \mu^-$ and $B \to X_s \gamma$ should be consistent with their experimental measurements at the $2\sigma$ level~\cite{pdg2018}. Note that the current measurement results~\cite{ParticleDataGroup:2024cfk} show no significant deviations from the values used in the present work. It has been verified that using the 2024 limits would not change the number of surviving samples or the parameter regions that can explain the di-photon and $b\bar{b}$ excesses at the $1\sigma$ level.
\item \textbf{Vacuum stability:} The vacuum state of the Higgs potential must be stable or long-lived~\cite{Hollik:2018wrr}. This condition is tested using \textsf{VevaciousPlusPlus}~\cite{VPP2014} (the C++ version of \textsf{Vevacious} ~\cite{Camargo-Molina:2013qva}).
\end{itemize}

Given the heavy mass ranges specified in Table.~\ref{tab:scan}, constraints from LHC searches for electroweakinos are expected to have no significant impact on the scan results, as suggested by recent statistical combination of ATLAS Run 2 searches in \cite{ATLAS:2024qxh}. To substantiate this, the obtained samples are evaluated using the code \texttt{SModelS-2.1.1}, which encoded various event-selection efficiencies by analyzing topologies of SUSY signals~\cite{Khosa:2020zar}. 
Moreover, the benchmark points presented in Table.~\ref{BP1BP2} and Table.~\ref{BP3BP4} undergo further investigated through simulation analyses listed in Table.~\ref{tab:LHC} using package \texttt{CheckMATE-2.0.26}~\cite{Drees:2013wra,Dercks:2016npn, Kim:2015wza} which is validated in a manner similar to the method described in~\cite{Cao:2022chy}. The $R$ values defined as $R \equiv max\{S_i/S_{i,obs}^{95}\}$ are calculated for all the involved analyses, where $S_i$ represents the simulated event number in the $i$-th signal region (SR), and $S_{i,obs}^{95}$ denotes the corresponding $95\%$ confidence level upper limit.  An $R$ value exceeding 1 indicates experimental exclusion, while an $R$ value below 1 suggests consistency with experimental analyses, assuming negligible uncertainties~\cite{Cao:2021tuh}.

\subsection{Numerical Results}
The scan process yields more than 28 thousand samples that are consistent with the experimental restrictions, among which about 21 thousand samples can explain both the $\gamma\gamma$ and $b\bar{b}$ excesses at a level of 2$\sigma$. All samples prefer Bino-dominated $\tilde{\chi}^0_1$s as DM candidates. They achieve the measured relic abundance primarily by co-annihilating with the Wino-like electroweakinos.

\begin{figure*}[h]
		\centering
		\resizebox{1.0\textwidth}{!}{
        \includegraphics{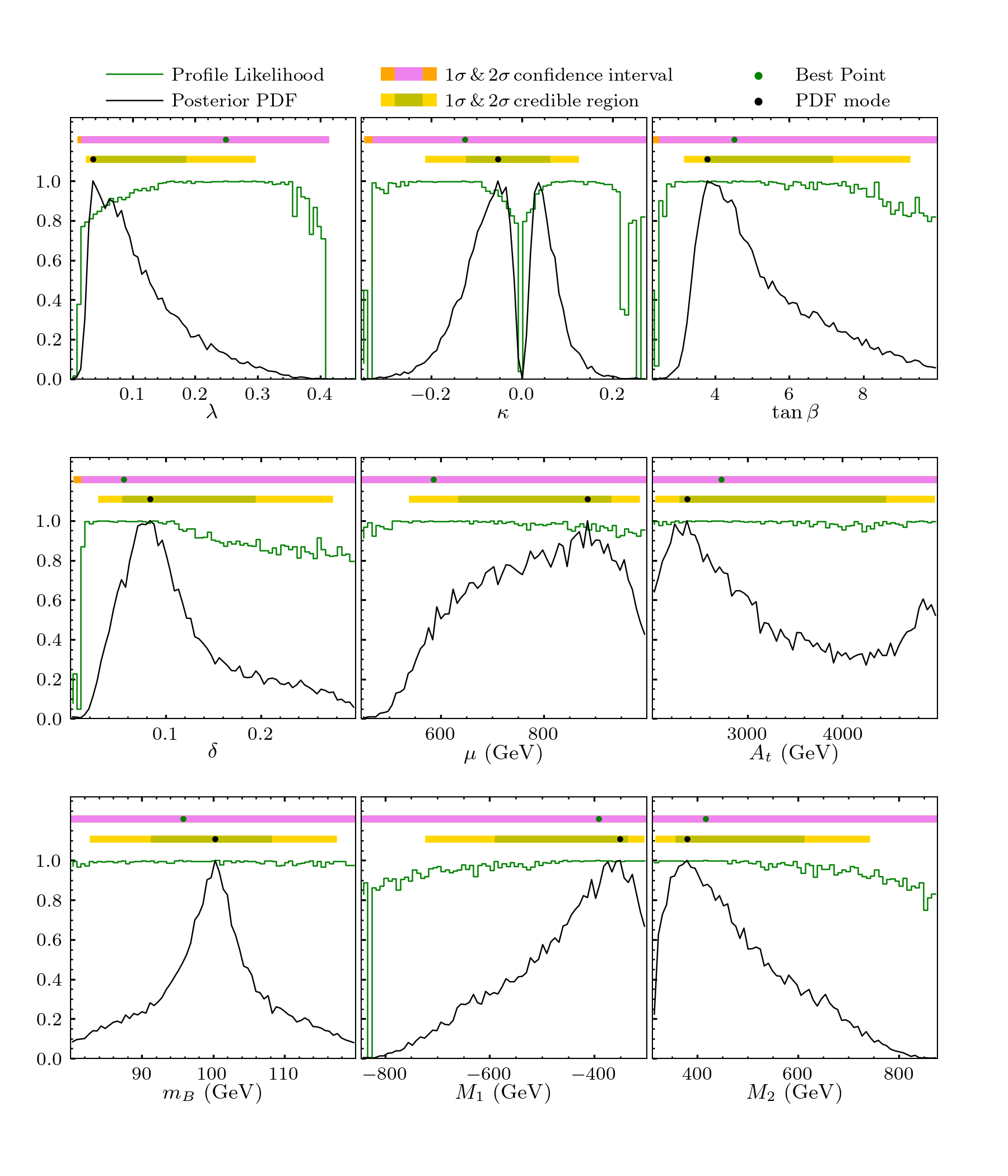}
        }
\vspace{-1.5cm}
\caption{One-dimensional profile likelihoods (green line) and marginalised posterior probability density functions (black line) of the input parameters $\lambda$, $\kappa$, $\tan \beta$, $\delta$, $\mu$, $A_t$, $m_B$, $M_1$ and $M_2$. The violet and orange bands show the $1\sigma$ and $2\sigma$ confidence intervals, respectively, and the green dot marks the best point corresponding to $\chi^2_{\gamma \gamma + b\bar{b}} \simeq 10^{-5}$. The yellow  and golden bands denote the $1\sigma$ and $2\sigma$ credible regions and the black dot denotes the PDF mode~\cite{Fowlie:2016hew}.  \label{Fig1}}
\end{figure*}	
Fig.~\ref{Fig1} presents the statistical distributions of the nine input parameters listed in Table~\ref{tab:scan}, including one-dimensional profile likelihoods (PL, delineated as green lines) and marginal posterior probability density functions (PDF, plotted as black lines).
 \footnote{The profile likelihood is the most significant likelihood value in a specific parameter space~\cite{Fowlie:2016hew}. Given a set of input parameters $\Theta \equiv (\Theta_1,\Theta_2,\cdots)$, one can acquire the one-dimensional PL by changing the other parameters to maximize the likelihood function, i.e.,
	\begin{eqnarray}
\mathcal{L}(\Theta_A)=\mathop{\max}_{\Theta_1,\cdots,\Theta_{A-1},\Theta_{A+1},\cdots}\mathcal{L}(\Theta), \nonumber
	\end{eqnarray}
The PL reflects the preference of a theory on the parameter space. For a given point $\Theta_A$, it represents the capability of the point in the theory to account for experimental data. By contrast, the one-dimensional marginal posterior probability density functions (PDF) of parameter $\Theta_A$  is obtained by integrating the posterior PDF from the Bayesian theorem, $P(\Theta)$, over the rest of the model inputs:
\begin{eqnarray}
	P(\Theta_A)&=&\int{P(\Theta) d\Theta_1 d\Theta_2 \cdots d\Theta_{A-1} d\Theta_{A+1} \cdots  \cdots }.  \nonumber
\end{eqnarray}
It represents the updated belief about the parameter values after taking the observed data into account. }
The violet and orange bands represent the $1\sigma$ and $2\sigma$ confidence intervals, respectively, while the green dots mark the best points corresponding to $\chi^2_{\gamma \gamma + b\bar{b}} \simeq 0$. The yellow and golden bands denote the $1\sigma$ and $2\sigma$ credible regions and the black dots denote the modes of the posterior probability density function.
The PL distributions indicate that the parameter space outlined in Table.~\ref{tab:scan} can readily account for the di-photon and $b\bar{b}$ excesses at a level of $2\sigma$ while satisfying other experimental constraints. 
The results support the previous analysis in Sec.\ref{subsection-signals} and exhibit the following features:
\begin{itemize}
\item  The signal rates of the excesses require the Higgs mixing angles $V_{h_s}^{\rm SM}$ in Eq.(\ref{VSM}) and $V_{h_s}^{\rm NSM}$ in Eq.(\ref{VNSM}) to be sizeable enough. Thus, $\delta$ should have a relatively small value, hence a prior range of $\delta \lesssim 0.3$. Consequently, Eq.(\ref{definition-delta}), Eq.(\ref{mB2}) together with $V_{h_s}^{\rm SM} \simeq 0.35$ and $V_{h_s}^{\rm NSM} \simeq 0.012$, impose limits on $\lambda$ and $\kappa$, i.e. $\lambda \lesssim 0.42$ and $|\kappa| \lesssim 0.3$, and constrain $m_B$ within a narrow range. 
\item The trilinear coefficient $A_t$ affects the SM-like Higgs boson mass $m_h$, the mixing angles $V_{h_s}^{\rm SM}$, and the $h_s g g$ and $h_s \gamma \gamma$ coupling strengths via $\tilde{t}$-mediated loops. Given that the soft-breaking masses of the squarks are fixed at $2~{\rm TeV}$, $A_t$ in the whole range of (2TeV, 5TeV) can render the theory capable of explaining the excesses. 
\item The lower boundary of $\mu \gtrsim 450 {\rm GeV}$ arises from the LZ experimental constraint~\cite{Cao:2019qng} as the $\tilde{B}$-dominated DM-nucleon scattering cross sections are inversely proportional to $\mu^2$~\cite{Cao:2019qng}. To suppress the SI scattering cross-sections, $M_1$ is assigned an opposite sign to $\mu$ ab initio to form a ``Blind Spot" scenario where different contributions counteract each other~\cite{Huang:2014xua, Crivellin:2015bva, Han:2016qtc, Carena:2018nlf}. 
\end{itemize}
Besides, the posterior PDF lines may significantly deviate from the PL lines.  This occurs because the PL function evaluates the likelihood at the best fit value of other parameters and reflects how well different values of the parameter fit the data. In contrast,  the posterior PDFs depend not only on the likelihood function but also involve integration over the prior volume. Hence, the narrow peaks of posterior PDF lines represent the most credible region under the current experimental data, whereas the lower regions reflect the parameter space suppression due to a certain degree of fine tuning.

\begin{figure*}[t]
		\resizebox{1.0\textwidth}{!}{
        \includegraphics{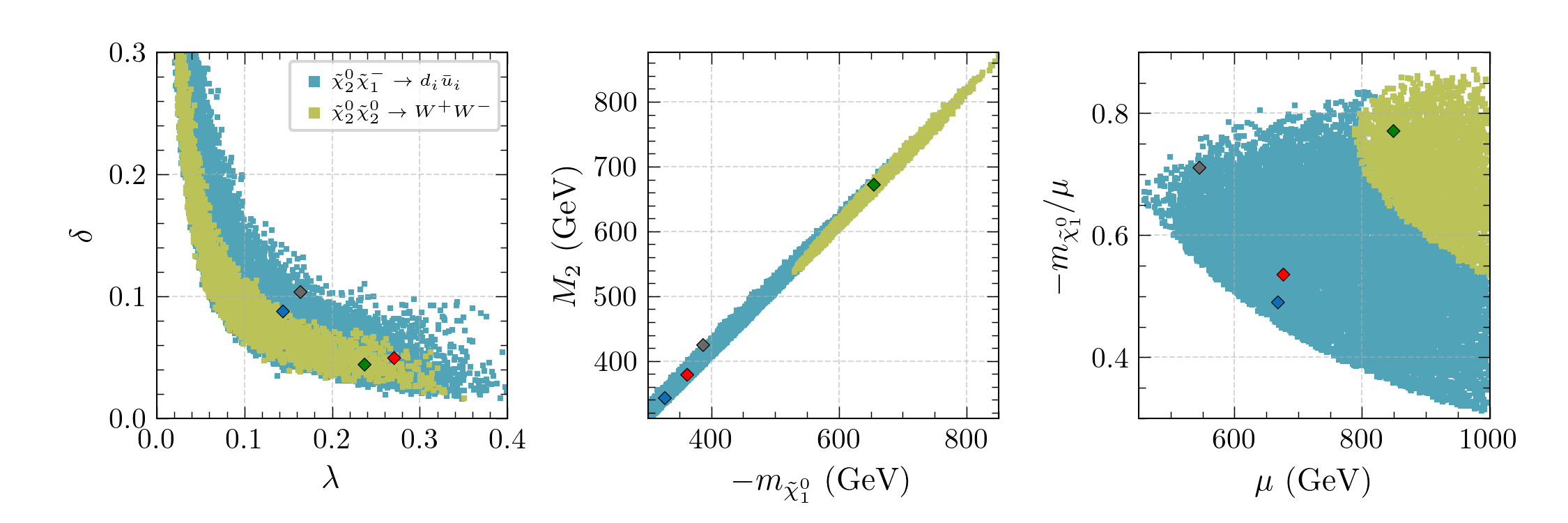}
        }
\caption{Scattering plots of samples satisfying the applied constraints and $\chi^2_{\gamma \gamma + b\bar{b}} \leq 2.3$, projected onto $(\lambda, \delta)$, $(-m_{\tilde{\chi}_1^0},M_2)$ and $ (\mu, - m_{\tilde{\chi}_1^0}/\mu)$ planes, respectively.  
The left panel shows that $\delta\lambda \lesssim 0.03$ is required for most cases to predict the excesses according to Eq.(\ref{delta_lam}).
The middle panel indicates that the DM achieves the measured relic abundance by co-annihilating with the Wino-like electroweakinos. 
The cyan squares represent samples where the co-annihilation process is dominated by the channel $\tilde{\chi}_2^0 \tilde{\chi}_1^- \to d_i \bar{u}_i$ (with $i=1,2,3$ denoting the quark generations), while the green squares indicate samples where the contribution from the channel $\tilde{\chi}_2^0 \tilde{\chi}_2^0 \to W^+ W^-$ is significant. The right panel implies that a lower limit of $\mu \gtrsim 450~{\rm GeV}$ for $|m_{\tilde{\chi}_1^0}| \simeq 300~{\rm GeV}$ is imposed by the DM-nucleon scattering cross sections from LZ results. The colored diamonds represent four benchmark points whose properties are given in Table.\ref{BP1BP2} and Table.\ref{BP3BP4}.  \label{Fig2}}
\end{figure*}

\begin{figure*}[t]
		\centering
		\resizebox{1.0\textwidth}{!}{
        \includegraphics{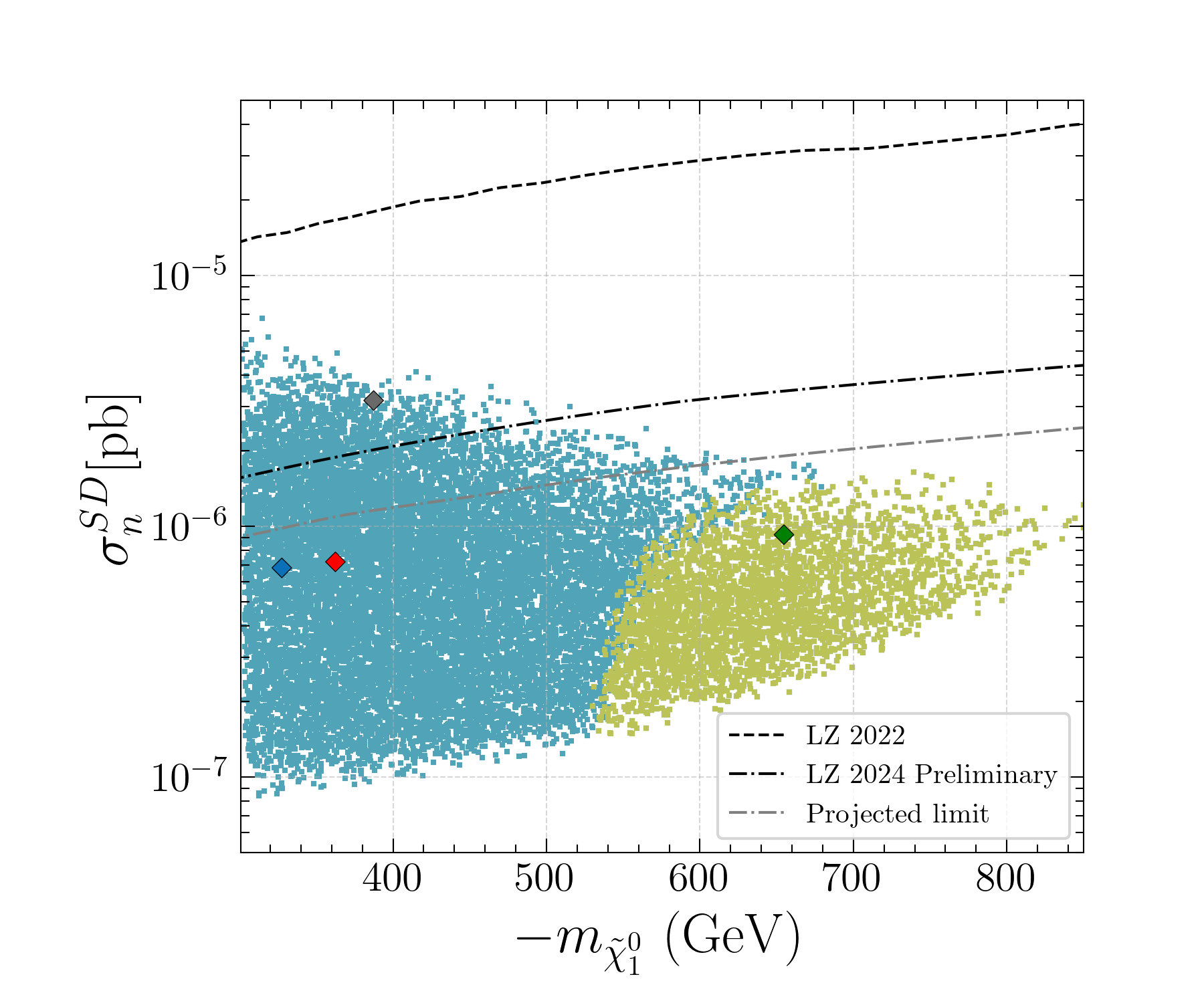}
        \includegraphics{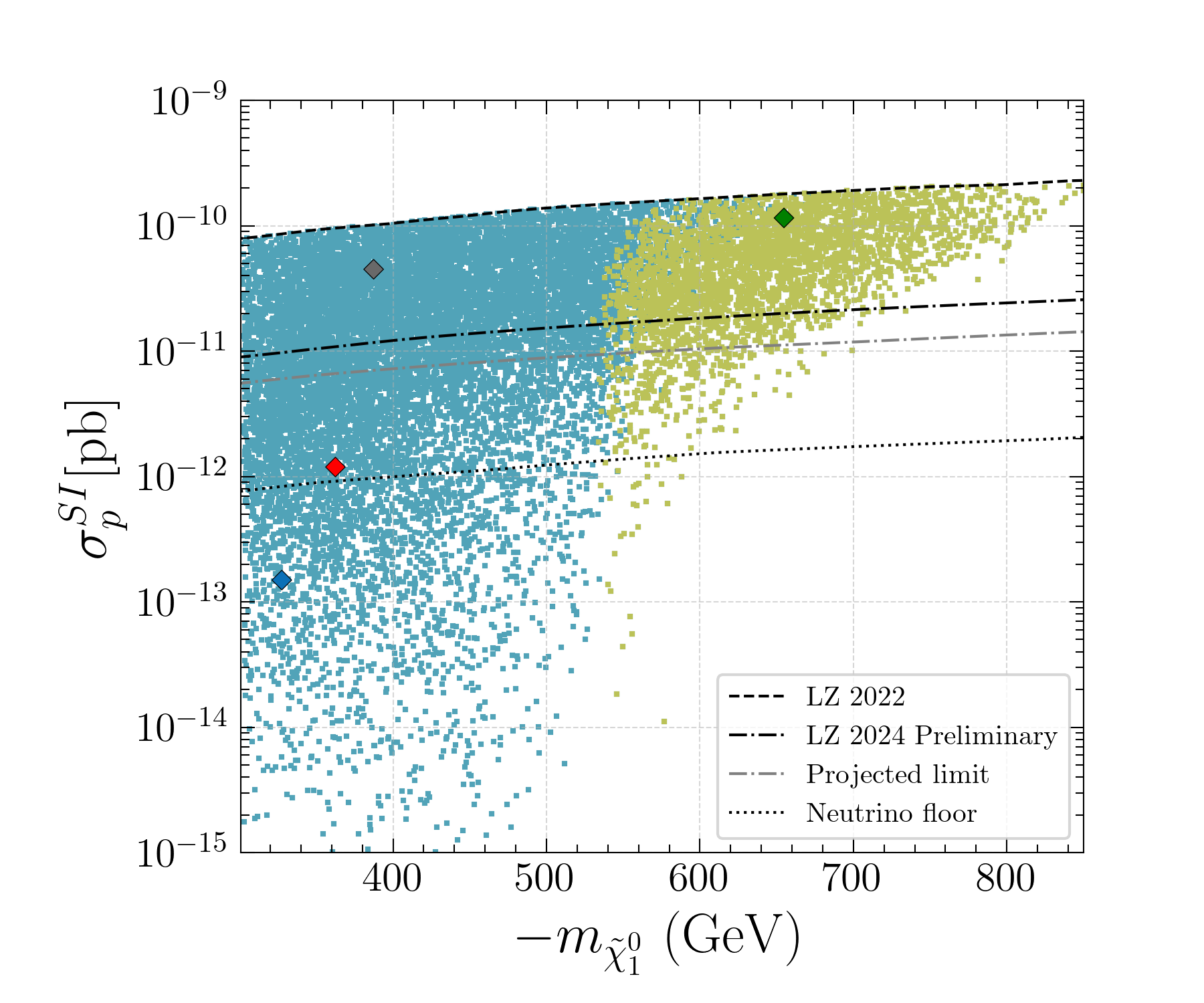}
        }
\caption{$\sigma^{SD}_n$ and $\sigma^{SI}_p$ as functions of $m_{\tilde{\chi}_1^0}$ for points in Fig.\ref{Fig2}. The dashed black lines denote the upper limits at 90\% confidence level from the 2022 results of the LZ experiment~\cite{LZ:2022lsv}, the dashed-dotted black lines represent the preliminary 2024 limits combined with the 2022 results~\cite{LZ2024slides},  the dashed-dotted grey lines indicate the projetced LZ sensitivity for future runs~\cite{LZ:2018qzl}, and the dotted lines correspond to the neutrino floor~\cite{Billard:2013qya}.   \label{Fig3}}
\end{figure*}

Fig.\ref{Fig2} and Fig.\ref{Fig3} display scattering plots for all samples that meet the applied constraints and explain the $\gamma\gamma$ and $b\bar{b}$ excesses at a level of $1\sigma$, i.e. $\chi^2_{\gamma \gamma + b\bar{b}} \lesssim 2.3$. The left panel in Fig.\ref{Fig2} demonstrates that explaining these excesses imposes an upper limit on the product of $\delta$ and $\lambda$. Specifically, neglecting cases with substantial deviations from the central values of the excesses, $\delta\lambda \lesssim 0.03$ is required for most samples, as indicated by Eq.(\ref{delta_lam}). 
The middle panel illustrates a rather small discrepancy between $m_{\tilde{\chi}1^0}$ and $M_2$, indicating that $\tilde{\chi}_1^0$ achieves the measured relic abundance by co-annihilating with the Wino-like electroweakinos. The largest contribution to the abundance could come from the channel $\tilde{\chi}_2^0 \tilde{\chi}_1^- \to d_i \bar{u}_i$ ($i=1,2,3$, represents the $i$-th quark generation), denoted as {\bf Case-I}, and the channel $\tilde{\chi}_2^0 \tilde{\chi}_2^0 \to W^+ W^-$, denoted as {\bf Case-II}, corresponding to Bayesian evidence of 79\% and 20.4\% respectively.~\footnote{For most samples, the total contribution from channels with $d_i \bar{u}_i$ final states for three generations usually dominate the co-annihilation cross section. The phrase of ``the largest contribution'' from a certain channel refers to the single channel with the highest weight among all channels.} {\bf Case-I} samples are marked with cyan squares, and {\bf Case-II} samples are marked with green squares. There are only less than 1\% of samples predicting other channels as largest contribution, e.g. $\tilde{\chi}_1^+ \tilde{\chi}_1^+ \to W^+ W^+$ and $\tilde{\chi}_1^0 \tilde{\chi}_1^0 \to t\bar{t}$. 
Based on this analysis, several key inferences can be made:
\begin{itemize}
\item   In {\bf Case-II}, The co-annihilation process $\tilde{\chi}_2^0 \tilde{\chi}_2^0 \to W^+ W^-$ proceeds mainly through $t/u$-channel exchanging $\tilde{H}$-dominated charginos. This process becomes significant at higher mass thresholds for the Higgsino and $\tilde{\chi}_1^0$, specifically when $\mu \gtrsim 780~\text{GeV}$ and $|m_{\tilde{\chi}_1^0}| \simeq 520~\text{GeV}$.   

\item  Given that $\tan\beta$ is lower bounded around 2.3, the largest value of $- m_{\tilde{\chi}_1^0}/\mu$ for  Blind Spot condition is $\sin2\beta \simeq 0.72$. This explains why $\mu$ can reach a minimum value of $450~\text{GeV}$ for {\bf Case-I} and $520~\text{GeV}$ for {\bf Case-II} when $-m_{\tilde{\chi}_1^0}/\mu \simeq 0.72$.

\item As shown in the left panel in Fig.\ref{Fig2}, {\bf Case-I} can exhibit a larger $\delta\lambda$ compared to {\bf Case-II}. This is because that the former case favors a lower range of $\mu$ and therefore requires an increased $\delta\lambda$ to predict the appropriate Higgs mixing angles $V_{h_s}^{\rm SM}$, as indicated by Eq.(\ref{VSM}). 

\item For the same reason,  {\bf Case-I} can predict a larger $C_{h_s \gamma \gamma}$ compared to {\bf Case-II} since the contribution from chargino loops is proportional to $\frac{\lambda v}{\mu}$. Under optimal conditions, the chargino's contribution in {\bf Case-I} can reach 0.041, whereas it is limited to 0.02 in {\bf Case-II}.
\end{itemize}

\begin{figure*}[t]
		\centering
		\resizebox{1.0\textwidth}{!}{
        \includegraphics{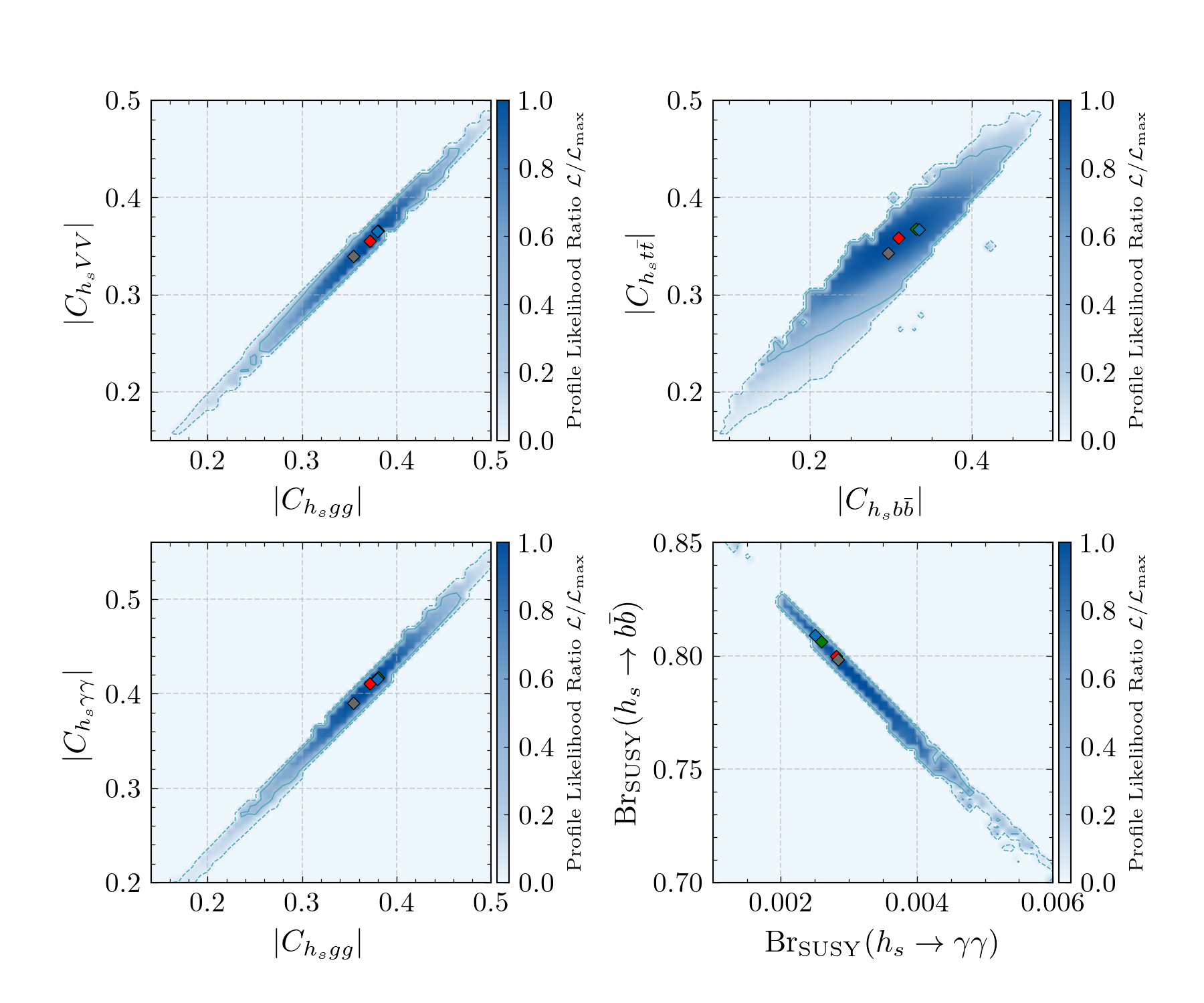}
        }
\caption{Two-dimensional profile likelihoods for all samples, projected onto $|C_{h_s V V}|-|C_{h_s g g}|$,  $|C_{h_s b \bar{b}}|-|C_{h_s t \bar{t}}|$, $|C_{h_s \gamma \gamma}|-|C_{h_s g g}|$, and ${\rm Br}(h_s \to \gamma \gamma)-{\rm Br}(h_s \to b \bar{b})$ planes, respectively.  The contours for $1\sigma$ and $2\sigma$ confidence intervals correspond to $\chi^2_{\gamma \gamma + b \bar{b}} = 2.3$ and $\chi^2_{\gamma \gamma + b \bar{b}} = 6.18$, labeled as solid and dashed lines, respectively. The colored diamonds denote the four benchmark points in Fig.\ref{Fig2}  and Fig.\ref{Fig3} with detailed information in Table.\ref{BP1BP2} and Table.\ref{BP3BP4}.    \label{Fig4}}
\end{figure*}

\begin{figure*}[t]
		\centering
		\resizebox{1.0\textwidth}{!}{
        \includegraphics{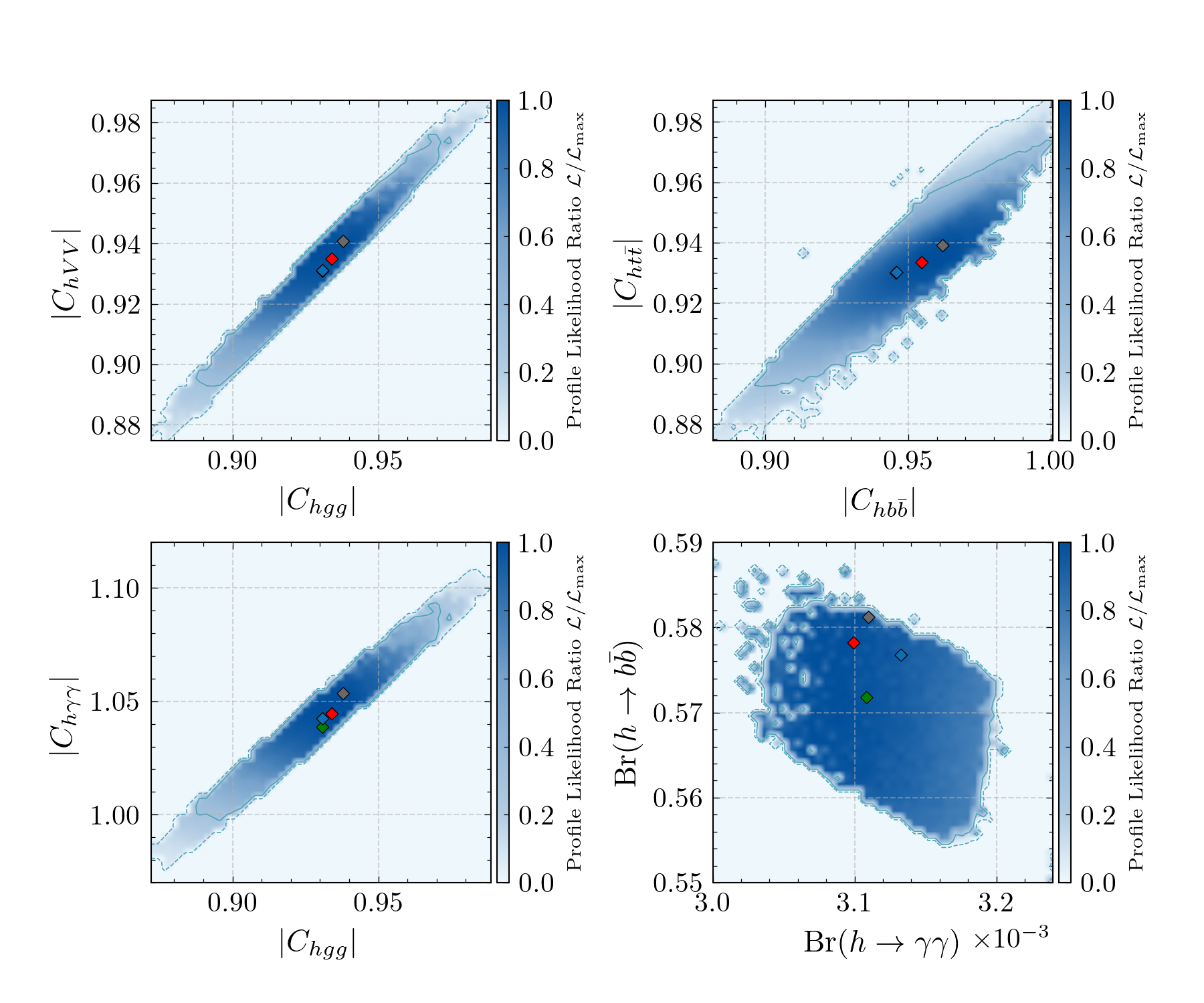}
        }
\caption{Same as Fig.~\ref{Fig4}, but projected onto $|C_{h VV}| - |C_{h gg}|$, $|C_{h b\bar{b}}| - |C_{h t\bar{t}}|$, $|C_{h \gamma\gamma}| - |C_{h gg}|$, and ${\rm Br}(h \to b\bar{b}) - {\rm Br}(h \to \gamma\gamma)$ planes.\label{Fig5}}
\end{figure*}

\begin{figure*}[t]
		\centering
		\resizebox{1.0\textwidth}{!}{
        \includegraphics{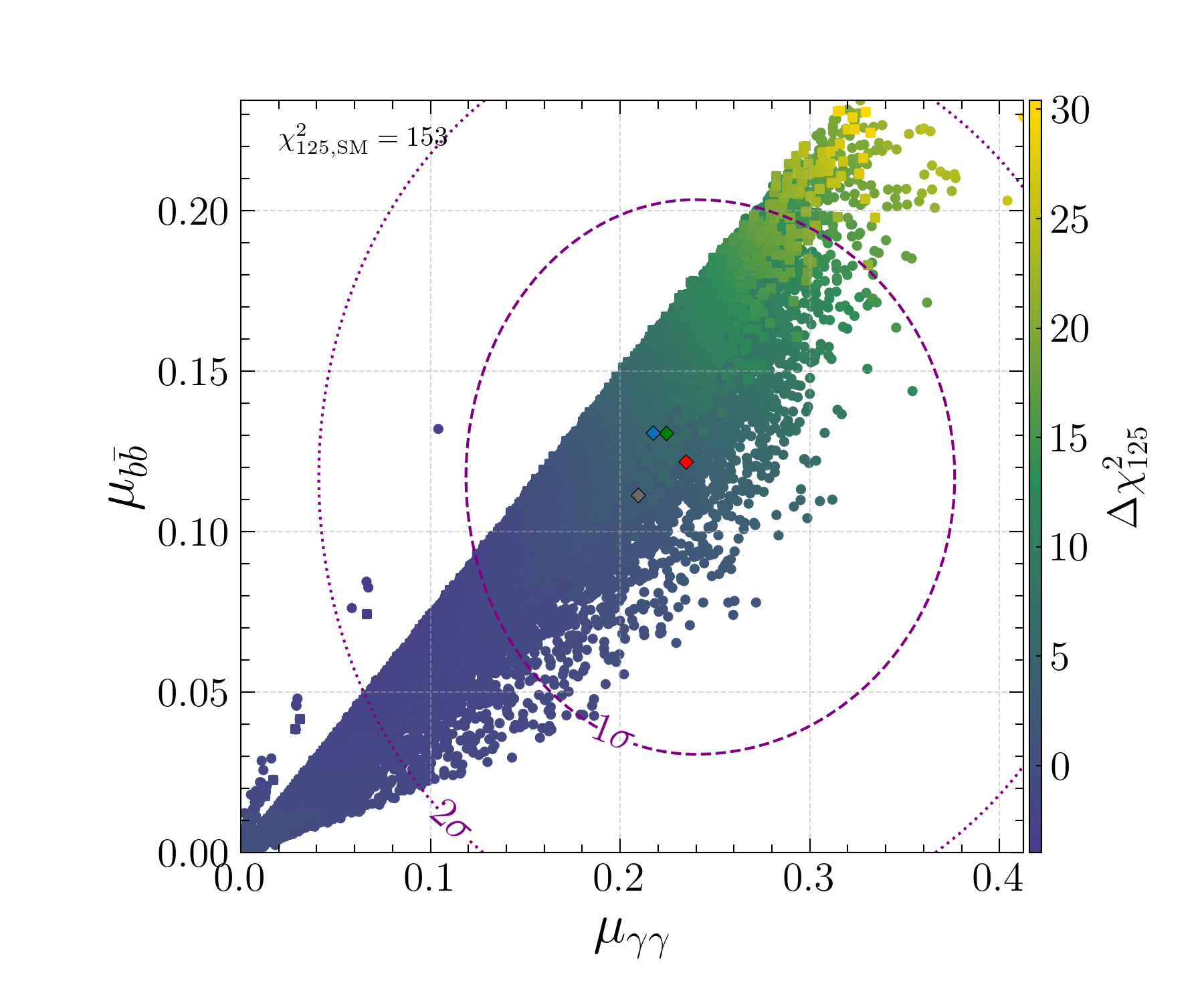}
        \includegraphics{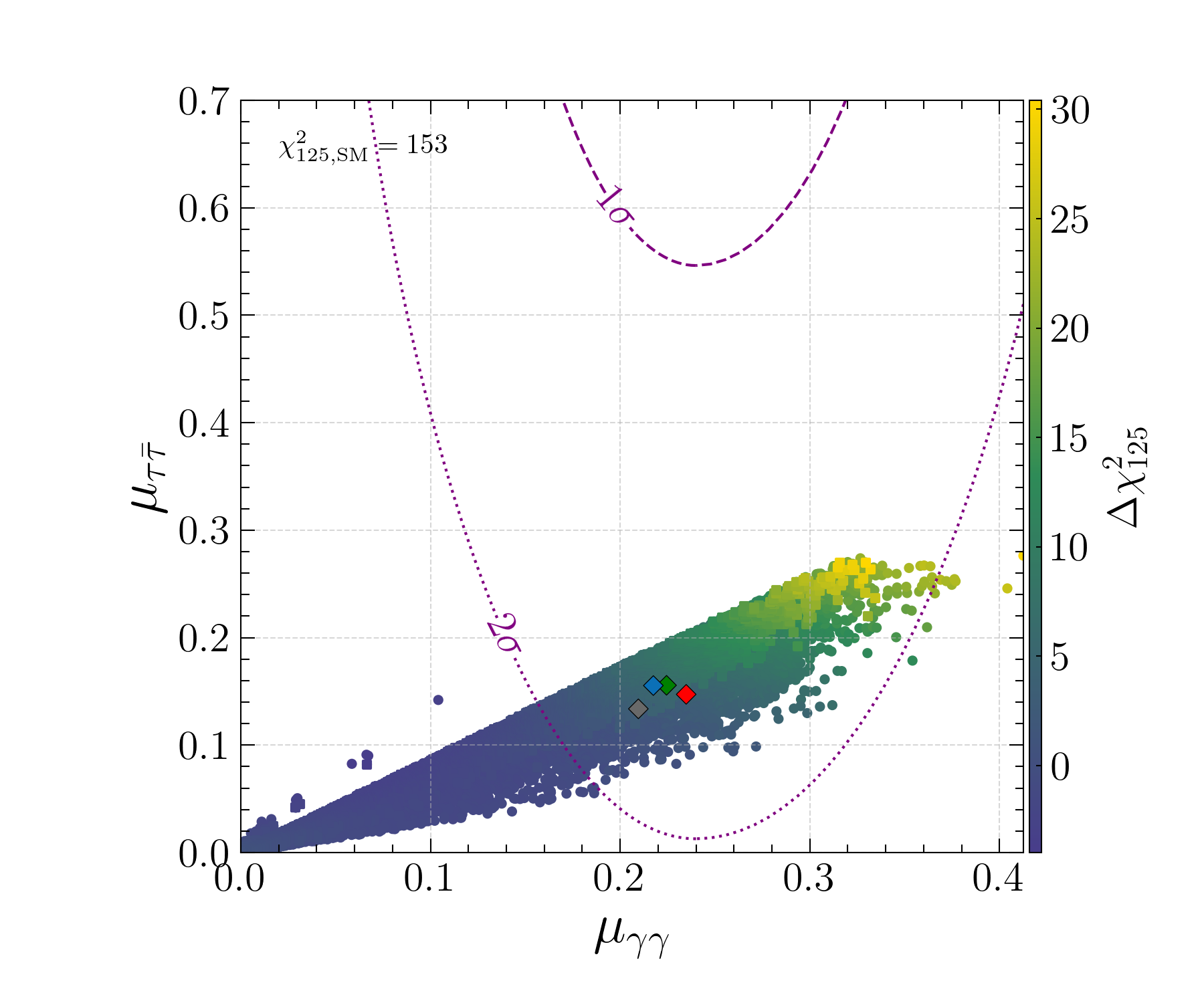}
        }
\caption{Projection of the samples surviving the experimental restrictions onto the $\mu_{\gamma\gamma}-\mu_{b\bar{b}}$ plane (left) and $\mu_{\gamma\gamma}-\mu_{\tau\bar{\tau}}$ (right). 
$\mu_{\tau\bar{\tau}}$  is the signal rate of the di-$\tau$ channel for $h_s$ near 95 GeV predicted by the samples.
The central value and uncertainty of $\mu_{\tau\bar{\tau}}$ observed by CMS in~\cite{CMS:2022goy} is taken as $\mu^{\rm CMS}_{\tau\bar{\tau}} = 1.38^{+0.69}_{-0.55}$~\cite{Ellwanger:2023zjc}.
The color coding indicates the value of $\Delta\chi^2_{125}$ that quantifies the Higgs data fit with respect to the SM result. 
The round and square points denote {\bf Case-I} and {\bf Case-II}, respectively.
Points enclosed within the dashed and dotted purple lines can explain the two excesses at  $1\sigma$ and $2\sigma$ levels, corresponding to $\chi^2_{\gamma\gamma + b\bar{b}(\tau\bar{\tau})} \leq 2.30$ and $\chi^2_{\gamma\gamma + b\bar{b}(\tau\bar{\tau})} \leq 6.18$, respectively.  \label{Fig6}}
\end{figure*}

The last two panels in Fig.\ref{Fig2} also suggest that the current LHC searches for electroweakinos pose minimal constraints on the presented samples. It is crucial to recognize that LHC constraints on the SUSY particle (sparticles) spectrum depend on the mass differences between the particles produced at the collider and the DM candidate.
 In scenarios with a compressed spectrum, Bino-Wino co-annihilation could be probed through the Wino-pair productions at the LHC. 
 However, this approach proves ineffective in this study. As illustrated in Figure.16 of Ref.~\cite{ATLAS:2021moa}, there are no restrictions on the co-annihilation region for $m_{\tilde{\chi}_1^0} \gtrsim 220~{\rm GeV}$.
For scenarios with significant mass splitting, Higgsino-pair production might be considered as a viable detection approach.
However, this method is also inefficient here, as inferred from Figure.12 in Ref.~\cite{ATLAS:2021yqv}, because both Higgsino and $m_{\tilde{\chi}_1^0}$ are predicted to be heavy. 
To solidify this conclusion, simulations of electroweakino production were performed for four benchmark points using the package \texttt{CheckMATE-2.0.26}~\cite{Drees:2013wra,Dercks:2016npn, Kim:2015wza}. The simulations consistently yielded low R-values, indicating weak signals at the LHC. 
The essential reason for these weak restrictions is that explaining the excesses in this study mainly hinges on the Higgs mixings rather than the presence of light sparticles needed to make significant SUSY contributions to $C_{h_s g g}$ and $C_{h_s \gamma \gamma}$. 

In Fig.\ref{Fig3}, the SD and SI DM-nucleon scattering cross sections as a function of the LSP mass $m_{\tilde{\chi}_1^0}$ are plotted on the left panel and the right panel, respectively. These cross sections are derived from the samples identified in Fig.\ref{Fig2}. For comparison, the figure also includes the upper limits from the 2022 results of the LZ experiment at 90\% confidence level~\cite{LZ:2022lsv}, the latest preliminary 2024 limits combined with the 2022 limits~\cite{LZ2024slides}, the projected sensitivities for future LZ runs~\cite{LZ:2018qzl} and the neutrino background expected in the experiment~\cite{Billard:2013qya}.  From the left panel, it is evident that numerous samples fall well below the current upper constraints and the projected limits for the SD scattering cross-sections. The right panel reveals that the SI scattering cross-sections are subject to stronger restrictions compared to the SD scattering cross-sections, especially in the high mass region of $m_{\tilde{\chi}_1^0}$. Despite the tight constraint, a significant number of samples remain below the current and projected upper limits. A few of these samples fall within the neutrino background, posing a challenge for future detections. Overall, the plots indicate that the interpretation of the observed excesses are consistent with the current DM direct detection experiments.    

Fig.~\ref{Fig4} presents the two-dimensional PL functions on the $|C_{h_s VV}| - |C_{h_s gg}|$, $|C_{h_s b\bar{b}}| - |C_{h_s t\bar{t}}|$, $|C_{h_s \gamma\gamma}| - |C_{h_s gg}|$, and ${\rm Br}(h_s\to b\bar{b}) - {\rm Br}(h_s\to \gamma\gamma)$ planes. The solid and dashed lines correspond to the contours for $1\sigma$ and $2\sigma$ confidence intervals, respectively.  
The figure reveals that $|C_{h_s V V}| \simeq |C_{h_s t \bar{t}}|$ and $|C_{h_s \gamma \gamma}| \gtrsim |C_{h_s g g}|$ (both slightly larger than $|C_{h_s t \bar{t}}|$). These relationships align with the predictions from Eqs.(\ref{hs-couplings}) and the associated discussions.
The figure also illustrates that explaining the excesses at the $1\sigma$ level requires the normalized couplings $|C_{h_s t \bar{t}}|$ to be within the range $0.23 \lesssim |C_{h_s t \bar{t}}| \lesssim 0.45$, $|C_{h_s b \bar{b}}|$ within $0.15 \lesssim |C_{h_s b \bar{b}}| \lesssim 0.45$, with a ratio $|C_{h_s t \bar{t}}|/|C_{h_s b \bar{b}}| \simeq 1.4$, $0.27 \lesssim |C_{h_s \gamma \gamma}| \lesssim 0.51$ and $0.23 \lesssim |C_{h_s g g}| \lesssim 0.45$. 
Additionally, the branching ratios for $h_s$ are constrained to $2.0 \times 10^{-3} \lesssim {\rm Br}_{\rm SUSY}(h_s \to \gamma \gamma) \lesssim 4.15 \times 10^{-3}$ and $ 74\% \lesssim {\rm Br}_{\rm SUSY} (h_s \to b \bar{b}) \lesssim 83 \%$.

Fig.~\ref{Fig5} depicts the normalized couplings (strengths of interaction) and branching ratios (decay probabilities) for the SM-like Higgs boson. 
As the figure shows, the normalized couplings $|C_{h V V}|$, $|C_{h t \bar{t}}|$, $|C_{h b \bar{b}}|$, $|C_{h g g}|$, and $|C_{h \gamma \gamma}|$ are centered around 0.926, 0.926, 0.952, 0.935, and 1.04, respectively, for a $1\sigma$ level interpretation of the excesses. 
These couplings agree well with the SM predictions within $10\%$ uncertainties. 
The branching ratio ${\rm Br}(h \to b \bar{b})$ ranges from $55.4 \%$ to $58.6 \%$, closely matching the SM prediction of $(57.7 \pm 1.8)\%$~\cite{LHCHiggsCrossSectionWorkingGroup:2013rie}. While ${\rm Br}(h \to \gamma \gamma)$  varies from $3.02 \times 10^{-3}$ to $3.20 \times 10^{-3}$, showing a significant deviation from the SM prediction of $(2.28 \pm 0.11) \times 10^{-3}$~\cite{LHCHiggsCrossSectionWorkingGroup:2013rie}, the di-photon signal of $h$ remains comparable with LHC Higgs data since the samples predict relatively low branching ratios of $h \to gg$ ($C_{hgg} < 1$).

The left panel of Fig.~\ref{Fig6} explores the compatibility of the di-photon and $b\bar{b}$ excesses within the $\mathbb{Z}_3$-NMSSM. It displays the surviving parameter points (after applying all experimental constraints described in Section~\ref{scanStrategy}) plotted in the $\mu_{\gamma\gamma}$-$\mu_{b\bar{b}}$ plane.
The dashed and dotted lines delineate the regions in which the two excesses can be accommodated at a level of $1\sigma$ and $2\sigma$, respectively. This plot illustrates that the model can adequately explain both excesses at the $1\sigma$ level.
The round and square points represent the dominant DM co-annihilation mechanisms. 
 Notably, {\bf Case-I} predicts signal strengths closer to the observed central values compared to {\bf Case-II}. This can be attributed to the larger contribution of charginos to the coupling $C_{h_s \gamma \gamma}$ in {\bf Case-I}.
 
To ensure a robust analysis that encompasses the 13 TeV LHC Higgs data, the latest version of HiggsTools~\cite{Bahl:2022igd} is employed to compute a $\chi^2$ value for each sample, quantifying the agreement with experimental measurements. Samples satisfying the condition $\Delta\chi^2_{125} \equiv \chi^2_{125} - \chi^2_{125, \rm SM} \lesssim 6.18$ where $\chi^2_{125, \rm SM} = 153$ is the SM fit result, would be considered valid as they provide a good description of the Higgs data at an approximate $2\sigma$ C.L. (assuming Gaussian uncertainties)~\cite{Muhlleitner:2020wwk}.
The analysis leads to the exclusion of about half of the samples that meet $\chi^2_{\gamma\gamma + b\bar{b}} \leq 2.30$ while simultaneously satisfying other experimental constraints. 
The color coding in Fig.~\ref{Fig6} indicates the value of $\Delta\chi^2_{125}$ and illustrates a strong tension between the interpretation of the 95 GeV excess with $\mu_{\gamma\gamma} \gtrsim 0.23$ and $\mu_{b\bar{b}} \gtrsim 0.11$ and the Higgs data fit. The four benchmark points with $\chi^2_{125} \simeq 159$ are thus facing stringent constraints from this. Consequently, the analysis restricts the normalized couplings of $h_s$ to be $|C_{h_s t \bar{t}}| \lesssim 0.39$, $|C_{h_s b \bar{b}}| \lesssim 0.39$, $|C_{h_s g g}| \lesssim 0.40$ and $|C_{h_s \gamma\gamma}| \lesssim 0.43$. Meanwhile, the lower limit of the normalized couplings of $h$ are raised to $|C_{h t \bar{t}}| \gtrsim 0.93$, $|C_{h b \bar{b}}| \gtrsim 0.93$, $|C_{h g g}| \gtrsim 0.92$ and $|C_{h \gamma\gamma}| \gtrsim 1.03$.

Additionally, the distribution of the samples are also projected onto the $\mu_{\gamma\gamma}-\mu_{\tau\bar{\tau}}$ plane in the right panel of Fig.~\ref{Fig6}. 
$\mu_{\tau\bar{\tau}}$  is the signal rate of the di-$\tau$ channel for $h_s$ near 95 GeV predicted by the samples.
The central value and uncertainty of $\mu_{\tau\bar{\tau}}$ observed by CMS in~\cite{CMS:2022goy} is taken as $\mu^{\rm CMS}_{\tau\bar{\tau}} = 1.38^{+0.69}_{-0.55}$~\cite{Ellwanger:2023zjc}. This plot illustrates that the di-photon and $\tau\bar{\tau}$ excesses can only be simultaneously accommodated at a $2\sigma$ level. The primary reason for this limitation is that $\mu_{\tau\bar{\tau}}$ approximates $\mu_{b\bar{b}}$ since the couplings of $h_s$ to tau leptons and bottom quarks are predicted to be equal, i.e. $C_{h_s \tau \bar{\tau}} = C_{h_s b \bar{b}}$. This finding aligns with the observation in Ref.~\cite{Iguro:2022dok} that the interpretation of the CMS di-$\tau$ excess as a CP-even scalar is excluded at the $1\sigma$ level by the ATLAS $t\bar{t}+\tau\bar{\tau}$ search~\cite{ATLAS:2022yrq}. Moreover, the study in~\cite{Banik:2023ecr} has highlighted that the possibility of accommodating the di-tau excess for a CP-even scalar is ruled out model-independently by CMS searches for di-tau resonances produced in association with top quarks~\cite{CMS:2024ulc}.

To further elucidate the interpretation of the observed excesses, four benchmark points (P1,  P2,  P3, P4) are selected to exemplify distinct scenarios within the model. These points are denoted as diamonds colored red, green, blue and grey, respectively, in the previous figures.  Detailed information for these benchmarks is provided in Table.~\ref{BP1BP2} and Table.~\ref{BP3BP4}. Specifically, P1 and P2 correspond to the {\bf Case-I} and {\bf Case-II} scenarios, respectively. While, P3 and  P4 represent two rare scenarios where the largest contributions to the DM annihilation cross section arise from the channels $\tilde{\chi}_1^+ \tilde{\chi}_1^+ \to W^+ W^+$ and $\tilde{\chi}_1^0 \tilde{\chi}_1^0 \to t\bar{t}$, respectively. 
As illustrated in Fig.~\ref{Fig3}, P1 and P3 could evade DM direct detections due to their prediction of $\sin2\beta + m_{\tilde{\chi}^0_1}/ \mu \simeq -0.03$, which aligns with the Blind Spot condition, thereby reducing the SI scattering cross-section to values approaching or below the neutrino floor. In contrast, P2 and P4 do not exhibit this mitigating effect, and hence, they are subject to exclusion by the projected sensitivity of the LZ experiment. 
Regarding constraints from the LHC experiments, given that all points predict compressed spectra and relatively large masses for the LSP, the LHC searches for SUSY would have no exclusion capability on them. Simulations of electroweakino productions using the package \texttt{CheckMATE-2.0.26} indicate that their R-values are all less than 0.6. The most significant signal regions and their experimental analyses are listed at the bottom row for each point.

\begin{table}[t]
\centering
\caption{\label{BP1BP2} Details of the benchmark points P1 and P2. Both of them can explain the di-photon and $b\bar{b}$ excesses at a $1\sigma$ level while satisfying the applied constraints. They correspond to the {\bf Case-I} scenario and the {\bf Case-II} scenario, respectively, and are represented as diamonds colored red and green in the previous figures. The notations $d_i$, $u_i$, and $\ell_i$ in the annihilation final state denote the $i$th generation of down-type quarks, up-type quarks, and leptons, respectively. }
\resizebox{1\textwidth}{!}{
 \begin{tabular}{lrlr|lrlr}
\hline \hline
\multicolumn{4}{c|}{\bf Benchmark Point P1}                                                                                                								& \multicolumn{4}{c}{\bf Benchmark Point P2}
 \\ \hline
$\mu$                			& 676.6~GeV 	&$\lambda$             		& 0.271				&$\mu$                		& 849.3~GeV		&$\lambda$             		& 0.237 \\
$m_B$					& 97.39~GeV	&$\kappa$              		& -0.192				&$m_B$				& 98.47~GeV		& $\kappa$              		&  -0.129\\
$A_t$					& 2962~GeV		&$\delta$         			& 0.049				&$A_t$				& 2881~GeV			& $\delta$         			& 0.044\\
$M_1$               			& -359.2~GeV 	&$\tan{\beta}$                	& 3.737				&$M_1$               		& -650.9~GeV		& $\tan{\beta}$                	& 4.221\\
$M_2$             		 	& 378.8~GeV	&	 					&					&$M_2$             		& 672.1~GeV		& 		         			& 	\\
\hline
$m_{\tilde{\chi}_1^0}$   	 & -362.5~GeV	&$m_{h_s}$                		& 95.02~GeV		&$m_{\tilde{\chi}_1^0}$   		& -654.8~GeV	&$m_{h_s}$                		& 95.55~GeV\\
$m_{\tilde{\chi}_2^0}$   	 & 386.3~GeV	&$m_{h}$              		& 124.8~GeV		&$m_{\tilde{\chi}_2^0}$   		& 678.7~GeV	&$m_{h}$              		& 125.0~GeV\\
$m_{\tilde{\chi}_3^0}$   	 & -683.3~GeV	&$m_{H}$                  		& 2900~GeV			&$m_{\tilde{\chi}_3^0}$   		& -847.9~GeV	&$m_{H}$                  		& 3948~GeV\\
$m_{\tilde{\chi}_4^0}$   	 & 703.8~GeV	&$m_{A_s}$                 	& 1655~GeV			&$m_{\tilde{\chi}_4^0}$   		& 885.1~GeV	&$m_{A_s}$                 	& 1601~GeV\\
$m_{\tilde{\chi}_5^0}$   	 & -961.7~GeV	&$m_{A_H}$                	& 2902~GeV			&$m_{\tilde{\chi}_5^0}$   		& -939.4~GeV	&$m_{A_H}$                	& 3947~GeV\\
$m_{\tilde{\chi}_1^\pm}$	 & 386.5~GeV	&$m_{H^\pm}$			& 2903~GeV			&$m_{\tilde{\chi}_1^\pm}$		& 678.9~GeV	&$m_{H^\pm}$			& 3947~GeV\\
$m_{\tilde{\chi}_2^\pm}$	 & 703.7~GeV	& 						& 					&$m_{\tilde{\chi}_2^\pm}$	 	& 885.2~GeV	&						& \\
\hline
\multicolumn{2}{l}{$\Omega h^2 $}			&\multicolumn{2}{l|}{0.101}
&\multicolumn{2}{l}{$\Omega h^2 $}		&\multicolumn{2}{l}{0.113}\\
\multicolumn{2}{l}{$\sigma^{SI}_p$,  ~$\sigma^{SD}_n$}		&\multicolumn{2}{l|}{$1.2\times 10^{-48}{\rm ~cm^2}, ~7.2\times 10^{-43}{\rm ~~cm^2}$}
&\multicolumn{2}{l}{$\sigma^{SI}_p$, ~$\sigma^{SD}_n$}		&\multicolumn{2}{l}{$1.0\times 10^{-46}{\rm ~cm^2},  ~9.3\times 10^{-43}{\rm ~~cm^2}$}\\
\hline
\multicolumn{2}{l}{$\mu_{\gamma\gamma}, ~\mu_{b\bar{b}}$}		&\multicolumn{2}{l|}{0.235, ~0.121}		
&\multicolumn{2}{l}{$\mu_{\gamma\gamma}, ~\mu_{b\bar{b}}$}	&\multicolumn{2}{l}{0.224, ~0.130} \\
\multicolumn{2}{l}{$V_{h_s}^{S}, ~V_{h_s}^{SM}, ~V_{h_s}^{NSM}$}  & \multicolumn{2}{l|}{0.935, -0.355, -0.012}
& \multicolumn{2}{l}{$V_{h_s}^{S}, ~V_{h_s}^{SM}, ~V_{h_s}^{NSM}$}  & \multicolumn{2}{l}{0.931, -0.365, -0.008} \\
\multicolumn{2}{l}{$C_{h_s gg}, ~C_{h_s VV}, ~C_{h_s \gamma\gamma}, ~C_{h_s t \bar{t}}, ~C_{h_s b\bar{b}}$} 	& \multicolumn{2}{l|}{0.372, ~~0.355, ~~0.410, ~~0.358, ~~0.310}
&\multicolumn{2}{l}{$C_{h_s gg}, ~C_{h_s VV}, ~C_{h_s \gamma\gamma}, ~C_{h_s t \bar{t}}, ~C_{h_s b\bar{b}}$} 	& \multicolumn{2}{l}{0.381,  ~~0.365, ~~0.417, ~~0.367, ~~0.331} \\
 \multicolumn{2}{l}{$C_{h gg}, ~~C_{h VV}, ~~C_{h \gamma\gamma}, ~~C_{h t \bar{t}}, ~~C_{h b\bar{b}}$} 			& \multicolumn{2}{l|}{0.934, ~~0.935, ~~1.044, ~~0.933, ~~0.955}
& \multicolumn{2}{l}{$C_{h gg}, ~~C_{h VV}, ~~C_{h \gamma\gamma}, ~~C_{h t \bar{t}}, ~~C_{h b\bar{b}}$} 		& \multicolumn{2}{l}{0.931,  ~~0.931, ~~1.039, ~~0.930, ~~0.946} \\
\hline
\multicolumn{2}{l}{$N_{11}, ~N_{12}, ~N_{13}, ~N_{14}, ~N_{15}$}   &\multicolumn{2}{l|}{~0.997,   	~~0.001,  ~~0.074,  ~~0.024,  ~-0.009}
& \multicolumn{2}{l}{$N_{11}, ~N_{12}, ~N_{13}, ~N_{14}, ~N_{15}$}   &\multicolumn{2}{l}{-0.010, 	~~0.044, ~-0.111,  ~~0.110, ~-0.987} \\
\multicolumn{2}{l}{$N_{21}, ~N_{22}, ~N_{23}, ~N_{24}, ~N_{25}$}   &\multicolumn{2}{l|}{-0.010, 	~-0.976, ~~0.179,  ~-0.129, ~-0.005}
& \multicolumn{2}{l}{$N_{21}, ~N_{22}, ~N_{23}, ~N_{24}, ~N_{25}$}   &\multicolumn{2}{l}{-0.025,	~-0.280,	 ~~0.680, ~-0.657,  ~-0.162} \\
\multicolumn{2}{l}{$N_{31}, ~N_{32}, ~N_{33}, ~N_{34}, ~N_{35}$}   &\multicolumn{2}{l|}{-0.070,	~~0.036, ~~0.693, ~~0.702,  ~-0.142}
& \multicolumn{2}{l}{$N_{31}, ~N_{32}, ~N_{33}, ~N_{34}, ~N_{35}$}   &\multicolumn{2}{l}{-0.122,	~-0.050,  ~-0.701,  ~-0.702,  ~-0.002} \\
\multicolumn{2}{l}{$N_{41}, ~N_{42}, ~N_{43}, ~N_{44}, ~N_{45}$}   &\multicolumn{2}{l|}{~0.035,	~-0.217, ~-0.685, ~~0.694, ~~0.013}
& \multicolumn{2}{l}{$N_{41}, ~N_{42}, ~N_{43}, ~N_{44}, ~N_{45}$}   &\multicolumn{2}{l}{~0.008,	~-0.958,	 ~-0.167,  ~~0.234, ~~0.002} \\
\multicolumn{2}{l}{$N_{51}, ~N_{52}, ~N_{53}, ~N_{54}, ~N_{55}$}   &\multicolumn{2}{l|}{~0.002,	~-0.003,  ~-0.110,  ~-0.091, ~-0.990}
& \multicolumn{2}{l}{$N_{51}, ~N_{52}, ~N_{53}, ~N_{54}, ~N_{55}$}   &\multicolumn{2}{l}{~0.992,	~-0.006,  ~-0.068,  ~-0.106, ~-0.000} \\
\hline
\multicolumn{2}{l}{ Coannihilations }                         & \multicolumn{2}{l|}{Fractions [\%]} 			& \multicolumn{2}{l}{Coannihilations}                                       & \multicolumn{2}{l}{Fractions [\%]}                                      \\
\multicolumn{2}{l}{$\tilde{\chi}_2^0\tilde{\chi}_1^- \to d_i \bar{u}_i /  Z W^- / \nu_{\ell_i} \ell_i^- /\gamma W^- $} 			& \multicolumn{2}{l|}{9.8/6.5/3.3/1.8}
& \multicolumn{2}{l}{$\tilde{\chi}_2^0 \tilde{\chi}_2^0 \to W^- W^+ $} 					& \multicolumn{2}{l}{9.0}  \\

\multicolumn{2}{l}{$\tilde{\chi}_2^0 \tilde{\chi}_2^0 \to W^- W^+ $} 					& \multicolumn{2}{l|}{7.8}
& \multicolumn{2}{l}{$\tilde{\chi}_2^0\tilde{\chi}_1^- \to d_i \bar{u}_i /  Z W^- / \nu_{\ell_i} \ell_i^- /\gamma W^- $} 		& \multicolumn{2}{l}{8.3/7.3/2.9/2.1}   \\

\multicolumn{2}{l}{$\tilde{\chi}_1^\pm  \tilde{\chi}_1^\pm \to W^\pm W^\pm $} 					& \multicolumn{2}{l|}{7.6}
& \multicolumn{2}{l}{$\tilde{\chi}_1^\pm  \tilde{\chi}_1^\pm \to W^\pm W^\pm $} 					& \multicolumn{2}{l}{8.4}   \\

\multicolumn{2}{l}{$\tilde{\chi}_1^\pm \tilde{\chi}_1^\mp \to  Z Z  /  W^\pm W^\mp /\gamma Z  / u_i \bar{u}_i / d_i \bar{d}_i $} 		& \multicolumn{2}{l|}{4.7/4.1/2.7/2.4/2.4}
& \multicolumn{2}{l}{$\tilde{\chi}_1^\pm \tilde{\chi}_1^\mp \to  Z Z  /  W^\pm W^\mp /\gamma Z  / u_i \bar{u}_i / d_i \bar{d}_i $} 	& \multicolumn{2}{l}{5.4/4.8/3.2/2.1/2.0} \\  \hline

\multicolumn{2}{l}{ Decays }                         & \multicolumn{2}{l|}{Branching ratios [\%]} 		& \multicolumn{2}{l}{Decays}                                       & \multicolumn{2}{l}{Branching ratios [\%]}\\

\multicolumn{2}{l}{$\tilde{\chi}^0_2 \to \tilde{\chi}^0_1 Z^\ast$}      & \multicolumn{2}{l|}{100}
&\multicolumn{2}{l}{$\tilde{\chi}^0_2 \to \tilde{\chi}^0_1 Z^\ast$}      & \multicolumn{2}{l}{100} \\

\multicolumn{2}{l}{$\tilde{\chi}^0_3 \to \tilde{\chi}^\pm_1 W^\mp / \tilde{\chi}^0_2 Z / \tilde{\chi}^0_1 h /  \tilde{\chi}^0_1 Z$}      & \multicolumn{2}{l|}{64.1/29.5 /3.15 /1.18}
&\multicolumn{2}{l}{$\tilde{\chi}^0_3 \to \tilde{\chi}^\pm_1 W^{\mp} / \tilde{\chi}^0_2 Z / \tilde{\chi}^0_1 h / \tilde{\chi}^0_1 h_s $}      & \multicolumn{2}{l}{63.7/27.1/5.75/2.91} \\

\multicolumn{2}{l}{$\tilde{\chi}^0_4 \to \tilde{\chi}^\pm_1 W^\mp / \tilde{\chi}^0_2 h / \tilde{\chi}^0_2 h_s /  \tilde{\chi}^0_1 Z$}      & \multicolumn{2}{l|}{66.3/20.8/7.04/3.75}
&\multicolumn{2}{l}{$\tilde{\chi}^0_4 \to \tilde{\chi}^\pm_1 W^\mp /\tilde{\chi}^0_2 h / \tilde{\chi}^0_2 h_s/  \tilde{\chi}^0_1 Z$}      & \multicolumn{2}{l}{69.7/15.6/8.80/5.46 } \\

\multicolumn{2}{l}{$\tilde{\chi}^0_5 \to \tilde{\chi}^\pm_2 W^\mp / \tilde{\chi}^0_3 h / \tilde{\chi}^0_4 Z / \tilde{\chi}^\pm_1 W^\mp$}      & \multicolumn{2}{l|}{37.8/20.3/17.6/12.8}
&\multicolumn{2}{l}{$\tilde{\chi}^0_5 \to \tilde{\chi}^\pm_2 W^\mp / \tilde{\chi}^0_2 Z / \tilde{\chi}^0_1 h / \tilde{\chi}^0_1 Z$}      & \multicolumn{2}{l}{67.1/31.7/0.51/0.35} \\

\multicolumn{2}{l}{$\tilde{\chi}^+_1 \to \tilde{\chi}^0_1 W^{+\ast}$}      & \multicolumn{2}{l|}{100}
&\multicolumn{2}{l}{$\tilde{\chi}^+_1 \to \tilde{\chi}^0_1 W^{+\ast}$}      & \multicolumn{2}{l}{100}  \\

\multicolumn{2}{l}{$\tilde{\chi}^+_2 \to \tilde{\chi}^0_2 W^+ / \tilde{\chi}^+_1 Z / \tilde{\chi}^+_1 h / \tilde{\chi}^+_1 h_s$}      & \multicolumn{2}{l|}{33.8/32.2/21.7/7.25}
&\multicolumn{2}{l}{$\tilde{\chi}^+_2 \to \tilde{\chi}^0_2 W^+ / \tilde{\chi}^+_1 Z / \tilde{\chi}^+_1 h / \tilde{\chi}^+_1 h_s$}      & \multicolumn{2}{l}{36.2/32.7/15.9/8.95}\\

\multicolumn{2}{l}{$h_s \to b \bar{b} / \tau^+ \tau^- / gg / c \bar{c} / W ^+W^{-\ast} / \gamma \gamma$}      & \multicolumn{2}{l|}{79.9/8.85/5.38/5.07/0.48/0.28}
&\multicolumn{2}{l}{$h_s \to b \bar{b} / \tau^+ \tau^- / gg / c \bar{c} / W ^+W^{-\ast} / \gamma \gamma$}      & \multicolumn{2}{l}{80.6/8.93/5.02/4.70/0.41/0.26}\\

\multicolumn{2}{l}{$h \to b \bar{b} / W ^+W^{-\ast} / \tau^+ \tau^- / gg / ZZ^{\ast} / \gamma \gamma$}      & \multicolumn{2}{l|}{57.8/25.6/6.71/4.58/2.32/0.31}
&\multicolumn{2}{l}{$h \to b \bar{b} / W ^+W^{-\ast} / \tau^+ \tau^- / gg / ZZ^{\ast} / \gamma \gamma$}      & \multicolumn{2}{l}{57.2/26.2/6.63/4.61/2.39/0.31}\\

\multicolumn{2}{l}{$H \to  \tilde{\chi}^\pm_1 \tilde{\chi}^\mp_2 /  \tilde{\chi}^0_2 \tilde{\chi}^0_{3,4}/ t\bar{t}  /   A_s Z$}      & \multicolumn{2}{l|}{39.2/18.2/17.3/6.52}
&\multicolumn{2}{l}{$H \to \tilde{\chi}^\pm_1 \tilde{\chi}^\mp_2 /  \tilde{\chi}^0_2 \tilde{\chi}^0_{3,4}/ t\bar{t}  / A_s Z$}      & \multicolumn{2}{l}{36.7/13.5/15.0/4.93}\\

\multicolumn{2}{l}{$A_s \to  \tilde{\chi}^\pm_2 \tilde{\chi}^\mp_2 / \tilde{\chi}^0_3 \tilde{\chi}^0_3 / \tilde{\chi}^0_4 \tilde{\chi}^0_4/ \tilde{\chi}^\pm_1 \tilde{\chi}^\mp_2 $}     & \multicolumn{2}{l|}{45.0/25.8/22.7/3.68}
&\multicolumn{2}{l}{$A_s \to  \tilde{\chi}^\pm_1 \tilde{\chi}^\mp_2 / \tilde{\chi}^0_2 \tilde{\chi}^0_4 / \tilde{\chi}^0_1 \tilde{\chi}^0_3 / \tilde{\chi}^\pm_1 \tilde{\chi}^\mp_1 $}    & \multicolumn{2}{l}{53.7/26.8/7.91/6.81}\\

\multicolumn{2}{l}{$A_H \to  \tilde{\chi}^\pm_1 \tilde{\chi}^\mp_2 /  t\bar{t}  /  \tilde{\chi}^0_2 \tilde{\chi}^0_{4,3}/ hA_s $}      & \multicolumn{2}{l|}{35.5/17.4/16.4/5.36}
&\multicolumn{2}{l}{$A_H \to \tilde{\chi}^\pm_1 \tilde{\chi}^\mp_2 / t\bar{t}  /  \tilde{\chi}^0_{2} \tilde{\chi}^0_4 / \tilde{\chi}^\pm_1 \tilde{\chi}^\mp_1$}      & \multicolumn{2}{l}{29.0/13.8/8.96/7.23}\\

\multicolumn{2}{l}{$H^+ \to  \tilde{\chi}^0_2 \tilde{\chi}^+_2 / \tilde{\chi}^0_4 \tilde{\chi}^+_1 / \tilde{\chi}^0_3 \tilde{\chi}^+_1 / t \bar{b} /A_s W^+$}      & \multicolumn{2}{l|}{20.6/19.6/18.6/18.1/6.6}
&\multicolumn{2}{l}{$H^+ \to  \tilde{\chi}^0_2 \tilde{\chi}^+_2 / \tilde{\chi}^0_4 \tilde{\chi}^+_1/ \tilde{\chi}^0_3 \tilde{\chi}^+_1 /t\bar{b} / A_s W^+$}      & \multicolumn{2}{l}{20.6/19.7/15.3/14.8//4.99}\\
\hline

\multicolumn{2}{l}{$R$ value: 0.32}  & \multicolumn{2}{l|}{Signal Region: E-high-ee-30 in Ref.~\cite{ATLAS:2019lng}}
& \multicolumn{2}{l}{$R$ value: 0.02}  & \multicolumn{2}{l}{Signal Region: SR-G05 in Ref.~\cite{CMS:2017moi}} \\
\hline \hline

\end{tabular}}
\end{table}

\begin{table}[t]
\centering
\caption{\label{BP3BP4} Details of the benchmark points P3 and P4. They can also explain the di-photon and $b\bar{b}$ excesses at a $1\sigma$ level while satisfying the applied constraints. They correspond to two rare scenarios where the largest contributions to the DM annihilation cross section arise from the channels $\tilde{\chi}_1^+ \tilde{\chi}_1^+ \to W^+ W^+$ and $\tilde{\chi}_1^0 \tilde{\chi}_1^0 \to t\bar{t}$, respectively. }
\resizebox{1\textwidth}{!}{
\begin{tabular}{lrlr|lrlr}
\hline \hline
\multicolumn{4}{c|}{\bf Benchmark Point P3}                                                                                                								& \multicolumn{4}{c}{\bf Benchmark Point P4}
 \\ \hline
$\mu$              	& 668.1~GeV	&$\lambda$             	& 0.144 		&$\mu$         	& 544.9~GeV	&$\lambda$             		& 0.164\\
$m_B$			& 91.59~GeV	&$\kappa$              	& 0.037		&$m_B$		& 95.51~GeV	&$\kappa$              		&  0.069\\
$A_t$			& 2961~GeV		&$\delta$         		& 0.088 		&$A_t$		& 3505~GeV		&$\delta$         			& 0.104 \\
$M_1$               	& -324.6~GeV	&$\tan{\beta}$             & 4.100 		&$M_1$        & -385.5~GeV	&$\tan{\beta}$                	& 3.611 \\
$M_2$             	& 342.9~GeV	& 					&			&$M_2$        & 424.7~GeV		& 		         			& 	\\
\hline
$m_{\tilde{\chi}_1^0}$   	 & -327.5~GeV	&$m_{h_s}$                		& 95.24~GeV		&$m_{\tilde{\chi}_1^0}$   		& -387.4~GeV	&$m_{h_s}$                		& 96.10~GeV\\
$m_{\tilde{\chi}_2^0}$   	 & 343.7~GeV	&$m_{h}$              		& 124.7~GeV		&$m_{\tilde{\chi}_2^0}$   		& 414.9~GeV	&$m_{h}$              		& 124.6~GeV\\
$m_{\tilde{\chi}_3^0}$   	 & 352.6~GeV	&$m_{H}$                  		& 2710~GeV			&$m_{\tilde{\chi}_3^0}$   		& 459.6~GeV	&$m_{H}$                  		& 1922~GeV\\
$m_{\tilde{\chi}_4^0}$   	 & -682.3~GeV	&$m_{A_s}$                 	& 576.7~GeV		&$m_{\tilde{\chi}_4^0}$   		& -558.2~GeV	&$m_{A_s}$                 	& 778.2~GeV\\
$m_{\tilde{\chi}_5^0}$   	 & 694.8~GeV	&$m_{A_H}$                	& 2710~GeV			&$m_{\tilde{\chi}_5^0}$   		& 589.7~GeV	&$m_{A_H}$                	& 1922~GeV\\
$m_{\tilde{\chi}_1^\pm}$	 & 351.2~GeV	&$m_{H^\pm}$			& 2714~GeV			&$m_{\tilde{\chi}_1^\pm}$		& 416.1~GeV	&$m_{H^\pm}$			& 1926~GeV\\
$m_{\tilde{\chi}_2^\pm}$	 & 694.8~GeV	& 						& 					&$m_{\tilde{\chi}_1^\pm}$	 	& 588.7~GeV	&						& \\
\hline
\multicolumn{2}{l}{$\Omega h^2 $}			&\multicolumn{2}{l|}{0.126}
&\multicolumn{2}{l}{$\Omega h^2 $}		&\multicolumn{2}{l}{0.122}\\
\multicolumn{2}{l}{$\sigma^{SI}_p$,  ~$\sigma^{SD}_n$}		&\multicolumn{2}{l|}{$1.5\times 10^{-49}{\rm ~cm^2}, ~6.8\times 10^{-43}{\rm ~~cm^2}$}
&\multicolumn{2}{l}{$\sigma^{SI}_p$, ~$\sigma^{SD}_n$}		&\multicolumn{2}{l}{$4.5\times 10^{-47}{\rm ~cm^2},  ~3.2\times 10^{-42}{\rm ~~cm^2}$}\\
\hline
\multicolumn{2}{l}{$\mu_{\gamma\gamma}, ~\mu_{b\bar{b}}$}		&\multicolumn{2}{l|}{0.217, ~0.131}		
&\multicolumn{2}{l}{$\mu_{\gamma\gamma}, ~\mu_{b\bar{b}}$}	&\multicolumn{2}{l}{0.209, ~0.111} \\
\multicolumn{2}{l}{$V_{h_s}^{S}, ~V_{h_s}^{SM}, ~V_{h_s}^{NSM}$}  & \multicolumn{2}{l|}{~0.931, -0.361, ~-0.008}
& \multicolumn{2}{l}{$V_{h_s}^{S}, ~V_{h_s}^{SM}, ~V_{h_s}^{NSM}$}  & \multicolumn{2}{l}{-0.941, -0.340, ~-0.012} \\
\multicolumn{2}{l}{$C_{h_s gg}, ~C_{h_s VV}, ~C_{h_s \gamma\gamma}, ~C_{h_s t \bar{t}}, ~C_{h_s b\bar{b}}$} 	& \multicolumn{2}{l|}{0.380, ~~0.365,~~~0.415,~~~0.367,~0.335}
&\multicolumn{2}{l}{$C_{h_s gg}, ~C_{h_s VV}, ~C_{h_s \gamma\gamma}, ~C_{h_s t \bar{t}}, ~C_{h_s b\bar{b}}$} 	& \multicolumn{2}{l}{0.354,~~~0.340,~~~0.390,~~~0.343,~~0.297} \\
 \multicolumn{2}{l}{$C_{h gg}, ~~C_{h VV}, ~~C_{h \gamma\gamma}, ~~C_{h t \bar{t}}, ~~C_{h b\bar{b}}$} 			& \multicolumn{2}{l|}{0.931,~~~0.931, ~~1.042,~~~0.930,~~~0.946}
& \multicolumn{2}{l}{$C_{h gg}, ~~C_{h VV}, ~~C_{h \gamma\gamma}, ~~C_{h t \bar{t}}, ~~C_{h b\bar{b}}$} 		& \multicolumn{2}{l}{0.938,~~~0.940, ~~1.053,~~~0.939,~~~0.962} \\
\hline
\multicolumn{2}{l}{$N_{11}, ~N_{12}, ~N_{13}, ~N_{14}, ~N_{15}$}   &\multicolumn{2}{l|}{~0.997,   	~~0.000,  ~-0.072,  ~-0.020,  ~-0.004}
& \multicolumn{2}{l}{$N_{11}, ~N_{12}, ~N_{13}, ~N_{14}, ~N_{15}$}   &\multicolumn{2}{l}{-0.991, 	~-0.003,   ~-0.112,  ~-0.062, ~-0.005} \\
\multicolumn{2}{l}{$N_{21}, ~N_{22}, ~N_{23}, ~N_{24}, ~N_{25}$}   &\multicolumn{2}{l|}{-0.008, 	~-0.425,   ~~0.082,  ~-0.085, ~~0.897}
& \multicolumn{2}{l}{$N_{21}, ~N_{22}, ~N_{23}, ~N_{24}, ~N_{25}$}   &\multicolumn{2}{l}{-0.020,	~-0.886,	  ~~0.331,  ~-0.287,  ~~0.152} \\
\multicolumn{2}{l}{$N_{31}, ~N_{32}, ~N_{33}, ~N_{34}, ~N_{35}$}   &\multicolumn{2}{l|}{-0.007,	~-0.883,   ~~0.145, ~-0.083,  ~-0.439}
& \multicolumn{2}{l}{$N_{31}, ~N_{32}, ~N_{33}, ~N_{34}, ~N_{35}$}   &\multicolumn{2}{l}{~0.002,	~-0.176,   ~~0.009,   ~~0.034,  ~-0.984} \\
\multicolumn{2}{l}{$N_{41}, ~N_{42}, ~N_{43}, ~N_{44}, ~N_{45}$}   &\multicolumn{2}{l|}{~0.065,	~-0.039,   ~-0.703,   ~-0.707, ~-0.020}
& \multicolumn{2}{l}{$N_{41}, ~N_{42}, ~N_{43}, ~N_{44}, ~N_{45}$}   &\multicolumn{2}{l}{-0.128,	~~0.039,  ~-0.696,  ~~0.705, ~~0.024} \\
\multicolumn{2}{l}{$N_{51}, ~N_{52}, ~N_{53}, ~N_{54}, ~N_{55}$}   &\multicolumn{2}{l|}{~-0.036,	~~0.197,  ~~0.688,  ~-0.697, ~-0.036}
& \multicolumn{2}{l}{$N_{51}, ~N_{52}, ~N_{53}, ~N_{54}, ~N_{55}$}   &\multicolumn{2}{l}{~0.035,	~-0.428,  ~-0.626,  ~~0.645, ~~0.093} \\
\hline
\multicolumn{2}{l}{ Coannihilations }                         & \multicolumn{2}{l|}{Fractions [\%]} 			& \multicolumn{2}{l}{Coannihilations}                                       & \multicolumn{2}{l}{Fractions [\%]}                                      \\
\multicolumn{2}{l}{$\tilde{\chi}_1^\pm  \tilde{\chi}_1^\pm \to W^\pm W^\pm $} 					& \multicolumn{2}{l|}{7.1}
&\multicolumn{2}{l}{$\tilde{\chi}_1^0\tilde{\chi}_1^0 \to t \bar{t} /  b \bar{b}$} 		& \multicolumn{2}{l}{26.7/ 1.2}  \\

\multicolumn{2}{l}{$\tilde{\chi}_3^0\tilde{\chi}_1^- \to d_i \bar{u}_i /  Z W^- / \nu_{\ell_i} \ell_i^- /\gamma W^- $} 		& \multicolumn{2}{l|}{6.4/4.3/2.2/1.1}
& \multicolumn{2}{l}{$\tilde{\chi}_2^0\tilde{\chi}_1^- \to d_i \bar{u}_i /  Z W^- / \nu_{\ell_i} \ell_i^- /\gamma W^- $} 		& \multicolumn{2}{l}{6.6/3.9/2.2/1.2}   \\

\multicolumn{2}{l}{$\tilde{\chi}_1^\pm \tilde{\chi}_1^\mp \to  Z Z  /  W^\pm W^\mp /\gamma Z  / u_i \bar{u}_i / d_i \bar{d}_i $} 		& \multicolumn{2}{l|}{4.3/3.8/2.5/2.3}
& \multicolumn{2}{l}{$\tilde{\chi}_2^0 \tilde{\chi}_3^0 \to  W^\pm W^\mp$} 					& \multicolumn{2}{l}{5.1}   \\

\multicolumn{2}{l}{$\tilde{\chi}_2^0 \tilde{\chi}_3^0 \to  W^\pm W^\mp $} 		& \multicolumn{2}{l|}{4.0}
& \multicolumn{2}{l}{$\tilde{\chi}_1^\pm \tilde{\chi}_1^\mp \to  Z Z  /  W^\pm W^\mp /\gamma Z  / u_i \bar{u}_i / d_i \bar{d}_i $} 	& \multicolumn{2}{l}{2.7/2.5/1.7/1.6/1.5} \\  \hline

\multicolumn{2}{l}{ Decays }                         & \multicolumn{2}{l|}{Branching ratios [\%]} 		& \multicolumn{2}{l}{Decays}                                       & \multicolumn{2}{l}{Branching ratios [\%]}\\

\multicolumn{2}{l}{$\tilde{\chi}^0_2 \to \tilde{\chi}^0_1 Z^\ast$}      & \multicolumn{2}{l|}{100}
&\multicolumn{2}{l}{$\tilde{\chi}^0_2 \to \tilde{\chi}^0_1 Z^\ast$}      & \multicolumn{2}{l}{100} \\

\multicolumn{2}{l}{$\tilde{\chi}^0_3 \to \tilde{\chi}^0_1 Z^\ast / \tilde{\chi}^\pm_1 W^{\mp\ast } / \tilde{\chi}^0_2 Z^\ast $}      & \multicolumn{2}{l|}{98.2/1.54 /0.2}
&\multicolumn{2}{l}{$\tilde{\chi}^0_3 \to \tilde{\chi}^\pm_1 W^{\mp} / \tilde{\chi}^0_2 Z $}      & \multicolumn{2}{l}{99.7/0.3} \\

\multicolumn{2}{l}{$\tilde{\chi}^0_4 \to \tilde{\chi}^\pm_1 W^\mp / \tilde{\chi}^0_3 Z / \tilde{\chi}^0_2 Z /  \tilde{\chi}^0_1 h/ \tilde{\chi}^0_1 Z$}      & \multicolumn{2}{l|}{63.1/19.3/10.9/3.02/1.41}
&\multicolumn{2}{l}{$\tilde{\chi}^0_4 \to \tilde{\chi}^\pm_1 W^\mp / \tilde{\chi}^0_2 Z / \tilde{\chi}^0_1 h / \tilde{\chi}^0_1 h_s$}      & \multicolumn{2}{l}{66.4/27.1/3.84/1.67 } \\

\multicolumn{2}{l}{$\tilde{\chi}^0_5 \to \tilde{\chi}^\pm_1 W^\mp / \tilde{\chi}^0_3 h / \tilde{\chi}^0_2 h / \tilde{\chi}^0_1 Z$}      & \multicolumn{2}{l|}{65.0/14.6/8.09/3.78}
&\multicolumn{2}{l}{$\tilde{\chi}^0_5 \to \tilde{\chi}^\pm_1 W^\mp / \tilde{\chi}^0_2 h / \tilde{\chi}^0_2 h_s / \tilde{\chi}^0_1 Z$}      & \multicolumn{2}{l}{76.7/11.3/6.12/5.03} \\

\multicolumn{2}{l}{$\tilde{\chi}^+_1 \to \tilde{\chi}^0_1 W^{+\ast}$}      & \multicolumn{2}{l|}{100}
&\multicolumn{2}{l}{$\tilde{\chi}^+_1 \to \tilde{\chi}^0_1 W^{+\ast}$}      & \multicolumn{2}{l}{100}  \\

\multicolumn{2}{l}{$\tilde{\chi}^+_2 \to  \tilde{\chi}^+_1 Z / \tilde{\chi}^0_3 W^+ / \tilde{\chi}^+_1 h / \tilde{\chi}^+_1 h_s/ \tilde{\chi}^0_1 W^+$}      & \multicolumn{2}{l|}{31.6/30.8/23.0/5.48/5.27}
&\multicolumn{2}{l}{$\tilde{\chi}^+_2 \to \tilde{\chi}^0_2 W^+ / \tilde{\chi}^+_1 Z / \tilde{\chi}^+_1 h / \tilde{\chi}^0_1 W^+ / \tilde{\chi}^+_1 h_s$}      & \multicolumn{2}{l}{38.5/34.9/11.4/6.37/6.30}\\

\multicolumn{2}{l}{$h_s \to b \bar{b} / \tau^+ \tau^- / gg / c \bar{c} / W ^+W^{-\ast} / \gamma \gamma$}      & \multicolumn{2}{l|}{80.9/8.96/4.87/4.59/0.37/0.25}
&\multicolumn{2}{l}{$h_s \to b \bar{b} / \tau^+ \tau^- / gg / c \bar{c} / W ^+W^{-\ast} / \gamma \gamma$}      & \multicolumn{2}{l}{79.8/8.85/5.43/5.04/0.51/0.28}\\

\multicolumn{2}{l}{$h \to b \bar{b} / W ^+W^{-\ast} / \tau^+ \tau^- / gg / ZZ^{\ast} / \gamma \gamma$}      & \multicolumn{2}{l|}{57.7/25.7/6.69/4.62/2.33/0.31}
&\multicolumn{2}{l}{$h \to b \bar{b} / W ^+W^{-\ast} / \tau^+ \tau^- / gg / ZZ^{\ast} / \gamma \gamma$}      & \multicolumn{2}{l}{58.1/25.3/6.74/4.57/2.29/0.31}\\

\multicolumn{2}{l}{$H \to  \tilde{\chi}^\pm_1 \tilde{\chi}^\mp_2 /  \tilde{\chi}^0_3 \tilde{\chi}^0_{4,5}/ t\bar{t}  /   \tilde{\chi}^0_2 \tilde{\chi}^0_4$}      & \multicolumn{2}{l|}{45.6/17.0/16.9/6.03}
&\multicolumn{2}{l}{$H \to \tilde{\chi}^\pm_1 t\bar{t}  / \tilde{\chi}^\mp_2 /  \tilde{\chi}^0_2 \tilde{\chi}^0_4/  \tilde{\chi}^0_4 \tilde{\chi}^0_5/\tilde{\chi}^\pm_1 \tilde{\chi}^\mp_1$}      & \multicolumn{2}{l}{38.9/22.5/13.5/4.61/4.22}\\

\multicolumn{2}{l}{$A_s \to   t\bar{t} /  b\bar{b}  / \tau\bar{\tau} / A_s Z / \gamma\gamma/ gg$}     & \multicolumn{2}{l|}{90.5/7.86/1.15/0.18/ 0.15/0.14}
&\multicolumn{2}{l}{$A_s \to t\bar{t} /  b\bar{b}/ \tilde{\chi}^0_1 \tilde{\chi}^0_1 / \tilde{\chi}^\pm_1 \tilde{\chi}^\mp_1 / Z h_s$}    & \multicolumn{2}{l}{91.3/4.47/2.22/0.7}\\

\multicolumn{2}{l}{$A_H \to  \tilde{\chi}^\pm_1 \tilde{\chi}^\mp_2 /  t\bar{t}  /  \tilde{\chi}^0_3 \tilde{\chi}^0_{5,4}/ \tilde{\chi}^0_5 \tilde{\chi}^0_{2,1} $}      & \multicolumn{2}{l|}{42.0/17.0/16.3/9.45}
&\multicolumn{2}{l}{$A_H \to t\bar{t} / \tilde{\chi}^\pm_1 \tilde{\chi}^\mp_2 / \tilde{\chi}^\pm_1 \tilde{\chi}^\mp_1 /\tilde{\chi}^\pm_2 \tilde{\chi}^\mp_2/ \tilde{\chi}^0_1 \tilde{\chi}^0_1$}      & \multicolumn{2}{l}{22.7/22.2/13.1/9.07/6.42}\\

\multicolumn{2}{l}{$H^+ \to  \tilde{\chi}^0_5 \tilde{\chi}^+_1/ \tilde{\chi}^0_4 \tilde{\chi}^+_1 / \tilde{\chi}^0_3 \tilde{\chi}^+_2 / t \bar{b} /\tilde{\chi}^0_1 \tilde{\chi}^+_2$} & \multicolumn{2}{l|}{22.9/22.0/20.2/18.2/6.11}
&\multicolumn{2}{l}{$H^+ \to  t\bar{b} / \tilde{\chi}^0_2 \tilde{\chi}^+_2 / \tilde{\chi}^0_5 \tilde{\chi}^+_1/ \tilde{\chi}^0_4 \tilde{\chi}^+_1 /  \tilde{\chi}^0_4 \tilde{\chi}^+_2$}      & \multicolumn{2}{l}{23.5/22.0/20.7/16.4/4.75}\\
\hline

\multicolumn{2}{l}{$R$ value: 0.54}  & \multicolumn{2}{l|}{Signal Region: SR2-stop-3high-pt-1 in Ref.~\cite{CMS:2018kag}}
& \multicolumn{2}{l}{$R$ value: 0.12}  & \multicolumn{2}{l}{Signal Region: SR-WZoff-high-njd in Ref.~\cite{ATLAS:2021moa}} \\
\hline \hline

\end{tabular}}
\end{table}

\section{Conclusion} \label{conclusion}
The recent analyses by CMS and ATLAS collaborations have revealed a deviation in the di-photon channel at about $95~{\rm GeV}$ with compatible signal strengths. The invariant mass of the di-photon signal coincides with that of the previously observed $b\bar{b}$ excess at LEP, suggesting they may originate from a same CP-even Higgs boson with its mass around $95~{\rm GeV}$.
Given the significant decrease in the di-photon rate and the notable improvement in sensitivities of DM direct detection experiments, the present research conducts a comprehensive investigation into the capacity of the well-established $\mathbb{Z}_3$-NMSSM to fully accommodate the observed excesses while adhering to various experimental constraints. 
By deriving analytic expressions and performing a specialized scan, the study demonstrates that the model can successfully explain the observed excesses through a singlet-dominated CP-even Higgs boson around 95 GeV. The investigation yields the following key insights:
\begin{enumerate}
    \item The model could effectively explain the di-photon and $b\bar{b}$ excesses at the $1\sigma$ level, which requires the normalized couplings to be within the range $0.23 \lesssim |C_{h_s t \bar{t}}| \lesssim 0.39$, $0.15 \lesssim |C_{h_s b \bar{b}}| \lesssim 0.39$, $0.27 \lesssim |C_{h_s \gamma \gamma}| \lesssim 0.43$ and $0.23 \lesssim |C_{h_s g g}| \lesssim 0.40$. 
    \item {\bf Case-I} can predict signal strengths closer to the central values compared to {\bf Case-II} due to larger contributions from chargino loops to $C_{h_s \gamma \gamma}$.
    \item  The interpretation favors the Bino-dominated $\tilde{\chi}^0_1$s as DM candidates. These particles achieve the measured relic abundance primarily by co-annihilating with the Wino-like electroweakinos. This interpretation adheres to the constraints from DM direct detection experiments, with many parameter points predicting scattering cross-sections below the current experimental limits and even within the neutrino background, rendering them challenging to detect.
    \item  Constraints from the LHC searches for electroweakinos have little impact on the compressed spectrum. Monte Carlo simulations for the four benchmark points consistently yield low R-values ($< 0.6$), indicating weak signals at the LHC. This result is attributed to the heavy electroweakino masses predicted in the study, which reduce the production rates for these particles.
    \item  The di-photon and $\tau\bar{\tau}$ excesses can only be jointly explained at the $2\sigma$ level. This limitation arises because $\mu_{\tau\bar{\tau}}$ closely follows $\mu_{b\bar{b}}$, given the equality of $C_{h_s \tau \bar{\tau}}$ and $C_{h_s b \bar{b}}$ in this model.
\end{enumerate}
These findings support the $\mathbb{Z}_3$-NMSSM as a viable extension of the Standard Model that could accommodate the new scalar particle, motivating further experimental investigations and more refined searches for potential signals in the upcoming collider experiments.

\bibliographystyle{CitationStyle}
\bibliography{LianRef}

\providecommand{\href}[2]{#2}\begingroup\raggedright\begin{thebibliography}{100}

\bibitem{ATLAS:2012yve}
{\scshape ATLAS} collaboration, G.~Aad et~al., \emph{{Observation of a new
  particle in the search for the Standard Model Higgs boson with the ATLAS
  detector at the LHC}},
  \href{https://doi.org/10.1016/j.physletb.2012.08.020}{\emph{Phys. Lett. B}
  {\bfseries 716} (2012) 1} [\href{https://arxiv.org/abs/1207.7214}{{\ttfamily
  1207.7214}}].

\bibitem{CMS:2012qbp}
{\scshape CMS} collaboration, S.~Chatrchyan et~al., \emph{{Observation of a New
  Boson at a Mass of 125 GeV with the CMS Experiment at the LHC}},
  \href{https://doi.org/10.1016/j.physletb.2012.08.021}{\emph{Phys. Lett. B}
  {\bfseries 716} (2012) 30} [\href{https://arxiv.org/abs/1207.7235}{{\ttfamily
  1207.7235}}].

\bibitem{ATLAS:2019nkf}
{\scshape ATLAS} collaboration, G.~Aad et~al., \emph{{Combined measurements of
  Higgs boson production and decay using up to $80$ fb$^{-1}$ of proton-proton
  collision data at $\sqrt{s}=$ 13 TeV collected with the ATLAS experiment}},
  \href{https://doi.org/10.1103/PhysRevD.101.012002}{\emph{Phys. Rev. D}
  {\bfseries 101} (2020) 012002}
  [\href{https://arxiv.org/abs/1909.02845}{{\ttfamily 1909.02845}}].

\bibitem{CMS:2023yay}
{\scshape CMS} collaboration, ``{Search for a standard model-like Higgs boson
  in the mass range between 70 and 110$~\mathrm{GeV}$ in the diphoton final
  state in proton-proton collisions at $\sqrt{s}=13~\mathrm{TeV}$}.''
  \href{https://cds.cern.ch/record/2852907}{CMS-PAS-HIG-20-002}, 2023.

\bibitem{Arcangeletti}
C.~Arcangeletti. {on behalf of ATLAS collaboration, LHC Seminar}
  \url{https://indico.cern.ch/event/1281604/attachments/2660420/4608571/LHCSeminarArcangeletti_final.pdf},
  7$^{th}$ of June, 2023.

\bibitem{Biekotter:2023oen}
T.~Biek\"otter, S.~Heinemeyer and G.~Weiglein, \emph{{The 95.4 GeV di-photon
  excess at ATLAS and CMS}},
  \href{https://arxiv.org/abs/2306.03889}{{\ttfamily 2306.03889}}.

\bibitem{LEPWorkingGroupforHiggsbosonsearches:2003ing}
{\scshape LEP Working Group for Higgs boson searches, ALEPH, DELPHI, L3, OPAL}
  collaboration, R.~Barate et~al., \emph{{Search for the standard model Higgs
  boson at LEP}},
  \href{https://doi.org/10.1016/S0370-2693(03)00614-2}{\emph{Phys. Lett. B}
  {\bfseries 565} (2003) 61}
  [\href{https://arxiv.org/abs/hep-ex/0306033}{{\ttfamily hep-ex/0306033}}].

\bibitem{Azatov:2012bz}
A.~Azatov, R.~Contino and J.~Galloway, \emph{{Model-Independent Bounds on a
  Light Higgs}}, \href{https://doi.org/10.1007/JHEP04(2012)127}{\emph{JHEP}
  {\bfseries 04} (2012) 127} [\href{https://arxiv.org/abs/1202.3415}{{\ttfamily
  1202.3415}}].

\bibitem{Cao:2016uwt}
J.~Cao, X.~Guo, Y.~He, P.~Wu and Y.~Zhang, \emph{{Diphoton signal of the light
  Higgs boson in natural NMSSM}},
  \href{https://doi.org/10.1103/PhysRevD.95.116001}{\emph{Phys. Rev. D}
  {\bfseries 95} (2017) 116001}
  [\href{https://arxiv.org/abs/1612.08522}{{\ttfamily 1612.08522}}].

\bibitem{CMS:2022goy}
{\scshape CMS} collaboration, A.~Tumasyan et~al., \emph{{Searches for
  additional Higgs bosons and for vector leptoquarks in $\tau\tau$ final states
  in proton-proton collisions at $\sqrt{s}$ = 13 TeV}},
  \href{https://doi.org/10.1007/JHEP07(2023)073}{\emph{JHEP} {\bfseries 07}
  (2023) 073} [\href{https://arxiv.org/abs/2208.02717}{{\ttfamily
  2208.02717}}].

\bibitem{Coloretti:2023wng}
G.~Coloretti, A.~Crivellin, S.~Bhattacharya and B.~Mellado, \emph{{Searching
  for low-mass resonances decaying into W bosons}},
  \href{https://doi.org/10.1103/PhysRevD.108.035026}{\emph{Phys. Rev. D}
  {\bfseries 108} (2023) 035026}
  [\href{https://arxiv.org/abs/2302.07276}{{\ttfamily 2302.07276}}].

\bibitem{Ashanujjaman:2023etj}
S.~Ashanujjaman, S.~Banik, G.~Coloretti, A.~Crivellin, B.~Mellado and A.-T.
  Mulaudzi, \emph{{$SU(2)_L$ triplet scalar as the origin of the 95 GeV
  excess?}},  \href{https://arxiv.org/abs/2306.15722}{{\ttfamily 2306.15722}}.

\bibitem{Aguilar-Saavedra:2020wrj}
J.~A. Aguilar-Saavedra and F.~R. Joaquim, \emph{{Multiphoton signals of a (96
  GeV?) stealth boson}},
  \href{https://doi.org/10.1140/epjc/s10052-020-7952-4}{\emph{Eur. Phys. J. C}
  {\bfseries 80} (2020) 403}
  [\href{https://arxiv.org/abs/2002.07697}{{\ttfamily 2002.07697}}].

\bibitem{Kundu:2019nqo}
A.~Kundu, S.~Maharana and P.~Mondal, \emph{{A 96 GeV scalar tagged to dark
  matter models}},
  \href{https://doi.org/10.1016/j.nuclphysb.2020.115057}{\emph{Nucl. Phys. B}
  {\bfseries 955} (2020) 115057}
  [\href{https://arxiv.org/abs/1907.12808}{{\ttfamily 1907.12808}}].

\bibitem{Fox:2017uwr}
P.~J. Fox and N.~Weiner, \emph{{Light Signals from a Lighter Higgs}},
  \href{https://doi.org/10.1007/JHEP08(2018)025}{\emph{JHEP} {\bfseries 08}
  (2018) 025} [\href{https://arxiv.org/abs/1710.07649}{{\ttfamily
  1710.07649}}].

\bibitem{Belyaev:2023xnv}
A.~Belyaev, R.~Benbrik, M.~Boukidi, M.~Chakraborti, S.~Moretti and S.~Semlali,
  \emph{{Explanation of the Hints for a 95 GeV Higgs Boson within a 2-Higgs
  Doublet Model}},  \href{https://arxiv.org/abs/2306.09029}{{\ttfamily
  2306.09029}}.

\bibitem{Azevedo:2023zkg}
D.~Azevedo, T.~Biek\"otter and P.~M. Ferreira, \emph{{2HDM interpretations of
  the CMS diphoton excess at 95 GeV}},
  \href{https://arxiv.org/abs/2305.19716}{{\ttfamily 2305.19716}}.

\bibitem{Benbrik:2022dja}
R.~Benbrik, M.~Boukidi and B.~Manaut, \emph{{$W$-mass and 96 GeV excess in
  type-III 2HDM}},  \href{https://arxiv.org/abs/2204.11755}{{\ttfamily
  2204.11755}}.

\bibitem{Benbrik:2022azi}
R.~Benbrik, M.~Boukidi, S.~Moretti and S.~Semlali, \emph{{Explaining the 96 GeV
  Di-photon anomaly in a generic 2HDM Type-III}},
  \href{https://doi.org/10.1016/j.physletb.2022.137245}{\emph{Phys. Lett. B}
  {\bfseries 832} (2022) 137245}
  [\href{https://arxiv.org/abs/2204.07470}{{\ttfamily 2204.07470}}].

\bibitem{Haisch:2017gql}
U.~Haisch and A.~Malinauskas, \emph{{Let there be light from a second light
  Higgs doublet}}, \href{https://doi.org/10.1007/JHEP03(2018)135}{\emph{JHEP}
  {\bfseries 03} (2018) 135}
  [\href{https://arxiv.org/abs/1712.06599}{{\ttfamily 1712.06599}}].

\bibitem{Biekotter:2019mib}
T.~Biek\"otter, M.~Chakraborti and S.~Heinemeyer, \emph{{An N2HDM Solution for
  the possible 96 GeV Excess}},
  \href{https://doi.org/10.22323/1.347.0015}{\emph{PoS} {\bfseries CORFU2018}
  (2019) 015} [\href{https://arxiv.org/abs/1905.03280}{{\ttfamily
  1905.03280}}].

\bibitem{Biekotter:2019kde}
T.~Biek\"otter, M.~Chakraborti and S.~Heinemeyer, \emph{{A 96 GeV Higgs boson
  in the N2HDM}},
  \href{https://doi.org/10.1140/epjc/s10052-019-7561-2}{\emph{Eur. Phys. J. C}
  {\bfseries 80} (2020) 2} [\href{https://arxiv.org/abs/1903.11661}{{\ttfamily
  1903.11661}}].

\bibitem{Biekotter:2020cjs}
T.~Biek\"otter, M.~Chakraborti and S.~Heinemeyer, \emph{{The
  \textquotedblleft{}96 GeV excess\textquotedblright{} at the LHC}},
  \href{https://doi.org/10.1142/S0217751X21420185}{\emph{Int. J. Mod. Phys. A}
  {\bfseries 36} (2021) 2142018}
  [\href{https://arxiv.org/abs/2003.05422}{{\ttfamily 2003.05422}}].

\bibitem{Biekotter:2021ovi}
T.~Biek\"otter and M.~O. Olea-Romacho, \emph{{Reconciling Higgs physics and
  pseudo-Nambu-Goldstone dark matter in the S2HDM using a genetic algorithm}},
  \href{https://doi.org/10.1007/JHEP10(2021)215}{\emph{JHEP} {\bfseries 10}
  (2021) 215} [\href{https://arxiv.org/abs/2108.10864}{{\ttfamily
  2108.10864}}].

\bibitem{Heinemeyer:2021msz}
S.~Heinemeyer, C.~Li, F.~Lika, G.~Moortgat-Pick and S.~Paasch,
  \emph{{Phenomenology of a 96~GeV Higgs boson in the 2HDM with an additional
  singlet}}, \href{https://doi.org/10.1103/PhysRevD.106.075003}{\emph{Phys.
  Rev. D} {\bfseries 106} (2022) 075003}
  [\href{https://arxiv.org/abs/2112.11958}{{\ttfamily 2112.11958}}].

\bibitem{Biekotter:2022jyr}
T.~Biek\"otter, S.~Heinemeyer and G.~Weiglein, \emph{{Mounting evidence for a
  95 GeV Higgs boson}},
  \href{https://doi.org/10.1007/JHEP08(2022)201}{\emph{JHEP} {\bfseries 08}
  (2022) 201} [\href{https://arxiv.org/abs/2203.13180}{{\ttfamily
  2203.13180}}].

\bibitem{Li:2023hsr}
C.~Li, \emph{{Phenomenology of extended Two-Higgs-Doublets models}}, Ph.D.
  thesis, Hamburg U., 2023.
\newblock 10.3204/PUBDB-2023-03151.

\bibitem{Biekotter:2023jld}
T.~Biek\"otter, S.~Heinemeyer and G.~Weiglein, \emph{{The CMS di-photon excess
  at 95 GeV in view of the LHC Run 2 results}},
  \href{https://arxiv.org/abs/2303.12018}{{\ttfamily 2303.12018}}.

\bibitem{Banik:2023ecr}
S.~Banik, A.~Crivellin, S.~Iguro and T.~Kitahara, \emph{{Asymmetric di-Higgs
  signals of the next-to-minimal 2HDM with a U(1) symmetry}},
  \href{https://doi.org/10.1103/PhysRevD.108.075011}{\emph{Phys. Rev. D}
  {\bfseries 108} (2023) 075011}
  [\href{https://arxiv.org/abs/2303.11351}{{\ttfamily 2303.11351}}].

\bibitem{Dutta:2023cig}
J.~Dutta, J.~Lahiri, C.~Li, G.~Moortgat-Pick, S.~F. Tabira and J.~A. Ziegler,
  \emph{{Dark Matter Phenomenology in 2HDMS in light of the 95 GeV excess}},
  \href{https://arxiv.org/abs/2308.05653}{{\ttfamily 2308.05653}}.

\bibitem{Sachdeva:2019hvk}
D.~Sachdeva and S.~Sadhukhan, \emph{{Discussing 125 GeV and 95 GeV excess in
  light radion model}},
  \href{https://doi.org/10.1103/PhysRevD.101.055045}{\emph{Phys. Rev. D}
  {\bfseries 101} (2020) 055045}
  [\href{https://arxiv.org/abs/1908.01668}{{\ttfamily 1908.01668}}].

\bibitem{Vega:2018ddp}
R.~Vega, R.~Vega-Morales and K.~Xie, \emph{{Light (and darkness) from a light
  hidden Higgs}}, \href{https://doi.org/10.1007/JHEP06(2018)137}{\emph{JHEP}
  {\bfseries 06} (2018) 137}
  [\href{https://arxiv.org/abs/1805.01970}{{\ttfamily 1805.01970}}].

\bibitem{Borah:2023hqw}
D.~Borah, S.~Mahapatra, P.~K. Paul and N.~Sahu, \emph{{Scotogenic
  $U(1)_{L_{\mu}-L_{\tau}}$ origin of $(g-2)_\mu$, W-mass anomaly and 95 GeV
  excess}},  \href{https://arxiv.org/abs/2310.11953}{{\ttfamily 2310.11953}}.

\bibitem{Arcadi:2023smv}
G.~Arcadi, G.~Busoni, D.~Cabo-Almeida and N.~Krishnan, \emph{{Is there a
  (Pseudo)Scalar at 95 GeV?}},
  \href{https://arxiv.org/abs/2311.14486}{{\ttfamily 2311.14486}}.

\bibitem{Ahriche:2023wkj}
A.~Ahriche, \emph{{The 95 GeV Excess in the Georgi-Machacek Model: Single or
  Twin Peak Resonance}},  \href{https://arxiv.org/abs/2312.10484}{{\ttfamily
  2312.10484}}.

\bibitem{Chen:2023bqr}
T.-K. Chen, C.-W. Chiang, S.~Heinemeyer and G.~Weiglein, \emph{{A 95 GeV Higgs
  Boson in the Georgi-Machacek Model}},
  \href{https://arxiv.org/abs/2312.13239}{{\ttfamily 2312.13239}}.

\bibitem{Dev:2023kzu}
P.~S.~B. Dev, R.~N. Mohapatra and Y.~Zhang, \emph{{Explanation of the 95 GeV
  $\gamma\gamma$ and $b\bar{b}$ excesses in the Minimal Left-Right Symmetric
  Model}},  \href{https://arxiv.org/abs/2312.17733}{{\ttfamily 2312.17733}}.

\bibitem{Wang:2024bkg}
K.~Wang and J.~Zhu, \emph{{A 95 GeV light Higgs in the top-pair-associated
  diphoton channel at the LHC in the Minial Dilaton Model}},
  \href{https://arxiv.org/abs/2402.11232}{{\ttfamily 2402.11232}}.

\bibitem{Fan:2013gjf}
J.-W. Fan, J.-Q. Tao, Y.-Q. Shen, G.-M. Chen, H.-S. Chen, S.~Gascon-Shotkin
  et~al., \emph{{Study of diphoton decays of the lightest scalar Higgs boson in
  the Next-to-Minimal Supersymmetric Standard Model}},
  \href{https://doi.org/10.1088/1674-1137/38/7/073101}{\emph{Chin. Phys. C}
  {\bfseries 38} (2014) 073101}
  [\href{https://arxiv.org/abs/1309.6394}{{\ttfamily 1309.6394}}].

\bibitem{Biekotter:2017xmf}
T.~Biek\"otter, S.~Heinemeyer and C.~Mu\~noz, \emph{{Precise prediction for the
  Higgs-boson masses in the $\mu \nu $ SSM}},
  \href{https://doi.org/10.1140/epjc/s10052-018-5978-7}{\emph{Eur. Phys. J. C}
  {\bfseries 78} (2018) 504}
  [\href{https://arxiv.org/abs/1712.07475}{{\ttfamily 1712.07475}}].

\bibitem{Beskidt:2017dil}
C.~Beskidt, W.~de~Boer and D.~I. Kazakov, \emph{{Can we discover a light
  singlet-like NMSSM Higgs boson at the LHC?}},
  \href{https://doi.org/10.1016/j.physletb.2018.04.067}{\emph{Phys. Lett. B}
  {\bfseries 782} (2018) 69}
  [\href{https://arxiv.org/abs/1712.02531}{{\ttfamily 1712.02531}}].

\bibitem{Heinemeyer:2018wzl}
S.~Heinemeyer and T.~Stefaniak, \emph{{A Higgs Boson at 96 GeV?!}},
  \href{https://doi.org/10.22323/1.339.0016}{\emph{PoS} {\bfseries CHARGED2018}
  (2019) 016} [\href{https://arxiv.org/abs/1812.05864}{{\ttfamily
  1812.05864}}].

\bibitem{Heinemeyer:2018jcd}
S.~Heinemeyer, \emph{{A Higgs boson below 125 GeV?!}},
  \href{https://doi.org/10.1142/S0217751X18440062}{\emph{Int. J. Mod. Phys. A}
  {\bfseries 33} (2018) 1844006}.

\bibitem{Wang:2018vxp}
K.~Wang, F.~Wang, J.~Zhu and Q.~Jie, \emph{{The semi-constrained NMSSM in light
  of muon g-2, LHC, and dark matter constraints}},
  \href{https://doi.org/10.1088/1674-1137/42/10/103109}{\emph{Chin. Phys. C}
  {\bfseries 42} (2018) 103109}
  [\href{https://arxiv.org/abs/1811.04435}{{\ttfamily 1811.04435}}].

\bibitem{Domingo:2018uim}
F.~Domingo, S.~Heinemeyer, S.~Pa\ss{}ehr and G.~Weiglein, \emph{{Decays of the
  neutral Higgs bosons into SM fermions and gauge bosons in the
  $\mathcal{CP}$-violating NMSSM}},
  \href{https://doi.org/10.1140/epjc/s10052-018-6400-1}{\emph{Eur. Phys. J. C}
  {\bfseries 78} (2018) 942}
  [\href{https://arxiv.org/abs/1807.06322}{{\ttfamily 1807.06322}}].

\bibitem{Cao:2019ofo}
J.~Cao, X.~Jia, Y.~Yue, H.~Zhou and P.~Zhu, \emph{{96 GeV diphoton excess in
  seesaw extensions of the natural NMSSM}},
  \href{https://doi.org/10.1103/PhysRevD.101.055008}{\emph{Phys. Rev. D}
  {\bfseries 101} (2020) 055008}
  [\href{https://arxiv.org/abs/1908.07206}{{\ttfamily 1908.07206}}].

\bibitem{Biekotter:2019gtq}
T.~Biek\"otter, S.~Heinemeyer and C.~Mu\~noz, \emph{{Precise prediction for the
  Higgs-Boson masses in the $\mu \nu $SSM with three right-handed neutrino
  superfields}},
  \href{https://doi.org/10.1140/epjc/s10052-019-7175-8}{\emph{Eur. Phys. J. C}
  {\bfseries 79} (2019) 667}
  [\href{https://arxiv.org/abs/1906.06173}{{\ttfamily 1906.06173}}].

\bibitem{Choi:2019yrv}
K.~Choi, S.~H. Im, K.~S. Jeong and C.~B. Park, \emph{{Light Higgs bosons in the
  general NMSSM}},
  \href{https://doi.org/10.1140/epjc/s10052-019-7473-1}{\emph{Eur. Phys. J. C}
  {\bfseries 79} (2019) 956}
  [\href{https://arxiv.org/abs/1906.03389}{{\ttfamily 1906.03389}}].

\bibitem{Abdelalim:2020xfk}
A.~A. Abdelalim, B.~Das, S.~Khalil and S.~Moretti, \emph{{Di-photon decay of a
  light Higgs state in the BLSSM}},
  \href{https://doi.org/10.1016/j.nuclphysb.2022.116013}{\emph{Nucl. Phys. B}
  {\bfseries 985} (2022) 116013}
  [\href{https://arxiv.org/abs/2012.04952}{{\ttfamily 2012.04952}}].

\bibitem{Hollik:2020plc}
W.~G. Hollik, C.~Li, G.~Moortgat-Pick and S.~Paasch, \emph{{Phenomenology of a
  Supersymmetric Model Inspired by Inflation}},
  \href{https://doi.org/10.1140/epjc/s10052-021-08869-4}{\emph{Eur. Phys. J. C}
  {\bfseries 81} (2021) 141}
  [\href{https://arxiv.org/abs/2004.14852}{{\ttfamily 2004.14852}}].

\bibitem{Biekotter:2021qbc}
T.~Biek\"otter, A.~Grohsjean, S.~Heinemeyer, C.~Schwanenberger and G.~Weiglein,
  \emph{{Possible indications for new Higgs bosons in the reach of the LHC:
  N2HDM and NMSSM interpretations}},
  \href{https://doi.org/10.1140/epjc/s10052-022-10099-1}{\emph{Eur. Phys. J. C}
  {\bfseries 82} (2022) 178}
  [\href{https://arxiv.org/abs/2109.01128}{{\ttfamily 2109.01128}}].

\bibitem{Li:2022etb}
W.~Li, H.~Qiao and J.~Zhu, \emph{{Light Higgs boson in the NMSSM confronted
  with the CMS di-photon and di-tau excesses*}},
  \href{https://doi.org/10.1088/1674-1137/acfaf1}{\emph{Chin. Phys. C}
  {\bfseries 47} (2023) 123102}
  [\href{https://arxiv.org/abs/2212.11739}{{\ttfamily 2212.11739}}].

\bibitem{Ellwanger:2023zjc}
U.~Ellwanger and C.~Hugonie, \emph{{Additional Higgs Bosons near 95 and 650 GeV
  in the NMSSM}},
  \href{https://doi.org/10.1140/epjc/s10052-023-12315-y}{\emph{Eur. Phys. J. C}
  {\bfseries 83} (2023) 1138}
  [\href{https://arxiv.org/abs/2309.07838}{{\ttfamily 2309.07838}}].

\bibitem{Cao:2023gkc}
J.~Cao, X.~Jia, J.~Lian and L.~Meng, \emph{{95 GeV Diphoton and $b \bar{b}$
  Excesses in the General Next-to-Minimal Supersymmetric Standard Model}},
  \href{https://arxiv.org/abs/2310.08436}{{\ttfamily 2310.08436}}.

\bibitem{Ahriche:2023hho}
A.~Ahriche, M.~L. Bellilet, M.~O. Khojali, M.~Kumar and A.-T. Mulaudzi,
  \emph{{The scale invariant scotogenic model: CDF-II $W$-boson mass and the
  95\textasciitilde{}GeV excesses}},
  \href{https://arxiv.org/abs/2311.08297}{{\ttfamily 2311.08297}}.

\bibitem{Ellwanger:2024txc}
U.~Ellwanger and C.~Hugonie, \emph{{Nmssm with correct relic density and an
  additional 95~GeV Higgs boson}},
  \href{https://doi.org/10.1140/epjc/s10052-024-12886-4}{\emph{Eur. Phys. J. C}
  {\bfseries 84} (2024) 526}
  [\href{https://arxiv.org/abs/2403.16884}{{\ttfamily 2403.16884}}].

\bibitem{Liu:2024cbr}
C.-X. Liu, Y.~Zhou, X.-Y. Zheng, J.~Ma, T.-F. Feng and H.-B. Zhang, \emph{{95
  GeV excess in a $CP$-violating $\mu$-from-$\nu$ SSM}},
  \href{https://arxiv.org/abs/2402.00727}{{\ttfamily 2402.00727}}.

\bibitem{Ellwanger:2024vvs}
U.~Ellwanger, C.~Hugonie, S.~F. King and S.~Moretti, \emph{{NMSSM explanation
  for excesses in the search for neutralinos and charginos and a 95 GeV Higgs
  boson}}, \href{https://doi.org/10.1140/epjc/s10052-024-13129-2}{\emph{Eur.
  Phys. J. C} {\bfseries 84} (2024) 788}
  [\href{https://arxiv.org/abs/2404.19338}{{\ttfamily 2404.19338}}].

\bibitem{Ellwanger:2009dp}
U.~Ellwanger, C.~Hugonie and A.~M. Teixeira, \emph{{The Next-to-Minimal
  Supersymmetric Standard Model}},
  \href{https://doi.org/10.1016/j.physrep.2010.07.001}{\emph{Phys. Rept.}
  {\bfseries 496} (2010) 1} [\href{https://arxiv.org/abs/0910.1785}{{\ttfamily
  0910.1785}}].

\bibitem{Ellwanger:2014hia}
U.~Ellwanger and A.~M. Teixeira, \emph{{NMSSM with a singlino LSP: possible
  challenges for searches for supersymmetry at the LHC}},
  \href{https://doi.org/10.1007/JHEP10(2014)113}{\emph{JHEP} {\bfseries 10}
  (2014) 113} [\href{https://arxiv.org/abs/1406.7221}{{\ttfamily 1406.7221}}].

\bibitem{Cao:2021ljw}
J.~Cao, D.~Li, J.~Lian, Y.~Yue and H.~Zhou, \emph{{Singlino-dominated dark
  matter in general NMSSM}},
  \href{https://doi.org/10.1007/JHEP06(2021)176}{\emph{JHEP} {\bfseries 06}
  (2021) 176} [\href{https://arxiv.org/abs/2102.05317}{{\ttfamily
  2102.05317}}].

\bibitem{Cao:2019qng}
J.~Cao, L.~Meng, Y.~Yue, H.~Zhou and P.~Zhu, \emph{{Suppressing the scattering
  of WIMP dark matter and nucleons in supersymmetric theories}},
  \href{https://doi.org/10.1103/PhysRevD.101.075003}{\emph{Phys. Rev. D}
  {\bfseries 101} (2020) 075003}
  [\href{https://arxiv.org/abs/1910.14317}{{\ttfamily 1910.14317}}].

\bibitem{Ellwanger:2011aa}
U.~Ellwanger, \emph{{A Higgs boson near 125 GeV with enhanced di-photon signal
  in the NMSSM}}, \href{https://doi.org/10.1007/JHEP03(2012)044}{\emph{JHEP}
  {\bfseries 03} (2012) 044} [\href{https://arxiv.org/abs/1112.3548}{{\ttfamily
  1112.3548}}].

\bibitem{Badziak:2013bda}
M.~Badziak, M.~Olechowski and S.~Pokorski, \emph{{New Regions in the NMSSM with
  a 125 GeV Higgs}}, \href{https://doi.org/10.1007/JHEP06(2013)043}{\emph{JHEP}
  {\bfseries 06} (2013) 043} [\href{https://arxiv.org/abs/1304.5437}{{\ttfamily
  1304.5437}}].

\bibitem{Cao:2012fz}
J.-J. Cao, Z.-X. Heng, J.~M. Yang, Y.-M. Zhang and J.-Y. Zhu, \emph{{A SM-like
  Higgs near 125 GeV in low energy SUSY: a comparative study for MSSM and
  NMSSM}}, \href{https://doi.org/10.1007/JHEP03(2012)086}{\emph{JHEP}
  {\bfseries 03} (2012) 086} [\href{https://arxiv.org/abs/1202.5821}{{\ttfamily
  1202.5821}}].

\bibitem{Cao:2024axg}
J.~Cao, X.~Jia and J.~Lian, \emph{{Unified Interpretation of Muon g-2 anomaly,
  95 GeV Diphoton, and $b\bar{b}$ Excesses in the General Next-to-Minimal
  Supersymmetric Standard Model}},
  \href{https://arxiv.org/abs/2402.15847}{{\ttfamily 2402.15847}}.

\bibitem{Miller:2003ay}
D.~J. Miller, R.~Nevzorov and P.~M. Zerwas, \emph{{The Higgs sector of the
  next-to-minimal supersymmetric standard model}},
  \href{https://doi.org/10.1016/j.nuclphysb.2003.12.021}{\emph{Nucl. Phys. B}
  {\bfseries 681} (2004) 3}
  [\href{https://arxiv.org/abs/hep-ph/0304049}{{\ttfamily hep-ph/0304049}}].

\bibitem{Carena:2015moc}
M.~Carena, H.~E. Haber, I.~Low, N.~R. Shah and C.~E.~M. Wagner,
  \emph{{Alignment limit of the NMSSM Higgs sector}},
  \href{https://doi.org/10.1103/PhysRevD.93.035013}{\emph{Phys. Rev. D}
  {\bfseries 93} (2016) 035013}
  [\href{https://arxiv.org/abs/1510.09137}{{\ttfamily 1510.09137}}].

\bibitem{ATLAS:2022vkf}
{\scshape ATLAS} collaboration, \emph{{A detailed map of Higgs boson
  interactions by the ATLAS experiment ten years after the discovery}},
  \href{https://doi.org/10.1038/s41586-022-04893-w}{\emph{Nature} {\bfseries
  607} (2022) 52} [\href{https://arxiv.org/abs/2207.00092}{{\ttfamily
  2207.00092}}].

\bibitem{CMS:2022dwd}
{\scshape CMS} collaboration, A.~Tumasyan et~al., \emph{{A portrait of the
  Higgs boson by the CMS experiment ten years after the discovery}},
  \href{https://doi.org/10.1038/s41586-022-04892-x}{\emph{Nature} {\bfseries
  607} (2022) 60} [\href{https://arxiv.org/abs/2207.00043}{{\ttfamily
  2207.00043}}].

\bibitem{ATLAS:2020zms}
{\scshape ATLAS} collaboration, G.~Aad et~al., \emph{{Search for heavy Higgs
  bosons decaying into two tau leptons with the ATLAS detector using $pp$
  collisions at $\sqrt{s}=13$ TeV}},
  \href{https://doi.org/10.1103/PhysRevLett.125.051801}{\emph{Phys. Rev. Lett.}
  {\bfseries 125} (2020) 051801}
  [\href{https://arxiv.org/abs/2002.12223}{{\ttfamily 2002.12223}}].

\bibitem{Cao:2013gba}
J.~Cao, F.~Ding, C.~Han, J.~M. Yang and J.~Zhu, \emph{{A light Higgs scalar in
  the NMSSM confronted with the latest LHC Higgs data}},
  \href{https://doi.org/10.1007/JHEP11(2013)018}{\emph{JHEP} {\bfseries 11}
  (2013) 018} [\href{https://arxiv.org/abs/1309.4939}{{\ttfamily 1309.4939}}].

\bibitem{Baum:2017enm}
S.~Baum, M.~Carena, N.~R. Shah and C.~E.~M. Wagner, \emph{{Higgs portals for
  thermal Dark Matter. EFT perspectives and the NMSSM}},
  \href{https://doi.org/10.1007/JHEP04(2018)069}{\emph{JHEP} {\bfseries 04}
  (2018) 069} [\href{https://arxiv.org/abs/1712.09873}{{\ttfamily
  1712.09873}}].

\bibitem{Zhou:2021pit}
H.~Zhou, J.~Cao, J.~Lian and D.~Zhang, \emph{{Singlino-dominated dark matter in
  Z3-symmetric NMSSM}},
  \href{https://doi.org/10.1103/PhysRevD.104.015017}{\emph{Phys. Rev. D}
  {\bfseries 104} (2021) 015017}
  [\href{https://arxiv.org/abs/2102.05309}{{\ttfamily 2102.05309}}].

\bibitem{King:2012tr}
S.~F. King, M.~M\"uhlleitner, R.~Nevzorov and K.~Walz, \emph{{Natural NMSSM
  Higgs Bosons}},
  \href{https://doi.org/10.1016/j.nuclphysb.2013.01.020}{\emph{Nucl. Phys. B}
  {\bfseries 870} (2013) 323}
  [\href{https://arxiv.org/abs/1211.5074}{{\ttfamily 1211.5074}}].

\bibitem{Porod2003SPheno}
W.~Porod, \emph{{SPheno, a program for calculating supersymmetric spectra, SUSY
  particle decays and SUSY particle production at e+ e- colliders}},
  \href{https://doi.org/10.1016/S0010-4655(03)00222-4}{\emph{Comput. Phys.
  Commun.} {\bfseries 153} (2003) 275}
  [\href{https://arxiv.org/abs/hep-ph/0301101}{{\ttfamily hep-ph/0301101}}].

\bibitem{Porod2011SPheno3}
W.~Porod and F.~Staub, \emph{{SPheno 3.1: Extensions including flavour,
  CP-phases and models beyond the MSSM}},
  \href{https://doi.org/10.1016/j.cpc.2012.05.021}{\emph{Comput. Phys. Commun.}
  {\bfseries 183} (2012) 2458}
  [\href{https://arxiv.org/abs/1104.1573}{{\ttfamily 1104.1573}}].

\bibitem{Djouadi:2005gj}
A.~Djouadi, \emph{{The Anatomy of electro-weak symmetry breaking. II. The Higgs
  bosons in the minimal supersymmetric model}},
  \href{https://doi.org/10.1016/j.physrep.2007.10.005}{\emph{Phys. Rept.}
  {\bfseries 459} (2008) 1}
  [\href{https://arxiv.org/abs/hep-ph/0503173}{{\ttfamily hep-ph/0503173}}].

\bibitem{Djouadi:2005gi}
A.~Djouadi, \emph{{The Anatomy of electro-weak symmetry breaking. I: The Higgs
  boson in the standard model}},
  \href{https://doi.org/10.1016/j.physrep.2007.10.004}{\emph{Phys. Rept.}
  {\bfseries 457} (2008) 1}
  [\href{https://arxiv.org/abs/hep-ph/0503172}{{\ttfamily hep-ph/0503172}}].

\bibitem{Spira:1995rr}
M.~Spira, A.~Djouadi, D.~Graudenz and P.~M. Zerwas, \emph{{Higgs boson
  production at the LHC}},
  \href{https://doi.org/10.1016/0550-3213(95)00379-7}{\emph{Nucl. Phys. B}
  {\bfseries 453} (1995) 17}
  [\href{https://arxiv.org/abs/hep-ph/9504378}{{\ttfamily hep-ph/9504378}}].

\bibitem{Staub:2016dxq}
F.~Staub et~al., \emph{{Precision tools and models to narrow in on the 750 GeV
  diphoton resonance}},
  \href{https://doi.org/10.1140/epjc/s10052-016-4349-5}{\emph{Eur. Phys. J. C}
  {\bfseries 76} (2016) 516}
  [\href{https://arxiv.org/abs/1602.05581}{{\ttfamily 1602.05581}}].

\bibitem{Choi:2012he}
K.~Choi, S.~H. Im, K.~S. Jeong and M.~Yamaguchi, \emph{{Higgs mixing and
  diphoton rate enhancement in NMSSM models}},
  \href{https://doi.org/10.1007/JHEP02(2013)090}{\emph{JHEP} {\bfseries 02}
  (2013) 090} [\href{https://arxiv.org/abs/1211.0875}{{\ttfamily 1211.0875}}].

\bibitem{LHCHiggsCrossSectionWorkingGroup:2013rie}
{\scshape LHC Higgs Cross Section Working Group} collaboration, J.~R. Andersen
  et~al., \emph{{Handbook of LHC Higgs Cross Sections: 3. Higgs Properties}},
  \href{https://arxiv.org/abs/1307.1347}{{\ttfamily 1307.1347}}.

\bibitem{Barbieri:2013nka}
R.~Barbieri, D.~Buttazzo, K.~Kannike, F.~Sala and A.~Tesi, \emph{{One or more
  Higgs bosons?}},
  \href{https://doi.org/10.1103/PhysRevD.88.055011}{\emph{Phys. Rev. D}
  {\bfseries 88} (2013) 055011}
  [\href{https://arxiv.org/abs/1307.4937}{{\ttfamily 1307.4937}}].

\bibitem{ATLAS:2021moa}
{\scshape ATLAS} collaboration, G.~Aad et~al., \emph{{Search for
  chargino--neutralino pair production in final states with three leptons and
  missing transverse momentum in $\sqrt{s} = 13$ TeV $pp$ collisions with the
  ATLAS detector}},
  \href{https://doi.org/10.1140/epjc/s10052-021-09749-7}{\emph{Eur. Phys. J. C}
  {\bfseries 81} (2021) 1118}
  [\href{https://arxiv.org/abs/2106.01676}{{\ttfamily 2106.01676}}].

\bibitem{SARAH_Staub2008}
F.~Staub, \emph{{SARAH}},  \href{https://arxiv.org/abs/0806.0538}{{\ttfamily
  0806.0538}}.

\bibitem{SARAH3_Staub2012}
F.~Staub, \emph{{SARAH 3.2: Dirac Gauginos, UFO output, and more}},
  \href{https://doi.org/10.1016/j.cpc.2013.02.019}{\emph{Comput. Phys. Commun.}
  {\bfseries 184} (2013) 1792}
  [\href{https://arxiv.org/abs/1207.0906}{{\ttfamily 1207.0906}}].

\bibitem{SARAH4_Staub2013}
F.~Staub, \emph{{SARAH 4 : A tool for (not only SUSY) model builders}},
  \href{https://doi.org/10.1016/j.cpc.2014.02.018}{\emph{Comput. Phys. Commun.}
  {\bfseries 185} (2014) 1773}
  [\href{https://arxiv.org/abs/1309.7223}{{\ttfamily 1309.7223}}].

\bibitem{SARAH_Staub2015}
F.~Staub, \emph{{Exploring new models in all detail with SARAH}},
  \href{https://doi.org/10.1155/2015/840780}{\emph{Adv. High Energy Phys.}
  {\bfseries 2015} (2015) 840780}
  [\href{https://arxiv.org/abs/1503.04200}{{\ttfamily 1503.04200}}].

\bibitem{Porod:2014xia}
W.~Porod, F.~Staub and A.~Vicente, \emph{{A Flavor Kit for BSM models}},
  \href{https://doi.org/10.1140/epjc/s10052-014-2992-2}{\emph{Eur. Phys. J. C}
  {\bfseries 74} (2014) 2992}
  [\href{https://arxiv.org/abs/1405.1434}{{\ttfamily 1405.1434}}].

\bibitem{Belanger2002}
G.~Belanger, F.~Boudjema, A.~Pukhov and A.~Semenov, \emph{{MicrOMEGAs: A
  Program for calculating the relic density in the MSSM}},
  \href{https://doi.org/10.1016/S0010-4655(02)00596-9}{\emph{Comput. Phys.
  Commun.} {\bfseries 149} (2002) 103}
  [\href{https://arxiv.org/abs/hep-ph/0112278}{{\ttfamily hep-ph/0112278}}].

\bibitem{Belanger2004}
G.~Bélanger, F.~Boudjema, A.~Pukhov and A.~Semenov, \emph{{micrOMEGAs: Version
  1.3}}, \href{https://doi.org/10.1016/j.cpc.2005.12.005}{\emph{Comput. Phys.
  Commun.} {\bfseries 174} (2006) 577}
  [\href{https://arxiv.org/abs/hep-ph/0405253}{{\ttfamily hep-ph/0405253}}].

\bibitem{Belanger2005}
G.~Bélanger, F.~Boudjema, C.~Hugonie, A.~Pukhov and A.~Semenov, \emph{{Relic
  density of dark matter in the NMSSM}},
  \href{https://doi.org/10.1088/1475-7516/2005/09/001}{\emph{JCAP} {\bfseries
  09} (2005) 001} [\href{https://arxiv.org/abs/hep-ph/0505142}{{\ttfamily
  hep-ph/0505142}}].

\bibitem{Belanger2006}
G.~Bélanger, F.~Boudjema, A.~Pukhov and A.~Semenov, \emph{{MicrOMEGAs 2.0: A
  Program to calculate the relic density of dark matter in a generic model}},
  \href{https://doi.org/10.1016/j.cpc.2006.11.008}{\emph{Comput. Phys. Commun.}
  {\bfseries 176} (2007) 367}
  [\href{https://arxiv.org/abs/hep-ph/0607059}{{\ttfamily hep-ph/0607059}}].

\bibitem{BelangerRD2006qa}
G.~Bélanger, F.~Boudjema, S.~Kraml, A.~Pukhov and A.~Semenov, \emph{{Relic
  density of neutralino dark matter in the MSSM with CP violation}},
  \href{https://doi.org/10.1103/PhysRevD.73.115007}{\emph{Phys. Rev. D}
  {\bfseries 73} (2006) 115007}
  [\href{https://arxiv.org/abs/hep-ph/0604150}{{\ttfamily hep-ph/0604150}}].

\bibitem{Belanger2008}
G.~Bélanger, F.~Boudjema, A.~Pukhov and A.~Semenov, \emph{{Dark matter direct
  detection rate in a generic model with micrOMEGAs 2.2}},
  \href{https://doi.org/10.1016/j.cpc.2008.11.019}{\emph{Comput. Phys. Commun.}
  {\bfseries 180} (2009) 747}
  [\href{https://arxiv.org/abs/0803.2360}{{\ttfamily 0803.2360}}].

\bibitem{Belanger2010pz}
G.~Belanger, F.~Boudjema, A.~Pukhov and A.~Semenov, \emph{{micrOMEGAs: A Tool
  for dark matter studies}},
  \href{https://doi.org/10.1393/ncc/i2010-10591-3}{\emph{Nuovo Cim. C}
  {\bfseries 033N2} (2010) 111}
  [\href{https://arxiv.org/abs/1005.4133}{{\ttfamily 1005.4133}}].

\bibitem{Belanger2013}
{Bélanger, G. and Boudjema, F. and Pukhov, A. and Semenov, A.},
  \emph{{micrOMEGAs$\_$3: A program for calculating dark matter observables}},
  \href{https://doi.org/10.1016/j.cpc.2013.10.016}{\emph{Comput. Phys. Commun.}
  {\bfseries 185} (2014) 960}
  [\href{https://arxiv.org/abs/1305.0237}{{\ttfamily 1305.0237}}].

\bibitem{Barducci2016pcb}
D.~Barducci, G.~Belanger, J.~Bernon, F.~Boudjema, J.~Da~Silva, S.~Kraml et~al.,
  \emph{{Collider limits on new physics within micrOMEGAs$\_$4.3}},
  \href{https://doi.org/10.1016/j.cpc.2017.08.028}{\emph{Comput. Phys. Commun.}
  {\bfseries 222} (2018) 327}
  [\href{https://arxiv.org/abs/1606.03834}{{\ttfamily 1606.03834}}].

\bibitem{Belanger2018}
G.~Bélanger, F.~Boudjema, A.~Goudelis, A.~Pukhov and B.~Zaldivar,
  \emph{{micrOMEGAs5.0 : Freeze-in}},
  \href{https://doi.org/10.1016/j.cpc.2018.04.027}{\emph{Comput. Phys. Commun.}
  {\bfseries 231} (2018) 173}
  [\href{https://arxiv.org/abs/1801.03509}{{\ttfamily 1801.03509}}].

\bibitem{Fowlie:2016hew}
A.~Fowlie and M.~H. Bardsley, \emph{{Superplot: a graphical interface for
  plotting and analysing MultiNest output}},
  \href{https://doi.org/10.1140/epjp/i2016-16391-0}{\emph{Eur. Phys. J. Plus}
  {\bfseries 131} (2016) 391}
  [\href{https://arxiv.org/abs/1603.00555}{{\ttfamily 1603.00555}}].

\bibitem{MultiNest2009}
F.~Feroz, M.~P. Hobson and M.~Bridges, \emph{{MultiNest: an efficient and
  robust Bayesian inference tool for cosmology and particle physics}},
  \href{https://doi.org/10.1111/j.1365-2966.2009.14548.x}{\emph{Mon. Not. Roy.
  Astron. Soc.} {\bfseries 398} (2009) 1601}
  [\href{https://arxiv.org/abs/0809.3437}{{\ttfamily 0809.3437}}].

\bibitem{Importance2019}
F.~Feroz, M.~P. Hobson, E.~Cameron and A.~N. Pettitt, \emph{{Importance Nested
  Sampling and the MultiNest Algorithm}},
  \href{https://doi.org/10.21105/astro.1306.2144}{\emph{Open J. Astrophys.}
  {\bfseries 2} (2019) 10} [\href{https://arxiv.org/abs/1306.2144}{{\ttfamily
  1306.2144}}].

\bibitem{CMS:2020bfa}
{\scshape CMS} collaboration, A.~M. Sirunyan et~al., \emph{{Search for
  supersymmetry in final states with two oppositely charged same-flavor leptons
  and missing transverse momentum in proton-proton collisions at $\sqrt{s} =$
  13 TeV}}, \href{https://doi.org/10.1007/JHEP04(2021)123}{\emph{JHEP}
  {\bfseries 04} (2021) 123}
  [\href{https://arxiv.org/abs/2012.08600}{{\ttfamily 2012.08600}}].

\bibitem{CMS:2018szt}
{\scshape CMS} collaboration, A.~M. Sirunyan et~al., \emph{{Combined search for
  electroweak production of charginos and neutralinos in proton-proton
  collisions at $\sqrt{s} =$ 13 TeV}},
  \href{https://doi.org/10.1007/JHEP03(2018)160}{\emph{JHEP} {\bfseries 03}
  (2018) 160} [\href{https://arxiv.org/abs/1801.03957}{{\ttfamily
  1801.03957}}].

\bibitem{CMS:2017moi}
{\scshape CMS} collaboration, A.~M. Sirunyan et~al., \emph{{Search for
  electroweak production of charginos and neutralinos in multilepton final
  states in proton-proton collisions at $\sqrt{s}=$ 13 TeV}},
  \href{https://doi.org/10.1007/JHEP03(2018)166}{\emph{JHEP} {\bfseries 03}
  (2018) 166} [\href{https://arxiv.org/abs/1709.05406}{{\ttfamily
  1709.05406}}].

\bibitem{ATLAS:2018ojr}
{\scshape ATLAS} collaboration, M.~Aaboud et~al., \emph{{Search for electroweak
  production of supersymmetric particles in final states with two or three
  leptons at $\sqrt{s}=13\,$TeV with the ATLAS detector}},
  \href{https://doi.org/10.1140/epjc/s10052-018-6423-7}{\emph{Eur. Phys. J. C}
  {\bfseries 78} (2018) 995}
  [\href{https://arxiv.org/abs/1803.02762}{{\ttfamily 1803.02762}}].

\bibitem{ATLAS:2018eui}
{\scshape ATLAS} collaboration, M.~Aaboud et~al., \emph{{Search for
  chargino-neutralino production using recursive jigsaw reconstruction in final
  states with two or three charged leptons in proton-proton collisions at
  $\sqrt{s}=13$ TeV with the ATLAS detector}},
  \href{https://doi.org/10.1103/PhysRevD.98.092012}{\emph{Phys. Rev. D}
  {\bfseries 98} (2018) 092012}
  [\href{https://arxiv.org/abs/1806.02293}{{\ttfamily 1806.02293}}].

\bibitem{ATLAS:2020pgy}
{\scshape ATLAS} collaboration, G.~Aad et~al., \emph{{Search for direct
  production of electroweakinos in final states with one lepton, missing
  transverse momentum and a Higgs boson decaying into two $b$-jets in $pp$
  collisions at $\sqrt{s}=13$ TeV with the ATLAS detector}},
  \href{https://doi.org/10.1140/epjc/s10052-020-8050-3}{\emph{Eur. Phys. J. C}
  {\bfseries 80} (2020) 691}
  [\href{https://arxiv.org/abs/1909.09226}{{\ttfamily 1909.09226}}].

\bibitem{ATLAS:2018qmw}
{\scshape ATLAS} collaboration, M.~Aaboud et~al., \emph{{Search for chargino
  and neutralino production in final states with a Higgs boson and missing
  transverse momentum at $\sqrt{s} = 13$ TeV with the ATLAS detector}},
  \href{https://doi.org/10.1103/PhysRevD.100.012006}{\emph{Phys. Rev. D}
  {\bfseries 100} (2019) 012006}
  [\href{https://arxiv.org/abs/1812.09432}{{\ttfamily 1812.09432}}].

\bibitem{CMS:2017kxn}
{\scshape CMS} collaboration, A.~M. Sirunyan et~al., \emph{{Search for new
  phenomena in final states with two opposite-charge, same-flavor leptons,
  jets, and missing transverse momentum in pp collisions at $ \sqrt{s}=13 $
  TeV}}, \href{https://doi.org/10.1007/s13130-018-7845-2}{\emph{JHEP}
  {\bfseries 03} (2018) 076}
  [\href{https://arxiv.org/abs/1709.08908}{{\ttfamily 1709.08908}}].

\bibitem{CMS:2017bki}
{\scshape CMS} collaboration, A.~M. Sirunyan et~al., \emph{{Search for
  supersymmetry with Higgs boson to diphoton decays using the razor variables
  at $\sqrt{s} = $ 13 TeV}},
  \href{https://doi.org/10.1016/j.physletb.2017.12.069}{\emph{Phys. Lett. B}
  {\bfseries 779} (2018) 166}
  [\href{https://arxiv.org/abs/1709.00384}{{\ttfamily 1709.00384}}].

\bibitem{ATLAS:2019lff}
{\scshape ATLAS} collaboration, G.~Aad et~al., \emph{{Search for electroweak
  production of charginos and sleptons decaying into final states with two
  leptons and missing transverse momentum in $\sqrt{s}=13$ TeV $pp$ collisions
  using the ATLAS detector}},
  \href{https://doi.org/10.1140/epjc/s10052-019-7594-6}{\emph{Eur. Phys. J. C}
  {\bfseries 80} (2020) 123}
  [\href{https://arxiv.org/abs/1908.08215}{{\ttfamily 1908.08215}}].

\bibitem{CMS:2018xqw}
{\scshape CMS} collaboration, A.~M. Sirunyan et~al., \emph{{Searches for pair
  production of charginos and top squarks in final states with two oppositely
  charged leptons in proton-proton collisions at $\sqrt{s}=$ 13 TeV}},
  \href{https://doi.org/10.1007/JHEP11(2018)079}{\emph{JHEP} {\bfseries 11}
  (2018) 079} [\href{https://arxiv.org/abs/1807.07799}{{\ttfamily
  1807.07799}}].

\bibitem{ATLAS:2021yqv}
{\scshape ATLAS} collaboration, G.~Aad et~al., \emph{{Search for charginos and
  neutralinos in final states with two boosted hadronically decaying bosons and
  missing transverse momentum in $pp$ collisions at $\sqrt {s}$ = 13\,\,TeV
  with the ATLAS detector}},
  \href{https://doi.org/10.1103/PhysRevD.104.112010}{\emph{Phys. Rev. D}
  {\bfseries 104} (2021) 112010}
  [\href{https://arxiv.org/abs/2108.07586}{{\ttfamily 2108.07586}}].

\bibitem{ATLAS:2018nud}
{\scshape ATLAS} collaboration, M.~Aaboud et~al., \emph{{Search for photonic
  signatures of gauge-mediated supersymmetry in 13 TeV $pp$ collisions with the
  ATLAS detector}},
  \href{https://doi.org/10.1103/PhysRevD.97.092006}{\emph{Phys. Rev. D}
  {\bfseries 97} (2018) 092006}
  [\href{https://arxiv.org/abs/1802.03158}{{\ttfamily 1802.03158}}].

\bibitem{ATLAS:2021yyr}
{\scshape ATLAS} collaboration, G.~Aad et~al., \emph{{Search for supersymmetry
  in events with four or more charged leptons in 139$fb^{-1}$ of $\sqrt{s}$ =
  13 TeV pp collisions with the ATLAS detector}},
  \href{https://doi.org/10.1007/JHEP07(2021)167}{\emph{JHEP} {\bfseries 07}
  (2021) 167} [\href{https://arxiv.org/abs/2103.11684}{{\ttfamily
  2103.11684}}].

\bibitem{ATLAS:2019lng}
{\scshape ATLAS} collaboration, G.~Aad et~al., \emph{{Searches for electroweak
  production of supersymmetric particles with compressed mass spectra in
  $\sqrt{s}=$ 13 TeV $pp$ collisions with the ATLAS detector}},
  \href{https://doi.org/10.1103/PhysRevD.101.052005}{\emph{Phys. Rev. D}
  {\bfseries 101} (2020) 052005}
  [\href{https://arxiv.org/abs/1911.12606}{{\ttfamily 1911.12606}}].

\bibitem{ATLAS:2017vat}
{\scshape ATLAS} collaboration, M.~Aaboud et~al., \emph{{Search for electroweak
  production of supersymmetric states in scenarios with compressed mass spectra
  at $\sqrt{s}=13$ TeV with the ATLAS detector}},
  \href{https://doi.org/10.1103/PhysRevD.97.052010}{\emph{Phys. Rev. D}
  {\bfseries 97} (2018) 052010}
  [\href{https://arxiv.org/abs/1712.08119}{{\ttfamily 1712.08119}}].

\bibitem{CMS:2018kag}
{\scshape CMS} collaboration, A.~M. Sirunyan et~al., \emph{{Search for new
  physics in events with two soft oppositely charged leptons and missing
  transverse momentum in proton-proton collisions at $\sqrt{s}=$ 13 TeV}},
  \href{https://doi.org/10.1016/j.physletb.2018.05.062}{\emph{Phys. Lett. B}
  {\bfseries 782} (2018) 440}
  [\href{https://arxiv.org/abs/1801.01846}{{\ttfamily 1801.01846}}].

\bibitem{HS2013xfa}
P.~Bechtle, S.~Heinemeyer, O.~St\r{a}l, T.~Stefaniak and G.~Weiglein,
  \emph{{$HiggsSignals$: Confronting arbitrary Higgs sectors with measurements
  at the Tevatron and the LHC}},
  \href{https://doi.org/10.1140/epjc/s10052-013-2711-4}{\emph{Eur. Phys. J. C}
  {\bfseries 74} (2014) 2711}
  [\href{https://arxiv.org/abs/1305.1933}{{\ttfamily 1305.1933}}].

\bibitem{HSConstraining2013hwa}
O.~St\r{a}l and T.~Stefaniak, \emph{{Constraining extended Higgs sectors with
  HiggsSignals}}, \href{https://doi.org/10.22323/1.180.0314}{\emph{PoS}
  {\bfseries EPS-HEP2013} (2013) 314}
  [\href{https://arxiv.org/abs/1310.4039}{{\ttfamily 1310.4039}}].

\bibitem{HS2014ewa}
P.~Bechtle, S.~Heinemeyer, O.~St\r{a}l, T.~Stefaniak and G.~Weiglein,
  \emph{{Probing the Standard Model with Higgs signal rates from the Tevatron,
  the LHC and a future ILC}},
  \href{https://doi.org/10.1007/JHEP11(2014)039}{\emph{JHEP} {\bfseries 11}
  (2014) 039} [\href{https://arxiv.org/abs/1403.1582}{{\ttfamily 1403.1582}}].

\bibitem{HS2020uwn}
P.~Bechtle, S.~Heinemeyer, T.~Klingl, T.~Stefaniak, G.~Weiglein and
  J.~Wittbrodt, \emph{{HiggsSignals-2: Probing new physics with precision Higgs
  measurements in the LHC 13 TeV era}},
  \href{https://doi.org/10.1140/epjc/s10052-021-08942-y}{\emph{Eur. Phys. J. C}
  {\bfseries 81} (2021) 145}
  [\href{https://arxiv.org/abs/2012.09197}{{\ttfamily 2012.09197}}].

\bibitem{HB2008jh}
P.~Bechtle, O.~Brein, S.~Heinemeyer, G.~Weiglein and K.~E. Williams,
  \emph{{HiggsBounds: Confronting Arbitrary Higgs Sectors with Exclusion Bounds
  from LEP and the Tevatron}},
  \href{https://doi.org/10.1016/j.cpc.2009.09.003}{\emph{Comput. Phys. Commun.}
  {\bfseries 181} (2010) 138}
  [\href{https://arxiv.org/abs/0811.4169}{{\ttfamily 0811.4169}}].

\bibitem{HB2011sb}
P.~Bechtle, O.~Brein, S.~Heinemeyer, G.~Weiglein and K.~E. Williams,
  \emph{{HiggsBounds 2.0.0: Confronting Neutral and Charged Higgs Sector
  Predictions with Exclusion Bounds from LEP and the Tevatron}},
  \href{https://doi.org/10.1016/j.cpc.2011.07.015}{\emph{Comput. Phys. Commun.}
  {\bfseries 182} (2011) 2605}
  [\href{https://arxiv.org/abs/1102.1898}{{\ttfamily 1102.1898}}].

\bibitem{HBHS2012lvg}
P.~Bechtle, O.~Brein, S.~Heinemeyer, O.~Stal, T.~Stefaniak, G.~Weiglein et~al.,
  \emph{{Recent Developments in HiggsBounds and a Preview of HiggsSignals}},
  \href{https://doi.org/10.22323/1.156.0024}{\emph{PoS} {\bfseries CHARGED2012}
  (2012) 024} [\href{https://arxiv.org/abs/1301.2345}{{\ttfamily 1301.2345}}].

\bibitem{HB2013wla}
P.~Bechtle, O.~Brein, S.~Heinemeyer, O.~St\r{a}l, T.~Stefaniak, G.~Weiglein
  et~al., \emph{{$\mathsf{HiggsBounds}-4$: Improved Tests of Extended Higgs
  Sectors against Exclusion Bounds from LEP, the Tevatron and the LHC}},
  \href{https://doi.org/10.1140/epjc/s10052-013-2693-2}{\emph{Eur. Phys. J. C}
  {\bfseries 74} (2014) 2693}
  [\href{https://arxiv.org/abs/1311.0055}{{\ttfamily 1311.0055}}].

\bibitem{HB2020pkv}
P.~Bechtle, D.~Dercks, S.~Heinemeyer, T.~Klingl, T.~Stefaniak, G.~Weiglein
  et~al., \emph{{HiggsBounds-5: Testing Higgs Sectors in the LHC 13 TeV Era}},
  \href{https://doi.org/10.1140/epjc/s10052-020-08557-9}{\emph{Eur. Phys. J. C}
  {\bfseries 80} (2020) 1211}
  [\href{https://arxiv.org/abs/2006.06007}{{\ttfamily 2006.06007}}].

\bibitem{Bahl:2022igd}
H.~Bahl, T.~Biek\"otter, S.~Heinemeyer, C.~Li, S.~Paasch, G.~Weiglein et~al.,
  \emph{{HiggsTools: BSM scalar phenomenology with new versions of HiggsBounds
  and HiggsSignals}},
  \href{https://doi.org/10.1016/j.cpc.2023.108803}{\emph{Comput. Phys. Commun.}
  {\bfseries 291} (2023) 108803}
  [\href{https://arxiv.org/abs/2210.09332}{{\ttfamily 2210.09332}}].

\bibitem{Planck:2018vyg}
{\scshape Planck} collaboration, N.~Aghanim et~al., \emph{{Planck 2018 results.
  VI. Cosmological parameters}},
  \href{https://doi.org/10.1051/0004-6361/201833910}{\emph{Astron. Astrophys.}
  {\bfseries 641} (2020) A6}
  [\href{https://arxiv.org/abs/1807.06209}{{\ttfamily 1807.06209}}].

\bibitem{LZ:2022lsv}
{\scshape LZ} collaboration, J.~Aalbers et~al., \emph{{First Dark Matter Search
  Results from the LUX-ZEPLIN (LZ) Experiment}},
  \href{https://doi.org/10.1103/PhysRevLett.131.041002}{\emph{Phys. Rev. Lett.}
  {\bfseries 131} (2023) 041002}
  [\href{https://arxiv.org/abs/2207.03764}{{\ttfamily 2207.03764}}].

\bibitem{LZ2024slides}
S.~Haselschwardt, ``New dark matter search results from the lux-zeplin (lz)
  experiment.''
  \url{https://indico.uchicago.edu/event/427/contributions/1325/attachments/359/548/lz_results_tevpa.pdf},
  2024.

\bibitem{LZ:2018qzl}
{\scshape LZ} collaboration, D.~S. Akerib et~al., \emph{{Projected WIMP
  sensitivity of the LUX-ZEPLIN dark matter experiment}},
  \href{https://doi.org/10.1103/PhysRevD.101.052002}{\emph{Phys. Rev. D}
  {\bfseries 101} (2020) 052002}
  [\href{https://arxiv.org/abs/1802.06039}{{\ttfamily 1802.06039}}].

\bibitem{Fermi-LAT:2015att}
{\scshape Fermi-LAT} collaboration, M.~Ackermann et~al., \emph{{Searching for
  Dark Matter Annihilation from Milky Way Dwarf Spheroidal Galaxies with Six
  Years of Fermi Large Area Telescope Data}},
  \href{https://doi.org/10.1103/PhysRevLett.115.231301}{\emph{Phys. Rev. Lett.}
  {\bfseries 115} (2015) 231301}
  [\href{https://arxiv.org/abs/1503.02641}{{\ttfamily 1503.02641}}].

\bibitem{pdg2018}
{\scshape Particle Data Group} collaboration, M.~Tanabashi et~al.,
  \emph{{Review of Particle Physics}},
  \href{https://doi.org/10.1103/PhysRevD.98.030001}{\emph{Phys. Rev. D}
  {\bfseries 98} (2018) 030001}.

\bibitem{ParticleDataGroup:2024cfk}
{\scshape Particle Data Group} collaboration, S.~Navas et~al., \emph{{Review of
  particle physics}},
  \href{https://doi.org/10.1103/PhysRevD.110.030001}{\emph{Phys. Rev. D}
  {\bfseries 110} (2024) 030001}.

\bibitem{Hollik:2018wrr}
W.~G. Hollik, G.~Weiglein and J.~Wittbrodt, \emph{{Impact of Vacuum Stability
  Constraints on the Phenomenology of Supersymmetric Models}},
  \href{https://doi.org/10.1007/JHEP03(2019)109}{\emph{JHEP} {\bfseries 03}
  (2019) 109} [\href{https://arxiv.org/abs/1812.04644}{{\ttfamily
  1812.04644}}].

\bibitem{VPP2014}
B.~O'Leary and J.~E. Camargo-Molina, ``$\textsf{VevaciousPlusPlus}$.''
  \url{https://github.com/JoseEliel/VevaciousPlusPlus}, 2014.

\bibitem{Camargo-Molina:2013qva}
J.~E. Camargo-Molina, B.~O'Leary, W.~Porod and F.~Staub,
  \emph{{$\mathbf{Vevacious}$: A Tool For Finding The Global Minima Of One-Loop
  Effective Potentials With Many Scalars}},
  \href{https://doi.org/10.1140/epjc/s10052-013-2588-2}{\emph{Eur. Phys. J. C}
  {\bfseries 73} (2013) 2588}
  [\href{https://arxiv.org/abs/1307.1477}{{\ttfamily 1307.1477}}].

\bibitem{ATLAS:2024qxh}
{\scshape ATLAS} collaboration, G.~Aad et~al., \emph{{A statistical combination
  of ATLAS Run 2 searches for charginos and neutralinos at the LHC}},
  \href{https://arxiv.org/abs/2402.08347}{{\ttfamily 2402.08347}}.

\bibitem{Khosa:2020zar}
C.~K. Khosa, S.~Kraml, A.~Lessa, P.~Neuhuber and W.~Waltenberger,
  \emph{{SModelS Database Update v1.2.3}},
  \href{https://doi.org/10.31526/lhep.2020.158}{\emph{LHEP} {\bfseries 2020}
  (2020) 158} [\href{https://arxiv.org/abs/2005.00555}{{\ttfamily
  2005.00555}}].

\bibitem{Drees:2013wra}
M.~Drees, H.~Dreiner, D.~Schmeier, J.~Tattersall and J.~S. Kim,
  \emph{{CheckMATE: Confronting your Favourite New Physics Model with LHC
  Data}}, \href{https://doi.org/10.1016/j.cpc.2014.10.018}{\emph{Comput. Phys.
  Commun.} {\bfseries 187} (2015) 227}
  [\href{https://arxiv.org/abs/1312.2591}{{\ttfamily 1312.2591}}].

\bibitem{Dercks:2016npn}
D.~Dercks, N.~Desai, J.~S. Kim, K.~Rolbiecki, J.~Tattersall and T.~Weber,
  \emph{{CheckMATE 2: From the model to the limit}},
  \href{https://doi.org/10.1016/j.cpc.2017.08.021}{\emph{Comput. Phys. Commun.}
  {\bfseries 221} (2017) 383}
  [\href{https://arxiv.org/abs/1611.09856}{{\ttfamily 1611.09856}}].

\bibitem{Kim:2015wza}
J.~S. Kim, D.~Schmeier, J.~Tattersall and K.~Rolbiecki, \emph{{A framework to
  create customised LHC analyses within CheckMATE}},
  \href{https://doi.org/10.1016/j.cpc.2015.06.002}{\emph{Comput. Phys. Commun.}
  {\bfseries 196} (2015) 535}
  [\href{https://arxiv.org/abs/1503.01123}{{\ttfamily 1503.01123}}].

\bibitem{Cao:2022chy}
J.~Cao, J.~Lian, Y.~Pan, Y.~Yue and D.~Zhang, \emph{{Impact of recent (g
  \ensuremath{-} 2)$_{\mu}$ measurement on the light CP-even Higgs scenario in
  general Next-to-Minimal Supersymmetric Standard Model}},
  \href{https://doi.org/10.1007/JHEP03(2022)203}{\emph{JHEP} {\bfseries 03}
  (2022) 203} [\href{https://arxiv.org/abs/2201.11490}{{\ttfamily
  2201.11490}}].

\bibitem{Cao:2021tuh}
J.~Cao, J.~Lian, Y.~Pan, D.~Zhang and P.~Zhu, \emph{{Improved (g \ensuremath{-}
  2)$_{\mu}$ measurement and singlino dark matter in \ensuremath{\mu}-term
  extended \ensuremath{\mathbb{Z}}$_{3}$-NMSSM}},
  \href{https://doi.org/10.1007/JHEP09(2021)175}{\emph{JHEP} {\bfseries 09}
  (2021) 175} [\href{https://arxiv.org/abs/2104.03284}{{\ttfamily
  2104.03284}}].

\bibitem{Huang:2014xua}
P.~Huang and C.~E.~M. Wagner, \emph{{Blind Spots for neutralino Dark Matter in
  the MSSM with an intermediate $m_A$}},
  \href{https://doi.org/10.1103/PhysRevD.90.015018}{\emph{Phys. Rev. D}
  {\bfseries 90} (2014) 015018}
  [\href{https://arxiv.org/abs/1404.0392}{{\ttfamily 1404.0392}}].

\bibitem{Crivellin:2015bva}
A.~Crivellin, M.~Hoferichter, M.~Procura and L.~C. Tunstall, \emph{{Light
  stops, blind spots, and isospin violation in the MSSM}},
  \href{https://doi.org/10.1007/JHEP07(2015)129}{\emph{JHEP} {\bfseries 07}
  (2015) 129} [\href{https://arxiv.org/abs/1503.03478}{{\ttfamily
  1503.03478}}].

\bibitem{Han:2016qtc}
T.~Han, F.~Kling, S.~Su and Y.~Wu, \emph{{Unblinding the dark matter blind
  spots}}, \href{https://doi.org/10.1007/JHEP02(2017)057}{\emph{JHEP}
  {\bfseries 02} (2017) 057}
  [\href{https://arxiv.org/abs/1612.02387}{{\ttfamily 1612.02387}}].

\bibitem{Carena:2018nlf}
M.~Carena, J.~Osborne, N.~R. Shah and C.~E.~M. Wagner, \emph{{Supersymmetry and
  LHC Missing Energy Signals}},
  \href{https://doi.org/10.1103/PhysRevD.98.115010}{\emph{Phys. Rev. D}
  {\bfseries 98} (2018) 115010}
  [\href{https://arxiv.org/abs/1809.11082}{{\ttfamily 1809.11082}}].

\bibitem{Billard:2013qya}
J.~Billard, L.~Strigari and E.~Figueroa-Feliciano, \emph{{Implication of
  neutrino backgrounds on the reach of next generation dark matter direct
  detection experiments}},
  \href{https://doi.org/10.1103/PhysRevD.89.023524}{\emph{Phys. Rev. D}
  {\bfseries 89} (2014) 023524}
  [\href{https://arxiv.org/abs/1307.5458}{{\ttfamily 1307.5458}}].

\bibitem{Muhlleitner:2020wwk}
M.~M\"uhlleitner, M.~O.~P. Sampaio, R.~Santos and J.~Wittbrodt,
  \emph{{ScannerS: parameter scans in extended scalar sectors}},
  \href{https://doi.org/10.1140/epjc/s10052-022-10139-w}{\emph{Eur. Phys. J. C}
  {\bfseries 82} (2022) 198}
  [\href{https://arxiv.org/abs/2007.02985}{{\ttfamily 2007.02985}}].

\bibitem{Iguro:2022dok}
S.~Iguro, T.~Kitahara and Y.~Omura, \emph{{Scrutinizing the 95\textendash{}100
  GeV di-tau excess in the top associated process}},
  \href{https://doi.org/10.1140/epjc/s10052-022-11028-y}{\emph{Eur. Phys. J. C}
  {\bfseries 82} (2022) 1053}
  [\href{https://arxiv.org/abs/2205.03187}{{\ttfamily 2205.03187}}].

\bibitem{ATLAS:2022yrq}
{\scshape ATLAS} collaboration, G.~Aad et~al., \emph{{Measurements of Higgs
  boson production cross-sections in the~$H\to\tau^{+}\tau^{-}$ decay channel
  in pp collisions at $ \sqrt{s} $ = 13 TeV with the ATLAS detector}},
  \href{https://doi.org/10.1007/JHEP08(2022)175}{\emph{JHEP} {\bfseries 08}
  (2022) 175} [\href{https://arxiv.org/abs/2201.08269}{{\ttfamily
  2201.08269}}].

\bibitem{CMS:2024ulc}
{\scshape CMS} collaboration, A.~Tumasyan et~al., \emph{{Search for a scalar or
  pseudoscalar dilepton resonance produced in association with a massive vector
  boson or top quark-antiquark pair in multilepton events at s=13\,\,TeV}},
  \href{https://doi.org/10.1103/PhysRevD.110.012013}{\emph{Phys. Rev. D}
  {\bfseries 110} (2024) 012013}
  [\href{https://arxiv.org/abs/2402.11098}{{\ttfamily 2402.11098}}].

\end{thebibliography}\endgroup

\end{document}